\documentclass[12pt]{article}
\usepackage{amsfonts}
\usepackage{amsmath}
\usepackage{amssymb}
\usepackage{graphicx}
\usepackage{color}
\usepackage[all, knot]{xy}
\usepackage{tikz}

\usepackage{ulem}
\usepackage{hyperref}

\usepackage[utf8]{inputenc}
\usepackage{epstopdf}
\usepackage[footnotesize]{caption}
\usepackage{amsthm}
\usepackage{enumitem}
\usepackage{mathrsfs}

\usepackage[margin=3cm]{geometry}

\def \be {\begin{equation}}
\def \ee {\end{equation}}
\def \bea {\begin{eqnarray}}
\def \eea {\end{eqnarray}}
\def \nn {\nonumber}

\def \rr {\raise.35ex\hbox{\small $\prime$}\kern-.17em{\mbox{\large $\imath$}}}

\def \dels {\partial\kern-.6em /\kern.1em}
\def \As {{A\kern-.5em / \kern.5em}}
\def \Ds {D\kern-.7em / \kern.5em}

\def \ks {k\kern-.5em /}
\def \ls {l\kern-.5em /}

\def \sgn {\mbox{\small sgn}}

\newcommand{\ci}[1]{}
\newcommand{\ba}{\begin{eqnarray}}
\newcommand{\ea}{\end{eqnarray}}
\newcommand{\bal}{\begin{align}}
\newcommand{\eal}{\end{align}}
\newcommand{\bay}[1]{\left(\begin{array}{#1}}
\newcommand{\eay}{\end{array}\right)}

\newcommand{\hide}[1]{}

\newlist{axioms}{enumerate}{2}
\setlist[axioms,1]{label=\textbf{A\arabic{axiomsi}.}, ref=A\arabic{axiomsi}}
\setlist[axioms,2]{label=\textbf{A\arabic{axiomsi}\rlap{\myEnumCounter{axiomsii}}.},%
                   ref=A\arabic{axiomsi}\myEnumCounter{axiomsii},%
                   align=parleft,%
                   leftmargin=0em,%
                   itemsep=1.4ex,%
                   before={\stepcounter{axiomsi}}}

  \usetikzlibrary{decorations.markings}

\begin{document}

\begin{titlepage}
\begin{center}

\textbf{\LARGE
Non-Commutative Geometry for D-Branes in Large R-R Field Background
\vskip.3cm
}
\vskip .5in
{\large
Chen-Te Ma \footnote{e-mail address: yefgst@gmail.com}
\\
\vskip 1mm
}
{\sl
Department of Physics, Great Bay University, Dongguan, Guangdong 52300, China. 
}\\
\vskip 1mm
\vspace{40pt}
\end{center}

%\newpage
\begin{abstract} 
\noindent
We examine the role of non-commutative geometry in D$p$-branes within large R-R field backgrounds.
In this context, the background of a significant ($p-1$)-form R-R field can be effectively described using a ($p-1$)-bracket, similar to the method used in the NS-NS case.
We begin by recalling how non-commutative geometry arises from the quantization of open string theory.
In this framework, the Seiberg-Witten map is a key element that establishes the equivalence between commutative and non-commutative descriptions in the low-energy effective theory.
The Poisson bracket characterizes non-commutative structures, with deformation achieved through the Moyal product.
Next, we show how the Nambu-Poisson bracket emerges in the context of a single D4-brane with the large R-R field background limit, starting from the BLG model.
The generalization to a D$p$-brane leads to the ($p-1$)-bracket description, which reveals a duality web relating NS-NS and R-R field backgrounds via T-duality and S-duality in the low-energy limit.
Finally, we extend the single D-brane construction to multiple D-branes by promoting the ordinary product in the bracket to a covariant derivative at the Poisson level.
\end{abstract}
\end{titlepage}

\section{Introduction}
\label{sec:1}
\noindent
String theory is a generalization of point-particle physics to one-dimensional objects called strings.
These strings propagate through a two-dimensional worldsheet.
At huge distance scales, strings behave effectively as point-like particles.
The vibrational modes of a string determine the properties of the corresponding particle, and one such mode corresponds to the graviton.
Since the graviton mediates the gravitational force, string theory provides a quantum mechanical framework that naturally incorporates gravity.
Hence, string theory is regarded as a candidate for a theory of quantum gravity.
\\

\noindent
A remarkable feature of string theory is that its seemingly different formulations are deeply interconnected.
One such connection is target-space duality (T-duality).
For example, when a string propagates around a circle of radius $R$, there exists an equivalent description in which it propagates around a circle of radius $1/R$.
Another fundamental relation among string theories is strong–weak duality (S-duality).
This duality asserts that a strongly coupled theory can admit an equivalent description in terms of a weakly coupled one.
S-duality generalizes the familiar electromagnetic duality that interchanges electric and magnetic fields.
More broadly, these dualities reveal that distinct physical descriptions may, in fact, represent the same underlying physics.
\\

\noindent
In string theory, strings can be either {\it closed}, forming loops, or {\it open}, forming line segments with two endpoints.
A fundamental class of extended objects, known as Dirichlet membranes (D-branes), provides surfaces on which open strings can end under {\it Dirichlet} boundary conditions.
As an open string propagates through spacetime, its endpoints are constrained to lie on a {\it D-brane}.
The "D" in D-brane explicitly refers to these Dirichlet boundary conditions.
\\

\noindent
Because T-duality interchanges Neumann and Dirichlet boundary conditions, it implies that D-branes are an {\it essential} feature of open string theory.
In $p$ spatial dimensions, these objects are called D$p$-branes.
More generally, a brane is a dynamical object that extends the notion of a point particle (zero-dimensional) and a string (one-dimensional) to higher dimensions.
A D$p$-brane sweeps out a ($p+1$)-dimensional worldvolume in spacetime, and its dynamics are governed by the Dirac–Born–Infeld (DBI) action \cite{Fradkin:1985qd,Callan:1986bc}, a highly symmetric theory that generalizes Maxwell’s electrodynamics.
\\

\noindent
On the D$p$-brane worldvolume, a gauge theory lives whose spectrum includes a photon-like excitation: a $p$-dimensional analogue of the electromagnetic field, satisfying a generalized version of Maxwell’s equations.
Because D-branes inherently support such gauge fields, they are indispensable in a consistent formulation of open string theory. Furthermore, a D$p$-brane embedded in a spacetime of $d$ spatial dimensions carries $d-p$ massless scalar fields.
These scalars correspond to Goldstone modes associated with the spontaneous breaking of translational symmetry, and they encode the brane’s {\it transverse} fluctuations.
\\

\noindent
At the quantum level, the gauge field living on a single D-brane is associated with a U(1) gauge symmetry.
When $N$ D-branes coincide, the system supports a U(N) gauge theory \cite{Tseytlin:1997csa,Terashima:2000ej}, which provides a natural realization of {\it non-Abelian} gauge symmetries \cite{Yang:1954ek,Cronstrom:2005wt} within string theory.
The discovery that D-branes are sources of electric and magnetic Ramond–Ramond (R–R) fields revealed their central role in the {\it non-perturbative} structure of string theory.
Another profound consequence arises from the quantization of open strings in the presence of background fields, which leads to a {\it non-commutative} structure of the target spacetime at the string {\it endpoints} \cite{Chu:1998qz}.
This motivates the study of {\it non-commutative geometry} in the context of string theory.
The Seiberg–Witten (SW) map \cite{Seiberg:1999vs,Cornalba:1999ah,Okawa:1999cm,Asakawa:1999cu,Ishibashi:1999vi,Terashima:1999tn,Andreev:1999pv,Chu:1999ij} establishes an explicit correspondence between {\it commutative} and {\it non-commutative} gauge theories, ensuring the consistency of gauge transformations in both frameworks.
In non-commutative gauge theory \cite{Seiberg:1999vs,Cornalba:1999ah,Okawa:1999cm,Asakawa:1999cu,Ishibashi:1999vi,Terashima:1999tn,Andreev:1999pv,Chu:1999ij}, the usual product of functions is replaced by the {\it Moyal} product, which systematically incorporates all-order $\alpha'$ corrections \cite{Cornalba:1999hn}.
This formalism provides a convenient setting for analyzing symmetries and exploring new {\it non-perturbative} aspects of string theory.
\\

\noindent
String theory consists of five consistent versions of superstring theory: Type I, Type IIA, Type IIB, E$_8$$\times$E$_8$ heterotic, and SO(32) heterotic.
A theory of strings that incorporates supersymmetry is known as a superstring theory, and it automatically provides a quantum framework that encompasses gravity.
Each version exhibits distinct spectra of low-energy particles and symmetries.
For example, Type I string theory includes both open and closed strings. In contrast, Type IIA and Type IIB contain only closed strings.
The realization that these five theories are different limits of a more fundamental eleven-dimensional theory, known as {\it M-theory}, provided a unifying framework that incorporates the dualities and extended objects (branes) central to string theory.
\\

\noindent
Theories arising from {\it different} limits of M-theory are deeply interconnected.
For instance, Type IIA and Type IIB string theories are related by T-duality, which also interchanges the two heterotic theories. Meanwhile, Type I string theory is related by S-duality to the SO(32) heterotic theory, and Type IIB is self-dual under S-duality \cite{Cornalba:2002cu}.
\\

\noindent
M-theory should provide a framework for developing a unified theory of all of the fundamental forces of nature.
At low energies, supersymmetric string theories can be approximated by 10-dimensional supergravities.
Similarly, M-theory is approximated by 11-dimensional supergravity at low energies, as this is the highest spacetime dimension in which a consistent supersymmetric theory can be formulated.
\\

\noindent
In 11-dimensional supergravity, the brane electrically charged under the supergravity $C$-field is the membrane (M2-brane), while its electromagnetic dual is the M5-brane.
A well-known feature of multiple M2-branes is that their entropy scales as $N^{3/2}$ in the large-$N$ limit.
By contrast, gauge symmetry described by an ordinary Lie algebra corresponds to coincident $N$ D-branes, whose entropy instead scales as $N^2$.
This mismatch indicates that a Lie algebra is {\it insufficient} to capture the gauge symmetry of multiple M2-branes.
A natural generalization is the Lie $n$-algebra, defined by a totally antisymmetric $n$-linear map.
For $n=2$, the Lie $n$-algebra reduces to the standard Lie algebra.
If we truncate the Lie 3-algebra such that each dimension has $\tilde{N}$ degrees of freedom, then the number of independent generators is $\tilde{N}^3$.
When the truncated Lie 3-algebra functions depend {\it only} on two variables, they generally commute, which implies that the number of M2-branes scales as $N = \tilde{N}^2$ at large $N$.
This yields the desired $N^{3/2}$ entropy scaling law, thereby justifying the Lie 3-algebra description.
\\

\noindent
The first connection between Lie 3-algebras and M2-branes arises in the\\
Bagger–Lambert–Gustavsson (BLG) model \cite{Bagger:2006sk,Gustavsson:2007vu,Bagger:2007jr,Bagger:2007vi}, which describes the worldvolume theory of two coincident {\it M2-branes} with 16 real supercharges \cite{Berman:2008be}.
M2-branes in flat 11-dimensional spacetime are expected to be {\it half} BPS states, preserving 16 out of the 32 supercharges of 11-dimensional supergravity. 
The conditions for supersymmetry are also examined in Ref.  \cite{Berman:2009xd}. 
By relaxing the requirement of a positive-definite metric, one can extend this construction to multiple M2-branes for arbitrary finite $N$ \cite{Ho:2008ei,Berman:2009kj}.
Moreover, an {\it infinite-dimensional} Lie 3-algebra provides a framework for constructing a consistent {\it single} M5-brane theory in a {\it non-commutative} setting, derived from an {\it infinite} number of M2-branes \cite{Ho:2008nn}. 
The non-commutative geometry of the open membrane can be discovered from the Hamiltonian formulation \cite{Berman:2004jv}. 
The non-commutative M-branes ' metric is the open membrane metric analogous to the role of the open string metric \cite{Bergshoeff:2000jn,Bergshoeff:2000ai}. 
\\

\begin{figure}[h]
\begin{center}
\includegraphics[width=1.\textwidth]{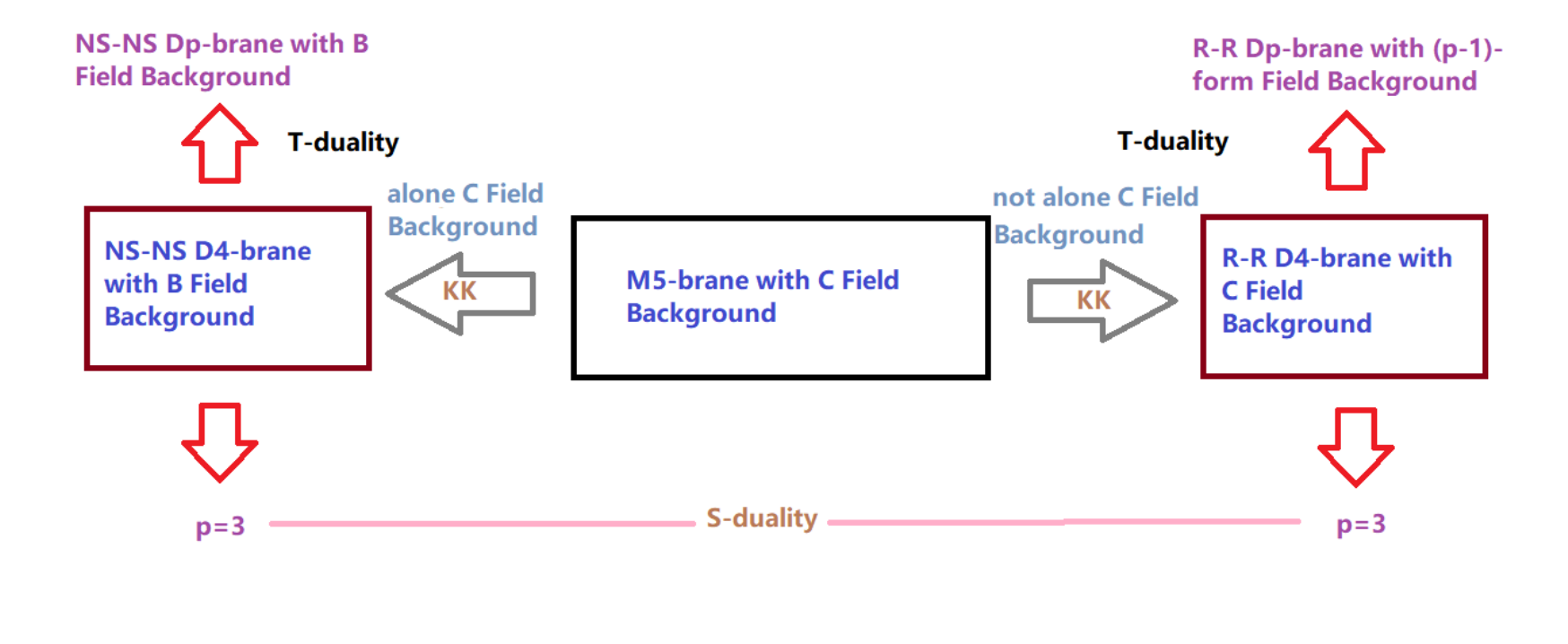}
\end{center}
\caption{The duality web for a single brane.}
\label{KK.pdf}
\end{figure}

\noindent
The dynamics of a single M5-brane in this framework are governed by the Nambu–Poisson bracket, which generates volume-preserving diffeomorphisms (VPDs) under a large $C$-field background \cite{Ho:2008nn}.
This is known as the Nambu–Poisson (NP) M5-brane theory.
The large $C$-field background {\it breaks} Lorentz symmetry, separating the worldvolume directions into those parallel and those orthogonal to the background \cite{Ho:2008nn}.
Compactifying one of the orthogonal directions yields the {\it non-commutative} D4-brane theory with a large NS–NS $B$-field background \cite{Ho:2008ve}.
Compactifying instead along a parallel direction gives the {\it non-commutative} D4-brane theory with a large R–R $C$-field background (the R–R D4-brane) \cite{Ho:2011yr,Ma:2012hc}.
Further dimensional reduction of the R–R D4-brane theory leads to the R–R D3-brane theory, which is {\it electromagnetic} dual to the NS–NS D3-brane theory \cite{Ho:2013opa,Ho:2015mfa}.
The generalization to higher dimensions is encoded in the {\it ($p-1$)-bracket}, which describes a large ($p-1$)-form field background and generates VPDs \cite{Ho:2013paa}.
The single-brane constructions summarized above form the duality web illustrated in Fig.~\ref{KK.pdf}.
Because a full Lagrangian formulation for multiple M5-branes is {\it not} yet available, multiple-brane dynamics are well-understood only in the case of D-branes \cite{Ma:2023hgi}.
There, the equivalence between commutative and non-commutative descriptions had already been established through the Seiberg–Witten (SW) map \cite{Ma:2020msx}. 
The open string parameters also had the generalized version to the R-R D-branes \cite{Berman:2001rka}. 

\subsection{Outline}
\noindent
The outline of this review is as follows.
In Sec.~\ref{sec:2} and Sec.~\ref{sec:3}, we introduce the worldsheet description of strings. This section covers the fundamental concepts and equations in string theory, including the emergence of non-commutative geometry from quantization, the derivation of the DBI action from the one-loop $\beta$-function, and dualities.
\\

\noindent
In Sec.~\ref{sec:4}, we examine the non-commutative geometry of NS–NS D-branes.
The discussion includes the Seiberg–Witten map and the formulation of non-commutative gauge theory using the Moyal product.
We then turn to the non-commutative geometry of the M5-brane in Sec.~\ref{sec:5}, beginning with the BLG model for multiple M2-branes. From this setup, we derive the non-commutative description of a single M5-brane in a large $C$-field background, constructed from infinitely many M2-branes, and show how its dynamics is governed by the Nambu–Poisson bracket.
\\

\noindent
Sec .~\ref {sec:6} presents the duality web of non-commutative brane theories.
We first demonstrate how to derive the D4-brane theory in the presence of large NS–NS and R–R field backgrounds, and then generalize the construction to D$p$-branes.
In this framework, the ($p-1$)-bracket naturally emerges from T-duality and generates volume-preserving diffeomorphisms (VPDs).
In particular, for $p=3$, we discuss the role of electromagnetic duality in connecting the NS–NS D3-brane theory with its R–R counterpart.
\\

\noindent
In Sec .~\ref {sec:7}, we extend the discussion to multiple R–R D-branes. The primary focus is to generalize the ($p-1$)-bracket to the case of non-Abelian gauge groups, while ensuring the consistent generation of VPDs required for the construction of R–R D-brane theories.
Finally, Sec .~\ref {sec:8} outlines possible future directions and open questions related to the topics covered in this review.

\section{String Theory}
\label{sec:2}
\noindent
We first introduce the necessary background knowledge of string theory.
Because this review only focuses on the bosonic sector, we will only review the bosonic string theory.
We begin with the Lagrangian formulation of the closed string and then introduce its spectrum and the T-duality transformation rule for the field background.
We then introduce the Dirichlet boundary condition to study the open string, which encompasses the non-commutativity of the target space \cite{Chu:1998qz}, the dynamics of the target space, as well as T-duality and S-duality.

\subsection{Closed String}
\noindent
We introduce the spectrum and the T-duality transformation rule in this section for bosonic string theory.
The Lagrangian description of the bosonic string theory is
\bea
S_{\mathrm{bs}}&=&-\frac{1}{4\pi\alpha^{\prime}}\int d^2\sigma\sqrt{|\det(h_{\rho\sigma})|}
\nn\\
&&\times
\bigg(h^{\alpha\beta}\partial_{\alpha}X^{\mu}g_{\mu\nu}(X)\partial_{\beta}X^{\nu}
-\epsilon^{\alpha\beta}\partial_{\alpha}X^{\mu}B_{\mu\nu}(X)\partial_{\beta}X^{\nu}
+\alpha^{\prime}R\phi(X)
\bigg),
\eea
where
\bea
d^2\sigma\equiv d\sigma^0d\sigma^1; \ \partial_{\alpha}\equiv\frac{\partial}{\partial\sigma^{\alpha}}.
\eea
The integration range of $\sigma^0$ is $(-\infty, \infty)$, and $\sigma^1$ is [0, 2$\pi$].
The $X^{\mu}$ is the target space with the periodic boundary condition
\bea
X^{\mu}[\sigma^0, \sigma^1]=X^{\mu}[\sigma^0, \sigma^1+2\pi],
\eea
$g_{\mu\nu}=g_{\nu\mu}$ is the metric field of the target space,
$h_{\alpha\beta}=h_{\beta\alpha}$ is the metric field of the worldsheet space,
$B_{\mu\nu}=-B_{\nu\mu}$ is the Kalb-Ramond field, and $\phi$ is the dilaton field.
The $\alpha^{\prime}$ is called the slope parameter, and it defines the string tension
\bea
T_s\equiv\frac{1}{2\pi\alpha^{\prime}}.
\eea
The indices of the worldsheet are denoted by the letters $\alpha$ and $\beta$, and the index of the target space are denoted by the letters $\mu$ and $\nu$.
We define the Levi-Civita symbol $\epsilon^{\alpha\beta}$ as in the following:
\bea
\epsilon^{01}=-\epsilon^{10}=1; \ \epsilon^{00}=\epsilon^{11}=0.
\eea
The measure is expressed as
\bea
\int {\cal D}X{\cal D}h.
\eea
The Ricci scalar is defined in terms of the metric tensor and the Ricci tensor.
The expression gives it
\bea
R\equiv h^{\alpha\beta}R_{\alpha\beta},
\eea
where $h^{\alpha\beta}$ is the inverse of the metric tensor and $R_{\alpha\beta}$ is the Ricci tensor,
\bea
R_{\alpha\beta}\equiv\partial_{\gamma}\Gamma^{\gamma}_{\beta\alpha}
-\partial_{\beta}\Gamma^{\gamma}_{\gamma\alpha}
+\Gamma^{\gamma}_{\gamma\delta}\Gamma^{\delta}_{\beta\alpha}
-\Gamma^{\gamma}_{\beta\delta}\Gamma^{\delta}_{\gamma\alpha},
\eea
where Christoffel symbol $\Gamma^{\alpha}_{\beta\delta}$ is
\bea
\Gamma^{\alpha}_{\beta\delta}&\equiv&\frac{1}{2}h^{\alpha\lambda}\bigg(\partial_{\delta}h_{\lambda\beta}+\partial_{\beta}h_{\lambda\delta}
-\partial_{\lambda}h_{\beta\delta}\bigg).
\eea

\subsubsection{Equations of Motion}
\noindent
In the 2D worldsheet, we can choose the Lorentzian flat metric with signature $(-, +)$.
We consider the constant background to demonstrate the quantization.
The equation of motion for the target is
\bea
\partial_0^2X^{\mu}=\partial_1^2X^{\mu}.
\eea
The generic solutions of this equation are
\bea
X=X_L(\sigma^0+\sigma^1)+X_R(\sigma^0-\sigma^1), \qquad
\tilde{X}=X_L(\sigma^0+\sigma^1)-X_R(\sigma^0-\sigma^1),
\eea
where $X$ is the target space, and $\tilde{X}$ is the dual target space. Now we define the new coordinates:
\bea
u\equiv \sigma^0+\sigma^1, \qquad v\equiv\sigma^0-\sigma^1.
\eea
Because we have the periodicity condition $\sigma^1\rightarrow\sigma^1+2\pi$ on the target space, we obtain
\bea
X_L(u)+X_R(v)&=&X_L(u+2\pi)+X_R(v-2\pi),
\nn\\
\tilde{X}_L(u)-\tilde{X}_R(v)&=&\tilde{X}_L(u+2\pi)-\tilde{X}_R(v-2\pi)
\eea
or
\bea
X_L(u+2\pi)-X_L(u)&=&X_R(v)-X_R(v-2\pi),
\nn\\
\tilde{X}_L(u+2\pi)-\tilde{X}_L(u)&=&\tilde{X}_R(v-2\pi)-\tilde{X}_R(v).
\eea
\\

\noindent
If we take the derivative with respect to $u$, we can obtain
\bea
\partial_uX_L(u+2\pi)=\partial_uX_L(u), \qquad \partial_u\tilde{X}_L(u+2\pi)=\partial_u\tilde{X}_L(u).
\eea
Similarly, we can also take the derivative with respect to $v$ to obtain
\bea
\partial_vX_R(v)=\partial_vX_R(v-2\pi), \qquad \partial_v\tilde{X}_R(v-2\pi)=\partial_v\tilde{X}_R(v-2\pi).
\eea
Therefore, the $\partial_uX_L(u)$, $\partial_vX_R(v)$, $\partial_u\tilde{X}_L(u)$, and $\partial_v\tilde{X}_R(v)$ are the periodic function with $2\pi$ periodicity. We can write the mode expansion as follows:
\bea
\partial_uX_L^{\mu}=\sqrt{\frac{\alpha^{\prime}}{2}}\sum_{n\in{\cal Z}}\bar{\alpha}_n^{\mu}e^{-inu},
\qquad
\partial_vX_R(v)=\sqrt{\frac{\alpha^{\prime}}{2}}\sum_{n\in{\cal Z}}\alpha_n^{\mu}e^{-inv},
\eea
where $\alpha_n^{\mu}$ and $\bar{\alpha}_n^{\mu}$ are constants.
Because we want the target space to be real, we impose the following condition on the constants
\bea
\alpha_{-n}^{\mu}=(\alpha_n^{\mu})^*, \qquad \bar{\alpha}_{-n}^{\mu}=(\bar{\alpha}_n^{\mu})^*.
\eea
Then we can do the integration to obtain the target space
\bea
X_L^{\mu}(u)&=&\frac{1}{2}x_{0, L}^{\mu}+\sqrt{\frac{\alpha^{\prime}}{2}}\bar{\alpha}_0^{\mu}u+i\sqrt{\frac{\alpha^{\prime}}{2}}\sum_{n\ne 0}\frac{\bar{\alpha}_n^{\mu}}{n}e^{-inu},
\nn\\
X_R^{\mu}(v)&=&\frac{1}{2}x_{0,R}^{\mu}+\sqrt{\frac{\alpha^{\prime}}{2}}\alpha_0^{\mu}v+i\sqrt{\frac{\alpha^{\prime}}{2}}\sum_{n\ne0}\frac{\alpha^{\mu}_n}{n}e^{-inv}.
\eea
Because we have a constraint on the target space
\bea
X_L(u+2\pi)-X_L(u)=X_R(v)-X_R(v-2\pi),
\eea
we get
\bea
2\pi\sqrt{\frac{\alpha^{\prime}}{2}}\bar{\alpha}_0^{\mu}=2\pi\sqrt{\frac{\alpha^{\prime}}{2}}\alpha_0^{\mu}.
\eea
Therefore, we obtain
\bea
\bar{\alpha}_0^{\mu}=\alpha_0^{\mu}.
\eea
Finally, we can obtain the target space, and the dual target space can be similarly obtained:
\bea
X^{\mu}
&=&\frac{1}{2}(x_{0,L}^{\mu}+x_{0,R}^{\mu})
+\sqrt{2\alpha^{\prime}}\alpha_0^{\mu}\sigma^0
+i\sqrt{\frac{\alpha^{\prime}}{2}}\sum_{n\ne 0}\frac{e^{-in\sigma^0}}{n}(\bar{\alpha}_n^{\mu}e^{-in\sigma^1}+\alpha_n^{\mu}e^{in\sigma^1}),
\nn\\
\tilde{X}^{\mu}
&=&\frac{1}{2}(x_{0,L}^{\mu}-x_{0,R}^{\mu})
+\sqrt{2\alpha^{\prime}}\alpha_0^{\mu}\sigma^1
+i\sqrt{\frac{\alpha^{\prime}}{2}}\sum_{n\ne 0}\frac{e^{-in\sigma^0}}{n}(\bar{\alpha}_n^{\mu}e^{-in\sigma^1}-\alpha_n^{\mu}e^{in\sigma^1}).
\eea
The dual target space should not be physical because it does not satisfy the periodicity condition.
\\

\noindent
The canonical momenta of the string sigma model are
\bea
P^{\mu}=\frac{1}{2\pi\alpha^{\prime}}\partial_{0}X^{\mu}.
\eea
Then we can find the momenta of the string:
\bea
p^{\mu}\equiv\int_0^{2\pi}d\sigma\ P^{\mu}=\sqrt{\frac{2}{\alpha^{\prime}}}\alpha_0^{\mu}.
\eea
Therefore, we find
\bea
\alpha_0^{\mu}=\sqrt{\frac{\alpha^{\prime}}{2}}p^{\mu}.
\eea
\\

\noindent
Here, we consider the Lorentzian metric
\bea
ds^2&=&-(dX^0)^2+(dX^1)^2+\cdots
=-dX^+dX^-+(dX^2)^2+(dX^3)^2+\cdots
\nn\\
&\equiv&-dX^+dX^-+dX^jdX^j,
\eea
where
\bea
X^+\equiv\frac{1}{\sqrt{2}}(X^0+X^1), \qquad X^-\equiv\frac{1}{\sqrt{2}}(X^0-X^1),
\nn
\eea
the index of transverse light-cone coordinates is labeled by $j=2, 3, \cdots, D-1$, where $D$ is the dimensions of the target space.
\\

\noindent
The equation of motion for $h_{\alpha\beta}$ is
\bea
\partial_{\alpha}X^{\mu}\partial_{\beta}X_{\mu}=\frac{1}{2}h_{\alpha\beta}\partial^{\gamma}X^{\mu}\partial_{\gamma}X_{\mu}.
\eea
When using the flat background, we find that the equation of motion implies
\bea
\partial_{0}X^{\mu}\partial_{1}X_{\mu}=0; \
\partial_{0}X^{\mu}\partial_{0}X_{\mu}+\partial_{1}X^{\mu}\partial_{1}X_{\mu}=0.
\eea
The first condition constraints that the tangent vectors along the direction $\sigma^0$,
\bea
\frac{\partial X^{\mu}}{\partial{\sigma^0}},
\eea
are orthogonal to the tangent vectors along the direction $\sigma^1$,
\bea
\frac{\partial X^{\mu}}{\partial{\sigma^1}}.
\eea
The second condition constrains the length of tangent vectors.
The above conditions are the sub-conditions of the following condition
\bea
\bigg(\frac{\partial X^{\mu}}{\partial{\sigma^0}}\frac{\partial X_{\mu}}{\partial{\sigma^0}}\pm
\frac{\partial X^{\mu}}{\partial{\sigma^1}}\frac{\partial X_{\mu}}{\partial{\sigma^1}}\bigg)^2
=0.
\label{lg1}
\eea
Therefore, we use the above condition to make the necessary adjustments.
Finally, we fix the $\sigma^0$-parameterization by choosing
\bea
X^+=\alpha^{\prime}p^{+}\sigma^0,
\label{lg2}
\eea
where $p^+$ is the momentum of the string in the $+$ direction.
This is known as the light-cone gauge.

\subsubsection{Spectrum}
\noindent
We compute the spectrum beginning from the quantization algebra related to the canonical position and momenta:
\bea
\lbrack P^{\mu}, X^{\nu}\rbrack=-i\delta^{\mu\nu}, \qquad
\lbrack P^{\mu}, P^{\nu}\rbrack=0, \qquad
\lbrack X^{\mu}, X^{\nu}\rbrack=0.
\eea
This results in the following algebraic expression:
\bea
\lbrack\alpha_m^{j_1}, \alpha_n^{j_2}\rbrack=m\delta_{m+n, 0}\eta^{j_1j_2}, \qquad
\lbrack\bar{\alpha}_m^{j_1}, \bar{\alpha}_n^{j_2}\rbrack=m\delta_{m+n, 0}\eta^{j_1j_2}, \qquad
\lbrack\alpha_m^{j_1}, \bar{\alpha}_n^{j_2}\rbrack=0.
\eea
By fixing the light-cone gauge, we obtain $\alpha_n^+=0$ for $n\neq 0$, and the $\alpha_n^-$ can be expressed by the transverse oscillators, $\alpha^j$.
From the algebra, we can find that the left-moving and right-moving oscillators are independent.
\\

\noindent
We first introduce the introduce the annihilation $a_n^{j}$ and creation operators $(a_n^{j})^{\dagger}$ by the following field redefinition:
\bea
\alpha_n^{j}=a_n^{j}\sqrt{n}, \qquad \alpha_{-n}^{j}=(a_n^{j})^{\dagger}\sqrt{n}, \qquad n\ge 1.
\eea
Therefore, we obtain:
\bea
\lbrack a_m^{j_1}, (a_n^{j_2})^{\dagger}\rbrack=\delta_{m,n}\eta^{j_1, j_2}, \qquad
\lbrack a_m^{j_1}, a_n^{j_2}\rbrack=0, \qquad
m, n\ge 1.
\eea
Note that the annihilation and creation operators are only defined on the positive integer $n>0$, not the same as the left-moving ($\alpha_m^{\mu}$) and right-moving operators ($\bar{\alpha}_m^{\mu}$).
\\

\noindent
We use the condition \eqref{lg1} written in terms of the light-cone coordinates
\bea
-2\bigg(\frac{\partial X^{+}}{\partial{\sigma^0}}\pm\frac{\partial X^{+}}{\partial \sigma^1}\bigg)
\bigg(\frac{\partial X^{-}}{\partial{\sigma^0}}\pm\frac{\partial X^{-}}{\partial \sigma^1}\bigg)
+
\bigg(\frac{\partial X^{\mu}}{\partial{\sigma^0}}\pm\frac{\partial X^{\mu}}{\partial \sigma^1}\bigg)
\bigg(\frac{\partial X_{\mu}}{\partial{\sigma^0}}\pm\frac{\partial X_{\mu}}{\partial \sigma^1}\bigg)
=0,
\eea
which this expression can be expressed in a different form as follows
\bea
\frac{\partial X^{-}}{\partial{\sigma^0}}\pm\frac{\partial X^{-}}{\partial \sigma^1}
=\frac{1}{2\alpha^{\prime}p^+}\bigg(\frac{\partial X^{\mu}}{\partial{\sigma^0}}\pm\frac{\partial X^{\mu}}{\partial \sigma^1}\bigg)
\bigg(\frac{\partial X_{\mu}}{\partial{\sigma^0}}\pm\frac{\partial X_{\mu}}{\partial \sigma^1}\bigg),
\eea
in which we used Eq. \eqref{lg2}.
From the computation results:
\bea
\frac{\partial X^{\mu}}{\partial\sigma^0}+\frac{\partial X^{\mu}}{\partial\sigma^1}
&=&\sqrt{2\alpha^{\prime}}\sum_{n\in\mathbf{Z}}\bar{\alpha}_n^{\mu}e^{-in(\sigma^0+\sigma^1)};
\nn\\
\frac{\partial X^{\mu}}{\partial\sigma^0}-\frac{\partial X^{\mu}}{\partial\sigma^1}
&=&\sqrt{2\alpha^{\prime}}\sum_{n\in\mathbf{Z}}\alpha_n^{\mu}e^{-in(\sigma^0-\sigma^1)},
\eea
we obtain:
\bea
\bigg(\frac{\partial X^{j}}{\partial{\sigma^0}}+\frac{\partial X^{j}}{\partial \sigma^1}\bigg)
\bigg(\frac{\partial X_{j}}{\partial{\sigma^0}}+\frac{\partial X_{j}}{\partial \sigma^1}\bigg)
&=&2\alpha^{\prime}\sum_{n_1,n_2\in\mathbf{Z}}\bar{\alpha}_{n_2}^{j}\bar{\alpha}_{n_1-n_2}^{j}
e^{-in_1(\sigma^0+\sigma^1)}
\nn\\
&\equiv& 4\alpha^{\prime}\sum_{n\in\mathbf{Z}}\bar{L}_ne^{-in(\sigma^0+\sigma^1)};
\nn\\
\bigg(\frac{\partial X^{j}}{\partial{\sigma^0}}-\frac{\partial X^{j}}{\partial \sigma^1}\bigg)
\bigg(\frac{\partial X_{j}}{\partial{\sigma^0}}-\frac{\partial X_{j}}{\partial \sigma^1}\bigg)
&=&2\alpha^{\prime}\sum_{n_1,n_2\in\mathbf{Z}}\alpha_{n_2}^{j}\alpha_{n_1-n_2}^{j}
e^{-in_1(\sigma^0-\sigma^1)}
\nn\\
&\equiv& 4\alpha^{\prime}\sum_{n\in\mathbf{Z}}L_ne^{-in(\sigma^0-\sigma^1)},
\eea
where
\bea
L_n\equiv\frac{1}{2}\sum_{m\in\mathbf{Z}}\alpha_m^j\alpha_{n-m}^j, \qquad
\bar{L}_n\equiv\frac{1}{2}\sum_{m\in\mathbf{Z}}\bar{\alpha}_m^j\bar{\alpha}_{n-m}^j.
\eea
The operators $L_n$ and $\bar{L}_n$ are called transverse Virasoro operators.
\\

\noindent
We can now explicitly express the $\bar{\alpha}^-$ in terms of the traversal oscillators:
\bea
\frac{\partial X^-}{\partial\sigma^0}+\frac{\partial X^-}{\partial\sigma^1}
&=&\frac{2}{p^+}\sum_{n\in\mathbf{Z}}\bar{L}_ne^{-in(\sigma^0+\sigma^1)}
=\sqrt{2\alpha^{\prime}}\sum_{n\in\mathbf{Z}}\bar{\alpha}_n^-e^{-in(\sigma^0+\sigma^1)};
\nn\\
\frac{\partial X^-}{\partial\sigma^0}-\frac{\partial X^-}{\partial\sigma^1}
&=&\frac{2}{p^+}\sum_{n\in\mathbf{Z}}L_ne^{-in(\sigma^0-\sigma^1)}
=\sqrt{2\alpha^{\prime}}\sum_{n\in\mathbf{Z}}\alpha_n^-e^{-in(\sigma^0-\sigma^1)},
\eea
\bea
\sqrt{2\alpha^{\prime}}\alpha_n^-=\frac{2}{p^+}L_n; \ \sqrt{2\alpha^{\prime}}\bar{\alpha}_n^-=\frac{2}{p^+}\bar{L}_n.
\eea
When we set $n=0$, we obtain:
\bea
\sqrt{2\alpha^{\prime}}\alpha_0^-=\alpha^{\prime}p^-=\frac{2}{p^+}L_0; \ \sqrt{2\alpha^{\prime}}\bar{\alpha}_0^-=\alpha^{\prime}p^-=\frac{2}{p^+}\bar{L}_0.
\eea
This establishes the level spacing condition
\bea
L_0 = \bar{L}_0.
\eea
This indicates that the action of $L_0$ and $\bar{L}_0$ on any state must be the same.
Nevertheless, we can also rewrite the $L_0$ and $\bar{L}_0$ in terms of the number operators:
\bea
N\equiv\sum_{n=1}^{\infty}n(a_n^j)^{\dagger}a_n^j; \
\bar{N}\equiv\sum_{n=1}^{\infty}n(\bar{a}_n^j)^{\dagger}\bar{a}_n^j,
\eea
\bea
L_0&=&\frac{1}{2}\sum_{m\in\mathbf{Z}}\alpha_m^j\alpha_{-m}^j
=\frac{1}{2}\alpha_0^j\alpha_0^j+N+\frac{D-2}{2}\sum_{n=1}^{\infty}n
\nn\\
&=&\frac{\alpha^{\prime}}{4}p^jp^j+N-\frac{D-2}{24}+\cdots;
\nn\\
\bar{L}_0&=&\frac{1}{2}\sum_{m\in\mathbf{Z}}\alpha_m^j\alpha_{-m}^j
=\frac{1}{2}\bar{\alpha}_0^j\bar{\alpha}_0^j+\bar{N}+\frac{D-2}{2}\sum_{n=1}^{\infty}n
\nn\\
&=&\frac{\alpha^{\prime}}{4}p^jp^j+\bar{N}-\frac{D-2}{24}+\cdots,
\eea
in which we used the zeta regularization result
\bea
\sum_{n=1}^{\infty}n=-\frac{1}{12}+\lim_{\epsilon\rightarrow}\frac{1}{\epsilon^2}.
\eea
The level spacing condition also implies that the number of left-moving oscillators is equal to the number of right-moving oscillators
\bea
N=\bar{N}.
\eea
\\

\noindent
Because we have
\bea
\alpha_0^-=\bar{\alpha}_0^-,
\eea
we obtain:
\bea
\frac{1}{2}\sqrt{2\alpha^{\prime}}(\alpha_0^-+\bar{\alpha}_0^-)=\frac{1}{p^+}(L_0+\bar{L}_0)=\alpha^{\prime}p^-.
\eea
Hence, the unregularized mass formula is:
\bea
M_u^2=2p^+p^--p^jp^j=\frac{2}{\alpha^{\prime}}(L_0+\bar{L}_0)-p^jp^j=\frac{2}{\alpha^{\prime}}\bigg(N+\bar{N}-\frac{D-2}{12}\bigg)+\cdots,
\eea
in which $\cdots$ only contains the divergent terms.
We define the regularized mass term
\bea
M^2\equiv\frac{2}{\alpha^{\prime}}\bigg(N+\bar{N}-\frac{D-2}{12}\bigg).
\eea
If we choose the ground state:
\bea
N=\bar{N}=0,
\eea
the state must have negative mass in any target-space dimension.
Therefore, this theory includes a tachyon, which is unstable.
Hence, the bosonic string theory is not self-contained, and we should remove the tachyon from some procedures similar to the superstring theory.
However, the analysis method of the superstring remains similar to that of bosonic string theories.
The study of the bosonic string is still meaningful.
If we assume that the first excited state:
\bea
N=\bar{N}=1
\eea
corresponds to the massless state or an irreducible representation of SO($D-2$), we can obtain a Lorentz invariant theory.
This condition shows the critical dimensions to the target space $D=26$.

\subsubsection{Compactification}
\noindent
Now we compactify a spatical target space $X^{25}$, which satifies the boundary condition
\bea
X^{25}(\sigma^0, \sigma^1+2\pi)=X^{25}(\sigma^0,\sigma^1)+m\cdot 2\pi R.
\eea
The string has the winding number $m$, which labels a new momentum, winding
\bea
w\equiv\frac{mR}{\alpha^{\prime}}.
\eea
Because $R$ is the length unit, $\alpha^{\prime}$ is the length squared unit, and $m$ is no unit, $w$ has the inverse length unit, which is a momentum unit.
Therefore, the boundary condition of the compact target space is
\bea
X^{25}(\sigma^0, \sigma^1+2\pi)=X^{25}(\sigma^0, \sigma^1)+2\pi\alpha^{\prime}w.
\eea
Hence we have
\bea
X^+,\ X^-,\ X^2,\ X^3, \cdots, X^{24},\ X^{25},
\eea
which satisfy
\bea
X^{+}(\sigma^0, \sigma^1+2\pi)&=&X^{+}(\sigma^0,\sigma^1);
\nn\\
X^{-}(\sigma^0, \sigma^1+2\pi)&=&X^{-}(\sigma^0,\sigma^1);
\nn\\
X^{\bar{j}}(\sigma^0, \sigma^1+2\pi)&=&X^{\bar{j}}(\sigma^0,\sigma^1);
\nn\\
X^{25}(\sigma^0, \sigma^1+2\pi)&=&X^{25}(\sigma^0,\sigma^1)+2\pi\alpha^{\prime}w.
\eea
The non-compact directions of the transverse light cone are labeled by $\bar{j} = 2, 3, \cdots, 24$.
Because the boundary condition of the compact target space is changed, we obtain:
\bea
X_L^{25}(u+2\pi)+X_R^{25}(v-2\pi)&=&X^{25}_L(u)+X^{25}_R(v)+2\pi\alpha^{\prime}w;
\nn\\
X_L^{25}(u+2\pi)-X_L^{25}(u)&=&X_R^{25}(v)-X_R^{25}(v-2\pi)+2\pi\alpha^{\prime}w.
\eea
Because $X^{25}_L$ only depends on $u$, and $X^{25}_R$ only depends on $v$, we still have the same expression:
\bea
X_L^{25}(u)&=&\frac{1}{2}x_{0, L}^{25}+\sqrt{\frac{\alpha^{\prime}}{2}}\bar{\alpha}_0^{25}u+i\sqrt{\frac{\alpha^{\prime}}{2}}\sum_{n\ne 0}\frac{\bar{\alpha}_n^{25}}{n}e^{-inu};
\nn\\
X_R^{25}(v)&=&\frac{1}{2}x_{0,R}^{25}+\sqrt{\frac{\alpha^{\prime}}{2}}\alpha_0^{25}v+i\sqrt{\frac{\alpha^{\prime}}{2}}\sum_{n\ne0}\frac{\alpha^{25}_n}{n}e^{-inv}.
\eea
\bea
&&X^{25}
\nn\\
&=&\frac{1}{2}(x_{0,L}^{25}+x_{0,R}^{25})
+\sqrt{\frac{\alpha^{\prime}}{2}}(\bar{\alpha}_0^{25}u+\alpha_0^{25}v)
+i\sqrt{\frac{\alpha^{\prime}}{2}}\sum_{n\ne 0}\frac{e^{-in\sigma^0}}{n}(\bar{\alpha}_n^{25}e^{-in\sigma^1}+\alpha_n^{25}e^{in\sigma^1});
\nn\\
&&\tilde{X}^{25}
\nn\\
&=&\frac{1}{2}(x_{0,L}^{25}-x_{0,R}^{25})
+\sqrt{\frac{\alpha^{\prime}}{2}}(\bar{\alpha}_0^{25}u-\alpha_0^{25}v)
+i\sqrt{\frac{\alpha^{\prime}}{2}}\sum_{n\ne 0}\frac{e^{-in\sigma^0}}{n}(\bar{\alpha}_n^{25}e^{-in\sigma^1}-\alpha_n^{25}e^{in\sigma^1}).
\nn\\
\eea
We can then get:
\bea
2\pi \sqrt{\frac{\alpha^{\prime}}{2}}\bar{\alpha}_0^{25}=
2\pi\sqrt{\frac{\alpha^{\prime}}{2}}\alpha_0^{25}+2\pi\alpha^{\prime}w; \
\bar{\alpha}_0^{25}=\alpha_0^{25}+\sqrt{2\alpha^{\prime}}w
\eea
from that
\bea
X_L^{25}(u+2\pi)-X_L^{25}(u)=X_R^{25}(v)-X_R^{25}(v-2\pi)+2\pi\alpha^{\prime}w.
\eea
\\

\noindent
The momentum is:
\bea
p^{25}=\int_0^{2\pi}d\sigma^1\ P^{25}
=\frac{1}{2\pi\alpha^{\prime}}\int_0^{2\pi}d\sigma^1\ \partial_0 X^{25}
=\frac{1}{\sqrt{2\alpha^{\prime}}}(\alpha_0^{25}+\bar{\alpha}_0^{25}).
\eea
The winding is
\bea
w=\frac{1}{\sqrt{2\alpha^{\prime}}}(\bar{\alpha}_0^{25}-\alpha_0^{25}).
\eea
Therefore, we find that the momentum $p^{25}$ has an equal role as the winding $w$.
The expression of zero-modes in the compact direction will be convenient:
\bea
\alpha_0^{25}=\sqrt{\frac{\alpha^{\prime}}{2}}(p^{25}-w); \
\bar{\alpha}_0^{25}=\sqrt{\frac{\alpha^{\prime}}{2}}(p^{25}+w).
\eea
Now we can rewrite the compact target space as:
\bea
X^{25}&=&x_0^{25}+\alpha^{\prime}(p^{25}\sigma^0+w\sigma^1)
+i\sqrt{\frac{\alpha^{\prime}}{2}}\sum_{n\ne 0}\frac{e^{-in\sigma^0}}{n}(\bar{\alpha}_n^{25}e^{-in\sigma^1}+\alpha_n^{25}e^{in\sigma^1});
\nn\\
\tilde{X}^{25}&=&\tilde{x}_0^{25}+\alpha^{\prime}(w\sigma^0+p^{25}\sigma^1)
+i\sqrt{\frac{\alpha^{\prime}}{2}}\sum_{n\ne 0}\frac{e^{-in\sigma^0}}{n}(\bar{\alpha}_n^{25}e^{-in\sigma^1}-\alpha_n^{25}e^{in\sigma^1}),
\eea
where
\bea
x_0^{25}\equiv\frac{1}{2}(x_{0, L}^{25}+x_{0, R}^{25}); \ \tilde{x}_0^{25}\equiv\frac{1}{2}(x_{0, L}^{25}-x_{0, R}^{25}).
\eea
We can find that the role of $p^{25}$ and $w$ is interchanged between the $X^{25}$ and $\tilde{X}^{25}$.
The dual target space satisfies the boundary condition
\bea
\tilde{X}^{25}(\sigma^0, \sigma^1+2\pi)=\tilde{X}(\sigma^0, \sigma^1)+2\pi\alpha^{\prime}p^{25}.
\eea
The translation along the 25th direction by an amount done by the operator $\exp(-iyp^{25})$.
Because we have the condition
\bea
e^{-ip^{25}y}=e^{-ip^{25}(y+2\pi R)}
\eea
in the compact direction, we obtain
\bea
e^{-i2\pi p^{25} R}=1,
\eea
which equivalently gives
\bea
p^{25}=\frac{n}{R}, \qquad n\in\mathbf{Z}.
\eea
Because we also have the winding
\bea
w=\frac{mR}{\alpha^{\prime}},
\eea
we can see the quantized momentum and winding in the compact direction.
\\

\noindent
It is easy to obtain:
\bea
&&L_0-\bar{L}_0
\nn\\
&=&N-\bar{N}+\frac{1}{2}(\alpha_0^{25}\alpha_0^{25}-\bar{\alpha}_0^{25}\bar{\alpha}_0^{25})
\nn\\
&=&
N-\bar{N}+\frac{\alpha^{\prime}}{4}\big((p^{25})^2+w^2-p^{25}w-wp^{25}-(p^{25})^2-w^2-p^{25}w-wp^{25}\big)
\nn\\
&=&
N-\bar{N}-\alpha^{\prime}p^{25}w.
\eea
By using the level spacing condition
\bea
L_0=\bar{L}_0,
\eea
we obtain:
\bea
N-\bar{N}=\alpha^{\prime}p^{25}w=nm.
\eea
Therefore, we find that the non-vanishing $n$ and $m$ yield non-equivalent left-moving and right-moving numbers when the target space is compactified.
\\

\noindent
Now we want to show the mass formula when an observer lives in a 25-dimensional Minkowski spacetime.
Because an observer cannot observe the compact direction, the mass formula from the viewpoint of the observer should not contain the momentum in the compact direction.
We then obtain:
\bea
M^2=\frac{1}{\alpha^{\prime}}(\alpha_0^{25}\alpha_0^{25}+\bar{\alpha}_0^{25}\bar{\alpha}_0^{25})
+\frac{2}{\alpha^{\prime}}(N+\bar{N}-2)
=(p^{25})^2+w^2+\frac{2}{\alpha^{\prime}}(N+\bar{N}-2).
\eea
The $(p^{25})^2$ is the rest mass from the compact direction because we cannot observe the compact direction.
For the interpretation of $w^2$, we can use the product of the string length $2\pi mR$ and the tension of the string
\bea
T_s\equiv\frac{1}{2\pi\alpha^{\prime}}
\eea
to obtain the energy:
\bea
2\pi m R\cdot\frac{1}{2\pi\alpha^{\prime}}=\frac{m R}{\alpha^{\prime}}=w.
\eea
Therefore, the $w^2$ is string energy.
We also explicitly see the T-duality, where we exchange the momentum number $n$ and the winding number $m$, to show the equivalent physics between $R$ and $\alpha^{\prime}/R$.
Note that exchanging $n$ and $m$ does not change the difference between the left-moving and right-moving numbers
\bea
N-\bar{N}=nm.
\eea

\subsubsection{T-Duality}
\noindent
We separate the coordinates as that
\bea
X^{\mu}=(X^j, \theta),
\eea
where $j=0, 1, \cdots, D-2$.
The string theory now has an isometry represented by a translation in the coordinate $\theta$.
The metric and dilaton fields do not depend on $\theta$.
We write the alternative description of the worldsheet theory
\bea
S_{\mathrm{bsa}}&=&-\frac{1}{4\pi\alpha^{\prime}}\int d^2\sigma\sqrt{|\det(h_{\rho\sigma})|}
\nn\\
&&\times
\bigg(h^{\alpha\beta}\partial_{\alpha}X^{j_1}g_{j_1j_2}(X)\partial_{\beta}X^{j_2}
+2h^{\alpha\beta}V_{\alpha}g_{25, j_2}(X)\partial_{\beta}X^{j_2}
+h^{\alpha\beta}V_{\alpha}g_{25, 25}(X)V_{\beta}
\nn\\
&&
-\epsilon^{\alpha\beta}\partial_{\alpha}X^{j_1}B_{j_1j_2}(X)\partial_{\beta}X^{j_2}
-2\epsilon^{\alpha\beta}V_{\alpha}B_{25, j_2}(X)\partial_{\beta}X^{j_2}
-2\epsilon^{\alpha\beta}\tilde{\theta}\partial_{\alpha}V_{\beta}
\nn\\
&&
+\alpha^{\prime}R\phi(X)
\bigg).
\eea
The periodicity of $\tilde{\theta}$ can be fixed
\bea
\tilde{\theta}\sim\tilde{\theta}+2\pi\alpha^{\prime}
\eea
by requiring that the theory is invariant under the boundary condition.
\\

\noindent
The integration of $\tilde{\theta}$ equivalently implies
\bea
\epsilon^{\alpha\beta}\partial_{\alpha}V_{\beta}=0.
\eea
It leads to the original worldsheet theory by substituting $V=d\theta$.
The dual theory is derived by integrating over the $V_{\mu}$ field along with the dual background fields:
\bea
\tilde{g}_{25, 25}=\frac{1}{g_{25, 25}}; \
\tilde{g}_{25, j}=\frac{B_{25, j}}{g_{25, 25}}; \
\tilde{g}_{jk}=g_{jk}-\frac{g_{25, j}g_{25, k}-B_{25, j}B_{25, k}}{g_{25, 25}};
\nn
\eea
\bea
\tilde{B}_{25, j}=\frac{g_{25, j}}{g_{25, 25}}; \
\tilde{B}_{j, k}=B_{j, k}-\frac{g_{25, j}B_{25, k}-B_{25, j}g_{25, k}}{g_{25, 25}}.
\eea
The integration on $V_{\mu}$ produces an additional a factor, $\det(g_{25, 25})$, which yields the dual dilaton field
\bea
\tilde{\phi}=\phi-\ln\sqrt{g_{25, 25}}.
\eea
The inverse of the $g_{25, 25}$ from the dual induces that
\bea
\tilde{g}_{25, 25}\partial_{\alpha}\tilde{\theta}\partial_{\beta}\tilde{\theta}=\partial_{\alpha}\bar{\theta}\partial_{\beta}\bar{\theta},
\eea
where
\bea
\tilde{g}_{25, 25}=\frac{1}{R^2}; \qquad \frac{1}{R}\tilde{\theta}\equiv\bar{\theta}.
\eea
The $\bar{\theta}$ satisfies the boundary condition
\bea
\bar{\theta}\sim \bar{\theta}+2\pi\frac{\alpha^{\prime}}{R}.
\eea
Hence, this duality is the T-duality.

\subsection{Open String}
\noindent
We introduce the open string from the quantization, the dynamics of the target space, and the duality.
Let us introduce the open string theory beginning from the flat worldsheet metric
\bea
L_{\mathrm{bsmc}}[X]&=&-\frac{1}{4\pi\alpha^{\prime}}\bigg(\partial_{\alpha}X^{\mu}g_{\mu\nu}(X)\partial^{\alpha}X^{\nu}
-\epsilon^{\alpha\beta}\partial_{\alpha}X^{\mu}B_{\mu\nu}(X)\partial_{\beta}X^{\nu}\bigg),
\eea
with the integration range of $\sigma^0$ is ($-\infty$, $\infty$) and $\sigma^1$ is (0, $\pi$).
When one imposes the Dirichlet boundary condition on the $\sigma^0$ direction:
\bea
\delta X^{\mu}=0; \ X^{\mu}=c; \ \sigma^0\rightarrow-\infty\ \mathrm{or}\ \infty,
\eea
where $c$ is a constant, and the Neumann boundary condition on the $\sigma^1$ direction
\bea
g_{\mu\nu}\partial_1X^{\nu}+B_{\mu\nu}\partial_0X^{\nu}=0, \qquad \sigma^1=0\ \mathrm{or}\ \pi,
\label{neubd}
\eea
which indicates that the target space $X$ can exhibit non-trivial fluctuations on the boundary of the $\sigma^1$ direction, but not on the boundary of the $\sigma^0$ direction, in the bosonic string.
\\

\noindent
The gauge transformation of the Kalb-Ramond field $B$ associated with the one-form gauge parameter
\bea
\delta_{\Lambda} B_{\mu\nu}=\partial_{\mu}\Lambda_{\nu}-\partial_{\nu}\Lambda_{\mu},
\eea
where $\Lambda_{\mu}$ and $\Lambda_{\nu}$ are a one-form gauge parameters, and $\delta_{\Lambda}$ is the transformation of this one-form parameter $\Lambda$, shows the non-gauge invariance on the boundary:
\bea
\delta_{\Lambda} S_{\mathrm{bsmc}}&=&\frac{1}{4\pi\alpha^{\prime}}\int d^2\sigma\ \epsilon^{\alpha\beta}\partial_{\alpha}X^{\mu}\delta_{\Lambda}B_{\mu\nu}\partial_{\beta}X^{\nu}
\nn\\
&=&\frac{1}{4\pi\alpha^{\prime}}\int d^2\sigma\ \epsilon^{\alpha\beta}\bigg(\partial_{\alpha}\Lambda_{\nu}\partial_{\beta}X^{\nu}
-\partial_{\alpha}X^{\mu}\partial_{\beta}\Lambda_{\mu}\bigg)
\nn\\
&=&\frac{1}{2\pi\alpha^{\prime}}\int d^2\sigma\ \epsilon^{\alpha\beta}\partial_{\alpha}\Lambda_{\nu}\partial_{\beta}X^{\nu}
\nn\\
&=&\frac{1}{2\pi\alpha^{\prime}}\int d^2\sigma\ \partial_{\alpha}\bigg(\epsilon^{\alpha\beta}\Lambda_{\nu}\partial_{\beta}X^{\nu}\bigg)
\nn\\
&=&-\frac{1}{2\pi\alpha^{\prime}}\int_{-\infty}^{\infty} d\sigma^0\ \Lambda_{\nu}\partial_0X^{\nu}\bigg|_{\sigma_1=0}^{\sigma_1=\pi}.
\eea
Hence, one needs to put the additional gauge field with the gauge transformation
\bea
\delta_{\Lambda} A_{\mu}(X)=\Lambda_{\mu}
\eea
on the boundary to cancel this non-gauge invariant boundary term.
The action of the open string theory is
\bea
S_{\mathrm{bsmco}}[X]&=&-\frac{1}{4\pi\alpha^{\prime}}\int d^2\sigma\ \bigg(\partial_{\alpha}X^{\mu}g_{\mu\nu}(X)\partial^{\alpha}X^{\nu}
-\epsilon^{\alpha\beta}\partial_{\alpha}X^{\mu}B_{\mu\nu}(X)\partial_{\beta}X^{\nu}\bigg)
\nn\\
&&+\frac{1}{2\pi\alpha^{\prime}}\int_{-\infty}^{\infty} d\sigma^0\ A_{\mu}(X)\partial_0 X^{\mu}.
\eea
The gauge field $A$ indicates the dynamics of the open string theory.
Thus, the open string only has non-trivial degrees on the boundary of the two-dimensional worldsheet.

\subsubsection{Non-Commutative Geometry}
\noindent
By using the Neumann boundary condition \eqref{neubd} and the momentum
\bea
2\pi\alpha^{\prime}P^{\mu}=\partial_0X^{\mu}+\partial_1X^{\nu}{\cal F}^{\mu}{}_{\nu},
\eea
where
\bea
{\cal F}_{\mu\nu}\equiv B_{\mu\nu}-F_{\mu\nu},
\eea
we can reach the relation \cite{Chu:1998qz}
\bea
2\pi\alpha^{\prime}P^{\nu}(\sigma^0, 0){\cal F}_{\mu\nu}
=-\partial_1X^{\nu}(\sigma^0, 0){\cal H}_{\mu\nu},
\eea
where
\bea
{\cal H}_{\mu\nu}\equiv g_{\mu\nu}-{\cal F}_{\mu\sigma}g^{\sigma\delta}{\cal F}_{\delta\nu}.
\eea
This relation implies the following equality \cite{Chu:1998qz}
\bea
2\pi\alpha^{\prime}\lbrack P^{\nu}(\sigma^0, 0), X^{\rho}(\sigma^0, \sigma^1)\rbrack {\cal F}_{\mu\nu}
=-\lbrack\partial_1 X^{\nu}(\sigma^0, 0), X^{\rho}(\sigma^0, \sigma^1)\rbrack{\cal H}_{\mu\nu},
\eea
which implies the target space should not be commutative \cite{Chu:1998qz}
\bea
\lbrack\partial_1 X^{\delta}(\sigma^0, 0), X^{\rho}(\sigma^0, \sigma^1)\rbrack
=2\pi i\alpha^{\prime}\delta(\sigma^1)({\cal H}^{-1} {\cal F})^{\delta\rho}.
\eea
\\

\noindent
In the closed string, we can use the following Fourier transformation to represent the delta function
\bea
\delta(\sigma_1-\sigma_1^{\prime})=\frac{1}{2\pi}\sum_{n\in\mathbb{Z}}e^{in(\sigma_1-\sigma_1^{\prime})},
\eea
which satisfies
\bea
\int^{\pi}_{-\pi}d\sigma_1^{\prime}\ \delta(\sigma_1-\sigma_1^{\prime})f(\sigma_1^{\prime})=f(\sigma_1)
\eea
for any periodic function
\bea
f(\sigma_1)=f(\sigma_1+2\pi).
\eea
For the open string, we can take $f(\sigma_1)$ as an even function in the interval $\lbrack-\pi, \pi\rbrack$, the equivalent relation can be given as
\bea
\int^{\pi}_0d\sigma_1^{\prime}\ \Delta(\sigma_1, \sigma_1^{\prime})f(\sigma_1^{\prime})=f(\sigma_1),
\eea
where
\bea
\Delta(\sigma_1, \sigma_1^{\prime})
\equiv \delta(\sigma_1-\sigma_1^{\prime})+\delta(\sigma_1+\sigma_1^{\prime})
=\frac{1}{\pi}+\frac{1}{\pi}\sum_{n\neq 0}\cos(n\sigma_1)\cos(n\sigma_1^{\prime}).
\eea
Hence, we can explicitly get the non-commutative relation to the target space \cite{Chu:1998qz}
\bea
\lbrack X^{\delta}(\sigma_0, \sigma_1), X^{\rho}(\sigma_0, \sigma_1^{\prime})\rbrack
=
2i\alpha^{\prime}({\cal H}^{-1}{\cal F})^{\delta\rho}
\bigg(\sigma_1+\sigma_1^{\prime}-\pi+\sum_{n\neq 0}\frac{1}{n}\sin\big(n(\sigma_1+\sigma_1^{\prime})\big)\bigg).
\eea
\\

\noindent
Now, we compute the summation explicitly to analyze the non-commutativity of the target space \cite{Chu:1998qz}.
The infinite summation of the series can be simplified as follows:
\bea
\sum_{n\neq 0}\frac{1}{n}\sin(n\theta)=
\sum_{n=1}^{\infty}\frac{1}{n}\sin(n\theta)+\sum^{-1}_{n=-\infty}\frac{1}{n}\sin(n\theta)=2\sum_{n=1}^{\infty}\frac{1}{n}\sin(n\theta),
\eea
where
\bea
\theta\equiv\sigma_1+\sigma_1^{\prime}.
\eea
We can calculate the infinite summation when $0<\theta<2\pi$ as in the following:
\bea
\sum_{n=1}^{\infty}\frac{\sin(n\theta)}{n}=\mathrm{Im}\bigg(\sum_{n=1}^{\infty}\frac{e^{in\theta}}{n}\bigg)=-\mathrm{Im}\big(\ln(1-e^{i\theta})\big)
=\frac{\pi-\theta}{2},
\eea
in which we use:
\bea
1-e^{i\theta}=2ie^{i\frac{\theta}{2}}\sin\frac{\theta}{2}=2\sin\bigg(\frac{\theta}{2}\bigg)e^{i\frac{\theta-\pi}{2}}.
\eea
When $\theta=0, 2\pi$, the infinite summation is zero \cite{Chu:1998qz}.
Hence, the non-commutativity only occurs at the ending point of the string \cite{Chu:1998qz}:
\bea
\lbrack X^{\delta}(\sigma_0, 0), X^{\rho}(\sigma_0, 0)\rbrack
&=&
-2\pi i\alpha^{\prime}({\cal H}^{-1}{\cal F})^{\delta\rho};
\nn\\
\lbrack X^{\delta}(\sigma_0, \pi), X^{\rho}(\sigma_0, \pi)\rbrack
&=&
2\pi i\alpha^{\prime}({\cal H}^{-1}{\cal F})^{\delta\rho}.
\eea

\section{Target Space}
\label{sec:3}
\noindent
We introduce a target space theory of the bosonic string, as well as T- and S-duality.
The relation between the worldsheet theory and the target space theory is built from the equations of motion of the target space theory through the vanishing $\beta$ function of the bosonic string \cite{Callan:1986bc}.
We first show the target space theory, introduce the Green's function in the bulk and on the boundary, and also reproduce an equation of motion from the vanishing one-loop $\beta$ function when the background fields are constant \cite{Fradkin:1985qd}.
The bosonic sigma model provides equations of motion of the target space theory through the vanishing one-loop $\beta$ function \cite{Callan:1986bc}.
Ref. \cite{Callan:1986bc} gives the target space theory
\bea
S_{\mathrm{let}}&\equiv&\frac{1}{16\pi G_N}\int d^Dx\sqrt{|\det g_{\mu\nu}|}\  e^{-2\phi}\bigg(R+4\partial_{\rho}\phi\partial^{\rho}\phi-\frac{1}{12}H_{\sigma\delta\gamma}H^{\sigma\delta\gamma}\bigg)
\nn\\
&&-T_{D-1}\int d^Dx\ e^{-\phi}\sqrt{\det(g_{\mu\nu}+B_{\mu\nu}-F_{\mu\nu})},
\eea
where
\bea
H_{\mu\nu\rho}&\equiv&\partial_{\mu}B_{\nu\rho}+\partial_{\nu}B_{\rho\mu}+\partial_{\rho}B_{\mu\nu}, \qquad F_{\mu\nu}\equiv\partial_{\mu}A_{\nu}-\partial_{\nu}A_{\mu}.
\eea
The $G_N$ is the 10D gravitational constant, and $T_p$ is the D$p$-brane tension
\bea
T_p\equiv\frac{1}{(2\pi)^p(\alpha^{\prime})^{\frac{p+1}{2}}}.
\eea
When $p=1$, the $T_p$ coincides with the string tension $T_s$.

\subsection{One-Loop $\beta$ Function}
\noindent
When one considers the closed string theory, the constant background fields should give exactly zero $\beta$ function without giving any constraint to the equations of motion of the target space theory.
Including the open string theory shows a constraint on the vanishing one-loop $\beta$ function from the boundary.
We demonstrate the equation of motion of the target space theory from the bosonic string theory with the constant background fields through the vanishing one-loop $\beta$ function.

\subsubsection{Fluctuation of Target Space}
\noindent
When one does the perturbation from $\xi$, $X=\bar{X}+\xi$ and $F_{\mu\nu}(\bar{X})$ is a constant, one obtains:
\bea
&&-2\pi\alpha^{\prime}S_{\mathrm{bsmc}}[X]
\nn\\
&=&\frac{1}{2}\int d^2\sigma\ \bigg(\partial_{\alpha}X^{\mu}g_{\mu\nu}(X)\partial^{\alpha}X^{\nu}
-\epsilon^{\alpha\beta}\partial_{\alpha}X^{\mu}(B_{\mu\nu}-F_{\mu\nu})(X)\partial_{\beta}X^{\nu}\bigg)
\nn\\
&=&\frac{1}{2}\int d^2\sigma\ \bigg(\partial_{\alpha}X^{\mu}g_{\mu\nu}\partial^{\alpha}X^{\nu}
-\epsilon^{\alpha\beta}\partial_{\alpha}X^{\mu}\big(B_{\mu\nu}-F_{\mu\nu}(X)\big)\partial_{\beta}X^{\nu}\bigg)
\nn\\
&=&\frac{1}{2}\int d^2\sigma\ \bigg(\partial_{\alpha}\bar{X}^{\mu}g_{\mu\nu}\partial^{\alpha}\bar{X}^{\nu}
-\epsilon^{\alpha\beta}\partial_{\alpha}\bar{X}^{\mu}\big(B_{\mu\nu}-F_{\mu\nu}(\bar{X})\big)\partial_{\beta}\bar{X}^{\nu}\bigg)
\nn\\
&&+\int d^2\sigma\ \bigg(\partial_{\alpha}\xi^{\mu}g_{\mu\nu}\partial^{\alpha}X^{\nu}
-\epsilon^{\alpha\beta}\partial_{\alpha}\xi^{\mu}(B_{\mu\nu}-F_{\mu\nu}(\bar{X}))\partial_{\beta}X^{\nu}
\nn\\
&&+\epsilon^{\alpha\beta}\xi^{\rho}\partial_{\alpha}\bar{X}^{\mu}\partial_{\rho}F_{\mu\nu}(\bar{X})\partial_{\beta}\bar{X}^{\nu}
\bigg)
\nn\\
&&+\frac{1}{2}\int d^2\sigma\ \bigg(\partial_{\alpha}\xi^{\mu}g_{\mu\nu}\partial^{\alpha}\xi^{\nu}
-\epsilon^{\alpha\beta}\partial_{\alpha}\xi^{\mu}\big(B_{\mu\nu}-F_{\mu\nu}(\bar{X})\big)\partial_{\beta}\xi^{\nu}
\nn\\
&&+\frac{1}{2}\epsilon^{\alpha\beta}\xi^{\rho}\xi^{\sigma}\partial_{\alpha}\bar{X}^{\mu}\partial_{\rho}\partial_{\sigma}F_{\mu\nu}(\bar{X})\partial_{\beta}\bar{X}^{\nu}+2\epsilon^{\alpha\beta}\xi^{\rho}\partial_{\alpha}\xi^{\mu}\partial_{\rho}F_{\mu\nu}(\bar{X})\partial_{\beta}\bar{X}^{\nu}
\bigg)+\cdots.
\nn\\
\eea
The partition function of this bosonic sigma model is given by
\bea
Z_{\mathrm{bsmc}}[X]=\int DX\ e^{iS_{\mathrm{bsmc}[X]}},
\eea
and one integrates out $\xi$ and treat $\bar{X}$ as a background, then one should obtain
\bea
Z_{\mathrm{2bsmc}}[X]=\int D\xi\ e^{iS_{\mathrm{2bsmc}}[\xi]},
\eea
where
\bea
&&-2\pi\alpha^{\prime}S_{\mathrm{2bsmc}}[\xi]
\nn\\
&\equiv&\frac{1}{2}\int d^2\sigma\ \bigg(\partial_{\alpha}\xi^{\mu}g_{\mu\nu}\partial^{\alpha}\xi^{\nu}
-\epsilon^{\alpha\beta}\partial_{\alpha}\xi^{\mu}\big(B_{\mu\nu}-F_{\mu\nu}(\bar{X})\big)\partial_{\beta}\xi^{\nu}
\nn\\
&&+\frac{1}{2}\epsilon^{\alpha\beta}\xi^{\rho}\xi^{\sigma}\partial_{\alpha}\bar{X}^{\mu}\partial_{\rho}\partial_{\sigma}F_{\mu\nu}(\bar{X})\partial_{\beta}\bar{X}^{\nu}
+2\epsilon^{\alpha\beta}\xi^{\rho}\partial_{\alpha}\xi^{\mu}\partial_{\rho}F_{\mu\nu}(\bar{X})\partial_{\beta}\bar{X}^{\nu}
\bigg),
\eea
from the second-order perturbation terms for $\xi$ in the action of the bosonic string.
When one integrates out $\xi$, the integration is equivalent to using the equation of motion $\xi$.
Hence, the first-order term vanishes.
The background terms do not affect dynamics.
\\

\noindent
Now we do integration by parts and obtain:
\bea
&&-2\pi\alpha^{\prime}S_{\mathrm{2bsmc}}[\xi]
\nn\\
&=&\frac{1}{2}\int d^2\sigma\ \bigg(\partial_{\alpha}\xi^{\mu}g_{\mu\nu}\partial^{\alpha}\xi^{\nu}
-\epsilon^{\alpha\beta}\partial_{\alpha}\xi^{\mu}\big(B_{\mu\nu}-F_{\mu\nu}(\bar{X})\big)\partial_{\beta}\xi^{\nu}
\nn\\
&&+\frac{1}{2}\epsilon^{\alpha\beta}\xi^{\rho}\xi^{\sigma}\partial_{\alpha}\bar{X}^{\mu}\partial_{\rho}\partial_{\sigma}F_{\mu\nu}(\bar{X})\partial_{\beta}\bar{X}^{\nu}+2\epsilon^{\alpha\beta}\xi^{\rho}\partial_{\alpha}\xi^{\mu}\partial_{\rho}F_{\mu\nu}(\bar{X})\partial_{\beta}\bar{X}^{\nu}
\bigg)
\nn\\
&=&\frac{1}{2}\int d^2\sigma\ \bigg(-\xi^{\mu}g_{\mu\nu}\partial_{\alpha}\partial^{\alpha}\xi^{\nu}
+2\partial_{1}\xi^{\mu}\big(B_{\mu\nu}-F_{\mu\nu}(\bar{X})\big)\partial_{0}\xi^{\nu}
\nn\\
&&+\epsilon^{\alpha\beta}\xi^{\rho}\xi^{\sigma}\partial_{\rho}\partial_{\sigma}\partial_{\alpha}A_{\mu}(\bar{X})\partial_{\beta}\bar{X}^{\mu}+2\epsilon^{\alpha\beta}\xi^{\rho}\partial_{\alpha}\xi^{\mu}\partial_{\rho}F_{\mu\nu}(\bar{X})\partial_{\beta}\bar{X}^{\nu}\bigg)
\nn\\
&&+\frac{1}{2}\int d\sigma^0\ \xi^{\mu}g_{\mu\nu}\partial_1\xi^{\nu}
\nn\\
&=&\frac{1}{2}\int d^2\sigma\ \bigg(-\xi^{\mu}g_{\mu\nu}\partial_{\alpha}\partial^{\alpha}\xi^{\nu}
+2\partial_{1}\xi^{\mu}\big(B_{\mu\nu}-F_{\mu\nu}(\bar{X})\big)\partial_{0}\xi^{\nu}
\nn\\
&&+\xi^{\rho}\xi^{\sigma}\partial_{\rho}\partial_{\sigma}\partial_0A_{\mu}(\bar{X})\partial_{1}\bar{X}^{\mu}
-\xi^{\rho}\xi^{\sigma}\partial_{\rho}\partial_{\sigma}\partial_1A_{\mu}(\bar{X})\partial_{0}\bar{X}^{\mu}
\nn\\
&&+2\xi^{\rho}\partial_{0}\xi^{\mu}\partial_{\rho}F_{\mu\nu}(\bar{X})\partial_{1}\bar{X}^{\nu}-2\xi^{\rho}\partial_{1}\xi^{\mu}\partial_{\rho}F_{\mu\nu}(\bar{X})\partial_{0}\bar{X}^{\nu}\bigg)
\nn\\
&&+\frac{1}{2}\int d\sigma^0\ \xi^{\mu}g_{\mu\nu}\partial_1\xi^{\nu}
\nn\\
&=&\frac{1}{2}\int d^2\sigma\ \bigg(-\xi^{\mu}g_{\mu\nu}\partial_{\alpha}\partial^{\alpha}\xi^{\nu}
+2\partial_{1}\xi^{\mu}\big(B_{\mu\nu}-F_{\mu\nu}(\bar{X})\big)\partial_{0}\xi^{\nu}
\nn\\
&&+\partial_0\big(\xi^{\rho}\xi^{\sigma}\partial_{\rho}\partial_{\sigma}A_{\mu}\partial_1\bar{X}^{\mu}
+2\xi^{\mu}\partial_1\xi^{\nu}\partial_{\mu}A_{\nu}\big)
-\partial_1\big(\xi^{\rho}\xi^{\sigma}\partial_{\rho}\partial_{\sigma}A_{\mu}\partial_0\bar{X}^{\mu}
+2\xi^{\mu}\partial_0\xi^{\nu}\partial_{\mu}A_{\nu}\big)
\nn\\
&&+2\partial_1\xi^{\mu}\partial_0\xi^{\nu}F_{\mu\nu}(\bar{X})\bigg)
\nn\\
&&+\frac{1}{2}\int d\sigma^0\ \xi^{\mu}g_{\mu\nu}\partial_1\xi^{\nu}
\nn\\
&=&\frac{1}{2}\int d^2\sigma\ \bigg(-\xi^{\mu}g_{\mu\nu}\partial_{\alpha}\partial^{\alpha}\xi^{\nu}
+2\partial_{1}\xi^{\mu}B_{\mu\nu}\partial_{0}\xi^{\nu}
\nn\\
&&+\partial_0\big(\xi^{\rho}\xi^{\sigma}\partial_{\rho}\partial_{\sigma}A_{\mu}\partial_1\bar{X}^{\mu}
+2\xi^{\mu}\partial_1\xi^{\nu}\partial_{\mu}A_{\nu}\big)
-\partial_1\big(\xi^{\rho}\xi^{\sigma}\partial_{\rho}\partial_{\sigma}A_{\mu}\partial_0\bar{X}^{\mu}
+2\xi^{\mu}\partial_0\xi^{\nu}\partial_{\mu}A_{\nu}\big)
\bigg)
\nn\\
&&+\frac{1}{2}\int d\sigma^0\ \xi^{\mu}g_{\mu\nu}\partial_1\xi^{\nu},
\nn\\
\eea
in which we use
\bea
\partial_{\alpha}\bar{X}^{\mu}\partial_{\mu}A_{\nu}(\bar{X})=\partial_{\alpha}A_{\nu}(\bar{X}).
\eea
Now we also integrate the $\sigma^1$ direction \cite{Fradkin:1985qd}:
\bea
&&-2\pi\alpha^{\prime}S_{\mathrm{2bsmc}}[\xi]
\nn\\
&=&\frac{1}{2}\int d^2\sigma\ \bigg(-\xi^{\mu}g_{\mu\nu}\partial_{\alpha}\partial^{\alpha}\xi^{\nu}
+2\partial_{1}\xi^{\mu}B_{\mu\nu}\partial_{0}\xi^{\nu}
\nn\\
&&+\partial_0\big(\xi^{\rho}\xi^{\sigma}\partial_{\rho}\partial_{\sigma}A_{\mu}\partial_1\bar{X}^{\mu}
+2\xi^{\mu}\partial_1\xi^{\nu}\partial_{\mu}A_{\nu}\big)
-\partial_1\big(\xi^{\rho}\xi^{\sigma}\partial_{\rho}\partial_{\sigma}A_{\mu}\partial_0\bar{X}^{\mu}
+2\xi^{\mu}\partial_0\xi^{\nu}\partial_{\mu}A_{\nu}\big)
\bigg)
\nn\\
&&+\frac{1}{2}\int d\sigma^0\ \xi^{\mu}g_{\mu\nu}\partial_1\xi^{\nu}
\nn\\
&=&\frac{1}{2}\int d^2\sigma\ \bigg(-\xi^{\mu}g_{\mu\nu}\partial_{\alpha}\partial^{\alpha}\xi^{\nu}
+\partial_{1}\big(\xi^{\mu}B_{\mu\nu}\partial_{0}\xi^{\nu}\big)-\partial_{0}\big(\xi^{\mu}B_{\mu\nu}\partial_{1}\xi^{\nu}\big)
\nn\\
&&+\partial_0\big(\xi^{\rho}\xi^{\sigma}\partial_{\rho}\partial_{\sigma}A_{\mu}\partial_1\bar{X}^{\mu}
+2\xi^{\mu}\partial_1\xi^{\nu}\partial_{\mu}A_{\nu}\big)
-\partial_1\big(\xi^{\rho}\xi^{\sigma}\partial_{\rho}\partial_{\sigma}A_{\mu}\partial_0\bar{X}^{\mu}
+2\xi^{\mu}\partial_0\xi^{\nu}\partial_{\mu}A_{\nu}\big)
\bigg)
\nn\\
&&+\frac{1}{2}\int d\sigma^0\ \xi^{\mu}g_{\mu\nu}\partial_1\xi^{\nu}
\nn\\
&=&\frac{1}{2}\int d^2\sigma\ \bigg(-\xi^{\mu}g_{\mu\nu}\partial_{\alpha}\partial^{\alpha}\xi^{\nu}
\bigg)
\nn\\
&&+\frac{1}{2}\int d\sigma^0\ \bigg(\xi^{\mu}g_{\mu\nu}\partial_1\xi^{\nu}+\xi^{\mu}B_{\mu\nu}\partial_0\xi^{\nu}-
\big(\xi^{\rho}\xi^{\sigma}\partial_{\rho}\partial_{\sigma}A_{\mu}\partial_0\bar{X}^{\mu}
+2\xi^{\mu}\partial_0\xi^{\nu}\partial_{\mu}A_{\nu}\big)
\bigg)
\nn\\
&=&\frac{1}{2}\int d^2\sigma\ \bigg(-\xi^{\mu}g_{\mu\nu}\partial_{\alpha}\partial^{\alpha}\xi^{\nu}
\bigg)
\nn\\
&&+\frac{1}{2}\int d\sigma^0\ \bigg(\xi^{\mu}g_{\mu\nu}\partial_1\xi^{\nu}+\xi^{\mu}\big(B_{\mu\nu}-F_{\mu\nu}(\bar{X})\big)\partial_0\xi^{\nu}
\nn\\
&&+\xi^{\mu}\xi^{\nu}\partial_{\mu}\big(B_{\nu\rho}-F_{\nu\rho}\big)(\bar{X})\partial_0\bar{X}^{\rho}
\bigg),
\eea
in which we use
\bea
&&-\frac{1}{2}\int d\sigma^0\ \xi^{\mu}F_{\mu\nu}(\bar{X})\partial_0\xi^{\nu}
\nn\\
&=&-\frac{1}{2}\int d\sigma^0\ \xi^{\mu}\partial_{\mu}A_{\nu}(\bar{X})\partial_0\xi^{\nu}+\frac{1}{2}\int d\sigma^0\ \xi^{\mu}\partial_{\nu}A_{\mu}(\bar{X})\partial_0\xi^{\nu};
\nn\\
&&\frac{1}{2}\int d\sigma^0\ \xi^{\mu}\xi^{\nu}\partial_{\mu}\big(B_{\nu\rho}-F_{\nu\rho}\big)(\bar{X})\partial_0\bar{X}^{\rho}
\nn\\
&=&
-\frac{1}{2}\int d\sigma^0\ \xi^{\mu}\xi^{\nu}\partial_{\mu}F_{\nu\rho}(\bar{X})\partial_0\bar{X}^{\rho}
\nn\\
&=&-\frac{1}{2}\int d\sigma^0\ \xi^{\mu}\xi^{\nu}\partial_{\mu}\partial_{\nu}A_{\rho}(\bar{X})\partial_0\bar{X}^{\rho}
+\frac{1}{2}\int d\sigma^0\ \xi^{\mu}\xi^{\nu}\partial_{\mu}\partial_{\rho}A_{\nu}(\bar{X})\partial_0\bar{X}^{\rho}
\nn\\
&=&-\frac{1}{2}\int d\sigma^0\ \xi^{\mu}\xi^{\nu}\partial_{\mu}\partial_{\nu}A_{\rho}(\bar{X})\partial_0\bar{X}^{\rho}
+\frac{1}{2}\int d\sigma^0\ \xi^{\mu}\xi^{\nu}\partial_{\mu}\partial_{0}A_{\nu}(\bar{X})
\nn\\
&=&-\frac{1}{2}\int d\sigma^0\ \xi^{\mu}\xi^{\nu}\partial_{\mu}\partial_{\nu}A_{\rho}(\bar{X})\partial_0\bar{X}^{\rho}
-\frac{1}{2}\int d\sigma^0\ \partial_0\xi^{\mu}\xi^{\nu}\partial_{\mu}A_{\nu}(\bar{X})-\frac{1}{2}\xi^{\mu}\partial_0\xi^{\nu}\partial_{\mu}A_{\nu}(\bar{X});
\nn\\
&&-\frac{1}{2}\int d\sigma^0\ \xi^{\mu}F_{\mu\nu}(\bar{X})\partial_0\xi^{\nu}+\frac{1}{2}\int d\sigma^0\ \xi^{\mu}\xi^{\nu}\partial_{\mu}\big(B_{\nu\rho}-F_{\nu\rho}(\bar{X})\big)\partial_0\bar{X}^{\rho}
\nn\\
&=&-\frac{1}{2}\int d\sigma^0\ \xi^{\mu}\partial_{\mu}A_{\nu}(\bar{X})\partial_0\xi^{\nu}+\frac{1}{2}\int d\sigma^0\ \xi^{\mu}\partial_{\nu}A_{\mu}(\bar{X})\partial_0\xi^{\nu}
\nn\\
&&-\frac{1}{2}\int d\sigma^0\ \xi^{\mu}\xi^{\nu}\partial_{\mu}\partial_{\nu}A_{\rho}(\bar{X})\partial_0\bar{X}^{\rho}
-\frac{1}{2}\int d\sigma^0\ \partial_0\xi^{\mu}\xi^{\nu}\partial_{\mu}A_{\nu}(\bar{X})-\frac{1}{2}\xi^{\mu}\partial_0\xi^{\nu}\partial_{\mu}A_{\nu}(\bar{X})
\nn\\
&=&-\frac{1}{2}\int d\sigma^0\ \xi^{\mu}\partial_{\mu}A_{\nu}(\bar{X})\partial_0\xi^{\nu}-\frac{1}{2}\int d\sigma^0\ \xi^{\mu}\xi^{\nu}\partial_{\mu}\partial_{\nu}A_{\rho}(\bar{X})\partial_0\bar{X}^{\rho}-\frac{1}{2}\xi^{\mu}\partial_0\xi^{\nu}\partial_{\mu}A_{\nu}(\bar{X})
\nn\\
&=&-\int d\sigma^0\ \xi^{\mu}\partial_{\mu}A_{\nu}(\bar{X})\partial_0\xi^{\nu}-\frac{1}{2}\int d\sigma^0\ \xi^{\mu}\xi^{\nu}\partial_{\mu}\partial_{\nu}A_{\rho}(\bar{X})\partial_0\bar{X}^{\rho}
\nn\\
&=&-\frac{1}{2}\int d\sigma^0\ \big(\xi^{\rho}\xi^{\sigma}\partial_{\rho}\partial_{\sigma}A_{\mu}\partial_0\bar{X}^{\mu}
+2\xi^{\mu}\partial_0\xi^{\nu}\partial_{\mu}A_{\nu}\big)
\eea
in the final equality.

\subsubsection{Green's Function in Bulk}
\noindent
From the fluctuation of the target space \cite{Fradkin:1985qd}
\bea
&&-2\pi\alpha^{\prime}S_{\mathrm{2bsmc}}[\xi]
\nn\\
&=&\frac{1}{2}\int d^2\sigma\ \bigg(-\xi^{\mu}g_{\mu\nu}\partial_{\alpha}\partial^{\alpha}\xi^{\nu}
\bigg)
\nn\\
&&+\frac{1}{2}\int d\sigma^0\ \bigg(\xi^{\mu}g_{\mu\nu}\partial_1\xi^{\nu}+\xi^{\mu}\big(B_{\mu\nu}-F_{\mu\nu}(\bar{X})\big)\partial_0\xi^{\nu}
\nn\\
&&+\xi^{\mu}\xi^{\nu}\partial_{\mu}\big(B_{\nu\rho}-F_{\nu\rho}\big)(\bar{X})\partial_0\bar{X}^{\rho}
\bigg),
\eea
the Green's function in the bulk satisfies the equation
\bea
g_{\mu\nu}\big(\partial_0^2-\partial_1^2\big)G^{\nu\rho}(\sigma-\sigma^{\prime})=-2\pi i\alpha^{\prime}\delta_{\mu}{}^{\rho}\delta^2(\sigma-\sigma^{\prime}).
\eea
We can use the field redefinition:
\bea
\frac{z+\bar{z}}{2}\equiv\sigma^1; \ \frac{z-\bar{z}}{2}\equiv\sigma^0
\eea
to obtain:
\bea
g_{\mu\nu}\big(\partial_0^2-\partial_1^2\big)G^{\nu\rho}(\sigma-\sigma^{\prime})=-2g_{\mu\nu}\partial_z\partial_{\bar{z}}G^{\nu\rho}(z-z^{\prime})=-4\pi i\alpha^{\prime}\delta_{\mu}{}^{\rho}\delta^2(z-z^{\prime}),
\eea
in which we use:
\bea
\partial_z=\frac{1}{2}(\partial_0+\partial_1); \
\partial_{\bar{z}}=\frac{1}{2}(\partial_1-\partial_0); \
\partial_z\partial_{\bar{z}}=\frac{1}{4}\big(\partial_1^2-\partial_0^2\big),
\eea
and
\bea
d^2\sigma&=&dzd\bar{z}\bigg|\frac{\partial(\sigma^0,\sigma^1)}{\partial(z,\bar{z})}\bigg|=\frac{1}{2}dzd\bar{z};
\nn\\
\int d^2\sigma\ \delta^2(\sigma-\sigma^{\prime})&=&1=\frac{1}{2}\int dzd\bar{z}\ \delta^2(\sigma-\sigma^{\prime});
\nn\\
\delta^2(\sigma-\sigma^{\prime})&=&2\delta^2(z-z^{\prime})
\eea
and assume
\bea
G^{\nu\rho}(\sigma-\sigma^{\prime})=\frac{1}{2}G^{\nu\rho}(z-z^{\prime}).
\eea
Hence, the Green's function in the bulk is given by
\bea
\frac{1}{4\pi\alpha^{\prime}}G^{\nu\rho}(z-z^{\prime})=-\frac{g^{\nu\rho}}{4\pi}\ln|z-z^{\prime}|^2,
\eea
which can be checked by
\bea
\int d^2z\ \bigg(\partial_zV^z+\partial_{\bar{z}}V^{\bar{z}}\bigg)=\oint\ \big(V^z d\bar{z}-V^{\bar{z}}dz\big),
\eea
in which the contour integral is counterclockwise,
\bea
\int  d^2z\ \bigg(\partial_z\partial_{\bar{z}}\big(\ln|z-z^{\prime}|^2\big)f(z-z^{\prime})\bigg)&=&\int d^2z\ \partial_{\bar{z}}\bigg(\frac{1}{z-z^{\prime}}f(z-z^{\prime})\bigg)
\nn\\
&=&-\oint\frac{1}{z-z^{\prime}}f(z-z^{\prime})dz
\nn\\
&=&-2\pi if(0);
\nn\\
\partial_z\partial_{\bar{z}}\big(\ln|z-z^{\prime}|^2\big)&=&-2\pi i\delta^2(z-z^{\prime}).
\eea

\subsubsection{Green's Function on Boundary}
\noindent
The corresponding equations follow the Green's function at the boundary \cite{Fradkin:1985qd}:
\bea
&&g_{\mu\nu}\partial_1G^{\nu\rho}+\big(B_{\mu\nu}-F_{\mu\nu}(\bar{X})\big)\partial_0G^{\nu\rho}
\nn\\
&=&\big(g_{\mu\nu}+(B_{\mu\nu}-F_{\mu\nu})\big)\partial_z G^{\nu\rho}+\big(g_{\mu\nu}-(B_{\mu\nu}-F_{\mu\nu})\big)\partial_z G^{\nu\rho}
\nn\\
&=&0
\eea
from the fluctuation of the target space \cite{Fradkin:1985qd}
\bea
&&-2\pi\alpha^{\prime}S_{\mathrm{2bsmc}}[\xi]
\nn\\
&=&\frac{1}{2}\int d^2\sigma\ \bigg(-\xi^{\mu}g_{\mu\nu}\partial_{\alpha}\partial^{\alpha}\xi^{\nu}
\bigg)
\nn\\
&&+\frac{1}{2}\int d\sigma^0\ \bigg(\xi^{\mu}g_{\mu\nu}\partial_1\xi^{\nu}+\xi^{\mu}\big(B_{\mu\nu}-F_{\mu\nu}(\bar{X})\big)\partial_0\xi^{\nu}
\nn\\
&&+\xi^{\mu}\xi^{\nu}\partial_{\mu}\big(B_{\nu\rho}-F_{\nu\rho}\big)(\bar{X})\partial_0\bar{X}^{\rho}
\bigg).
\eea
The Green's function solution at the boundary is obtained from the background solution, assuming the field strength remains constant, as detailed in Ref. \cite{Fradkin:1985qd}
\bea
&&\frac{1}{4\pi\alpha^{\prime}}G^{\mu\nu}\bigg|_{z=-\bar{z}, z^{\prime}=-\bar{z^{\prime}}}
\nn\\
&=&-\frac{1}{8\pi}g^{\mu\nu}\ln|z-z^{\prime}|^2-\frac{1}{8\pi}(g+B-F)^{\mu\rho}\big(g-(B-F)\big)_{\rho\sigma}g^{\sigma\nu}\ln(z+\bar{z}^{\prime})
\nn\\
&&-\frac{1}{8\pi}\big(g-(B-F)\big)^{\mu\rho}\big(g+(B-F)\big)_{\rho\sigma}g^{\sigma\nu}\ln(\bar{z}+z^{\prime})\bigg|_{z=-\bar{z}, z^{\prime}=-\bar{z^{\prime}}},
\eea
in which we used:
\bea
&&\big(g_{\mu\nu}+(B_{\mu\nu}-F_{\mu\nu})\big)\partial_z\bigg(\frac{1}{8\pi}g^{\nu\rho}\ln|z-z^{\prime}|^2\bigg)\bigg|_{z=-\bar{z}, z^{\prime}=-\bar{z^{\prime}}}
\nn\\
&=&\frac{1}{8\pi}\big(g_{\mu\nu}+(B_{\mu\nu}-F_{\mu\nu})\big)g^{\nu\rho}\frac{1}{z-z^{\prime}};
\nn\\
&&\big(g_{\mu\nu}-(B_{\mu\nu}-F_{\mu\nu})\big)\partial_{\bar{z}}\bigg(\frac{1}{8\pi}g^{\nu\rho}\ln|z-z^{\prime}|^2\bigg)\bigg|_{z=-\bar{z}, z^{\prime}=-\bar{z^{\prime}}}
\nn\\
&=&\frac{1}{8\pi}\big(g_{\mu\nu}-(B_{\mu\nu}-F_{\mu\nu})\big)g^{\nu\rho}\frac{1}{\bar{z}-\bar{z^{\prime}}}\bigg|_{z=-\bar{z}, z^{\prime}=-\bar{z^{\prime}}}
\nn\\
&=&-\frac{1}{8\pi}\big(g_{\mu\nu}-(B_{\mu\nu}-F_{\mu\nu})\big)g^{\nu\rho}\frac{1}{z-z^{\prime}};
\nn\\
&&\big(g_{\mu\nu}+(B_{\mu\nu}-F_{\mu\nu})\big)
\nn\\
&&\times\partial_z\bigg(\frac{1}{8\pi}(g+B-F)^{\nu\sigma}\big(g-(B-F)\big)_{\sigma\delta}g^{\delta\rho}\ln(z+\bar{z}^{\prime})\bigg)\bigg|_{z=-\bar{z}, z^{\prime}=-\bar{z^{\prime}}}
\nn\\
&=&\frac{1}{8\pi}\big(g_{\mu\delta}-(B_{\mu\delta}-F_{\mu\delta})\big)g^{\nu\rho}\frac{1}{z-z^{\prime}};
\nn\\
&&\big(g_{\mu\nu}-(B_{\mu\nu}-F_{\mu\nu})\big)
\nn\\
&&\times\partial_{\bar{z}}\bigg(\frac{1}{8\pi}\big(g-(B-F)\big)^{\nu\sigma}\big(g+(B-F)\big)_{\sigma\delta}g^{\delta\rho}\ln(\bar{z}+z^{\prime})\bigg)\bigg|_{z=-\bar{z}, z^{\prime}=-\bar{z^{\prime}}}
\nn\\
&=&\frac{1}{8\pi}\big(g_{\mu\delta}+(B_{\mu\delta}-F_{\mu\delta})\big)g^{\delta\rho}\frac{1}{\bar{z}-\bar{z^{\prime}}}\bigg|_{z=-\bar{z}, z^{\prime}=-\bar{z^{\prime}}}
\nn\\
&=&-\frac{1}{8\pi}\big(g_{\mu\delta}+(B_{\mu\delta}-F_{\mu\delta})\big)g^{\delta\rho}\frac{1}{z-z^{\prime}}.
\eea
When one sets $B-F=0$, the Green's function on the boundary reduces to the Green's function in the bulk:
\bea
&&\frac{1}{4\pi\alpha^{\prime}}G^{\mu\nu}\bigg|_{z=-\bar{z}, z^{\prime}=-\bar{z^{\prime}}}
\nn\\
&=&-\frac{1}{8\pi}g^{\mu\nu}\ln|z-z^{\prime}|^2-\frac{1}{8\pi}(g+B-F)^{\mu\rho}\big(g-(B-F)\big)_{\rho\sigma}g^{\sigma\nu}\ln(z+\bar{z}^{\prime})
\nn\\
&&-\frac{1}{8\pi}\big(g-(B-F)\big)^{\mu\rho}\big(g+(B-F)\big)_{\rho\sigma}g^{\sigma\nu}\ln(\bar{z}+z^{\prime})\bigg|_{z=-\bar{z}, z^{\prime}=-\bar{z^{\prime}}, B-F=0}
\nn\\
&=&-\frac{1}{8\pi}g^{\mu\nu}\ln|z-z^{\prime}|^2-\frac{1}{8\pi}g^{\mu\nu}\ln z-z^{\prime}-\frac{1}{8\pi}g^{\mu\nu}\ln\bar{z}-\bar{z}^{\prime}
\nn\\
&=&-\frac{1}{8\pi}g^{\mu\nu}\ln|z-z^{\prime}|^2-\frac{1}{8\pi}g^{\mu\nu}\ln|z-z^{\prime}|^2
\nn\\
&=&-\frac{1}{4\pi}g^{\mu\nu}\ln|z-z^{\prime}|^2.
\eea

\subsubsection{$\beta$ Function}
\noindent
From the result of the fluctuation of the target space \cite{Fradkin:1985qd}
\bea
&&-2\pi\alpha^{\prime}S_{\mathrm{2bsmc}}[\xi]
\nn\\
&=&\frac{1}{2}\int d^2\sigma\ \bigg(-\xi^{\mu}g_{\mu\nu}\partial_{\alpha}\partial^{\alpha}\xi^{\nu}
\bigg)
\nn\\
&&+\frac{1}{2}\int d\sigma^0\ \bigg(\xi^{\mu}g_{\mu\nu}\partial_1\xi^{\nu}+\xi^{\mu}\big(B_{\mu\nu}-F_{\mu\nu}(\bar{X})\big)\partial_0\xi^{\nu}
\nn\\
&&+\xi^{\mu}\xi^{\nu}\partial_{\mu}\big(B_{\nu\rho}-F_{\nu\rho}\big)(\bar{X})\partial_0\bar{X}^{\rho}
\bigg),
\eea
one can find the one-loop divergence term \cite{Fradkin:1985qd}
\bea
-\frac{1}{4\pi\alpha^{\prime}}\int d\sigma^0\ \xi^{\mu}\xi^{\nu}\partial_{\mu}\big(B_{\nu\rho}-F_{\nu\rho}\big)(\bar{X})\partial_0\bar{X}^{\rho}.
\eea
Thus, one should put a counter term $S_{\mathrm{ct}}$ to cancel the one-loop divergence term \cite{Fradkin:1985qd}
\bea
S_{\mathrm{ct}}\equiv-\int d\sigma^0\ \Gamma_{\mu}\partial_0\bar{X}^{\mu},
\eea
where
\bea
\Gamma_{\mu}\equiv-\frac{1}{4\pi\alpha^{\prime}}\lim_{\epsilon\rightarrow 0}G^{\nu\rho}_{\epsilon}\partial_{\nu}(B_{\rho\mu}-F_{\rho\mu})(\bar{X})=\frac{1}{4\pi\alpha^{\prime}}\lim_{\epsilon\rightarrow 0}G^{\nu\rho}_{\epsilon}\partial_{\nu}F_{\rho\mu}(\bar{X})
\eea
and $G^{\nu\rho}_{\epsilon}$ is the regularized Green's function by regularizing $\sigma^0-\sigma^{0\prime}$.
If one takes the solution of the Green's function on the boundary, one can obtain \cite{Fradkin:1985qd}
\bea
S_{\mathrm{ct}}&=&-\int d\sigma^0\ \Gamma_{\mu}\partial_0\bar{X}^{\mu}
\nn\\
&=&\int d\sigma^0\ \bigg(\frac{1}{4\pi}g^{\nu\rho}\ln\epsilon+\frac{1}{8\pi}\big(g+(B-F)(\bar{X})\big)^{\nu\sigma}\big(g-(B-F)(\bar{X})\big)_{\sigma\delta}g^{\delta\rho}\ln\epsilon
\nn\\
&&+\frac{1}{8\pi}\big(g-(B-F)(\bar{X})\big)^{\nu\sigma}\big(g+(B-F)(\bar{X})\big)_{\sigma\delta}g^{\delta\rho}\ln\epsilon\bigg)
\partial_{\nu}F_{\rho\mu}(\bar{X})
\partial_0\bar{X}^{\mu}
\nn\\
&=&\ln(\epsilon)\int d\sigma^0\ \bigg(\frac{1}{4\pi}g^{\nu\rho}+\frac{1}{8\pi}\big(g+(B-F)(\bar{X})\big)^{\nu\sigma}\big(g-(B-F)(\bar{X})\big)_{\sigma\delta}g^{\delta\rho}
\nn\\
&&+\frac{1}{8\pi}\big(g-(B-F)(\bar{X})\big)^{\nu\sigma}\big(g+(B-F)(\bar{X})\big)_{\sigma\delta}g^{\delta\rho}\bigg)
\partial_{\nu}F_{\rho\mu}(\bar{X})
\partial_0\bar{X}^{\mu}.
\eea
Hence, one can define the one-loop $\beta$ function $\beta_{\mu}$ as \cite{Fradkin:1985qd}:
\bea
&&\beta_{\mu}\ln\epsilon
\nn\\
&\equiv&\Gamma_{\mu}
\nn\\
&=&-\bigg(\frac{1}{4\pi}g^{\nu\rho}+\frac{1}{8\pi}\big(g+(B-F)(\bar{X})\big)^{\nu\sigma}\big(g-(B-F)(\bar{X})\big)_{\sigma\delta}g^{\delta\rho}
\nn\\
&&+\frac{1}{8\pi}\big(g-(B-F)(\bar{X})\big)^{\nu\sigma}\big(g+(B-F)(\bar{X})\big)_{\sigma\delta}g^{\delta\rho}\bigg)
\partial_{\nu}F_{\rho\mu}(\bar{X})\ln\epsilon;
\nn\\
&&\beta_{\mu}
\nn\\
&=&-\bigg(\frac{1}{4\pi}g^{\nu\rho}+\frac{1}{8\pi}\big(g+(B-F)(\bar{X})\big)^{\nu\sigma}\big(g-(B-F)(\bar{X})\big)_{\sigma\delta}g^{\delta\rho}
\nn\\
&&+\frac{1}{8\pi}\big(g-(B-F)(\bar{X})\big)^{\nu\sigma}\big(g+(B-F)(\bar{X})\big)_{\sigma\delta}g^{\delta\rho}\bigg)
\partial_{\nu}F_{\rho\mu}(\bar{X}).
\eea

\subsubsection{Equation of Motion}
\noindent
Now we want to show that the vanishing one-loop $\beta$ function leads to an equation of motion of the DBI theory.
The one-loop $\beta$ function can be rewritten as \cite{Fradkin:1985qd}:
\bea
&&\beta_{\mu}
\nn\\
&=&-\bigg(\frac{1}{4\pi}g^{\nu\rho}+\frac{1}{8\pi}\big(g+(B-F)(\bar{X})\big)^{\nu\sigma}\big(g-(B-F)(\bar{X})\big)_{\sigma\delta}g^{\delta\rho}
\nn\\
&&+\frac{1}{8\pi}\big(g-(B-F)(\bar{X})\big)^{\nu\sigma}\big(g+(B-F)(\bar{X})\big)_{\sigma\delta}g^{\delta\rho}\bigg)
\partial_{\nu}F_{\rho\mu}(\bar{X})
\nn\\
&=&-\frac{1}{4\pi}\bigg(g^{\nu\rho}+\frac{1}{2}\big(g+(B-F)(\bar{X})\big)^{\nu\sigma}\big(g-(B-F)(\bar{X})\big)_{\sigma\delta}g^{\delta\rho}
\nn\\
&&+\frac{1}{2}\big(g-(B-F)(\bar{X})\big)^{\nu\sigma}\big(g+(B-F)(\bar{X})\big)_{\sigma\delta}g^{\delta\rho}\bigg)
\partial_{\nu}F_{\rho\mu}(\bar{X})
\nn\\
&=&-\frac{1}{4\pi}\bigg(\big(g-(B-F)^2(\bar{X})\big)^{\nu\delta}\big(g-(B-F)^2(\bar{X})\big)_{\delta\sigma}g^{\sigma\rho}
\nn\\
&&+\frac{1}{2}\big(g-(B-F)^2(\bar{X})\big)^{\nu\delta}\big(g+(B-F)^2(\bar{X})-2g(B-F)(\bar{X})\big)_{\delta\sigma}g^{\sigma\rho}
\nn\\
&&+\frac{1}{2}\big(g-(B-F)^2(\bar{X})\big)^{\nu\delta}\big(g+(B-F)^2(\bar{X})+2g(B-F)(\bar{X})\big)_{\delta\sigma}g^{\sigma\rho}\bigg)\partial_{\nu}F_{\rho\mu}(\bar{X})
\nn\\
&=&-\frac{1}{2\pi}\big(g-(B-F)^2(\bar{X})\big)^{\nu\delta}\partial_{\nu}F_{\delta\mu}(\bar{X}).
\eea
Now we calculate:
\bea
&&\big(g-(B-F)^2(\bar{X})\big)^{\mu\nu}\beta_{\nu}
\nn\\
&=&-\frac{1}{2\pi}\big(g-(B-F)^2(\bar{X})\big)^{\mu\nu}\big(g-(B-F)^2(\bar{X})\big)^{\rho\sigma}\partial_{\rho}F_{\sigma\nu}(\bar{X})
\nn\\
&=&-\frac{1}{2\pi}\big(g-(B-F)^2(\bar{X})\big)^{\mu\nu}\big(g-(B-F)^2(\bar{X})\big)^{\rho\sigma}\partial_{\rho}F_{\sigma\nu}(\bar{X})
\nn\\
&=&-\frac{1}{2\pi}g^{\rho\sigma}\big(g-(B-F)^2(\bar{X})\big)^{\mu\nu}\partial_{\rho}F_{\sigma\nu}(\bar{X})
\nn\\
&&-\frac{1}{2\pi}\bigg((g-(B-F)^2(\bar{X}))^{-1}(B-F)^2(\bar{X})g^{-1}\bigg)^{\rho\sigma}\big(g-(B-F)^2(\bar{X})\big)^{\mu\nu}\partial_{\rho}F_{\sigma\nu}(\bar{X}),
\nn\\
\eea
in which we used:
\bea
(a+b)^{-1}&=&a^{-1}-a^{-1}(I_D+ba^{-1})^{-1}ba^{-1};
\nn\\
\big(g-(B-F)^2(\bar{X})\big)^{-1}&=&g^{-1}+(g-(B-F)^2(\bar{X}))^{-1}(B-F)^2(\bar{X})g^{-1}
\eea
in the third equality.
We have the following identities:
\bea
&&\partial^{\rho}\bigg(\big(g-(B-F)^2(\bar{X})\big)^{\mu\nu}(B-F)_{\nu\rho}(\bar{X})\bigg)
\nn\\
&=&-\big(g-(B-F)^2(\bar{X})\big)^{\mu\nu}\partial^{\rho}F_{\nu\rho}(\bar{X})
\nn\\
&&-\big(g-(B-F)^2(\bar{X})\big)^{\mu\rho}\bigg(\partial^{\rho}\big(g-(B-F)^2(\bar{X})\big)_{\rho\gamma}\bigg)\big(g-(B-F)^2(\bar{X})\big)^{\gamma\nu}
\nn\\
&&\times(B-F)_{\nu\rho}(\bar{X});
\eea
\bea
&&\partial^{\rho}\bigg(\big(g-(B-F)^2(\bar{X})\big)^{\mu\nu}(B-F)_{\nu\rho}(\bar{X})\bigg)
\nn\\
&&-\big(g-(B-F)^2(\bar{X})\big)^{\mu\nu}(B-F)_{\nu\rho}(\bar{X})\big(\partial^{\sigma}(B-F)^{\rho}{}_{\delta}(\bar{X})\big)
\nn\\
&&\times
\big(g-(B-F)^2(\bar{X})\big)^{\delta\gamma}(B-F)_{\gamma\sigma}(\bar{X})
\nn\\
&=&-\big(g-(B-F)^2(\bar{X})\big)^{\mu\nu}\partial^{\rho}F_{\nu\rho}(\bar{X})
\nn\\
&&+\big(g-(B-F)^2(\bar{X})\big)^{\mu\nu}\big(\partial^{\sigma}(B-F)_{\nu\rho}(\bar{X})\big)(B-F)^{\rho}{}_{\delta}(\bar{X})
\nn\\
&&\times
\big(g-(B-F)^2(\bar{X})\big)^{\delta\gamma}(B-F)_{\gamma\sigma}(\bar{X})
\nn\\
&=&g^{\rho\sigma}\big(g-(B-F)^2(\bar{X})\big)^{\mu\nu}\partial_{\rho}F_{\sigma\nu}(\bar{X})
\nn\\
&&+\bigg((g-(B-F)^2(\bar{X}))^{-1}(B-F)^2(\bar{X})g^{-1}\bigg)^{\rho\sigma}\big(g-(B-F)^2(\bar{X})\big)^{\mu\nu}\partial_{\rho}F_{\sigma\nu}(\bar{X}).
\nn\\
\eea
We also used the notation
\bea
\bigg(\frac{(B-F)(\bar{X})}{g-(B-F)^2(\bar{X})}\bigg)^{\mu}{}_{\nu}
\equiv\big(g-(B-F)^2(\bar{X})\big)^{\mu\sigma}\big((B-F)(\bar{X})\big)_{\sigma\nu}.
\eea
Hence, we obtain:
\bea
&&\big(g-(B-F)^2(\bar{X})\big)^{\mu\nu}\beta_{\nu}
\nn\\
&=&-\frac{1}{2\pi}\bigg\lbrack g^{\rho\sigma}\big(g-(B-F)^2(\bar{X})\big)^{\mu\nu}\partial_{\rho}F_{\sigma\nu}(\bar{X})
\nn\\
&&+\bigg((g-(B-F)^2(\bar{X}))^{-1}(B-F)^2(\bar{X})g^{-1}\bigg)^{\rho\sigma}\big(g-(B-F)^2(\bar{X})\big)^{\mu\nu}\partial_{\rho}F_{\sigma\nu}(\bar{X})\bigg\rbrack
\nn\\
&=&-\frac{1}{2\pi}\bigg\lbrack\partial^{\rho}\bigg(\big(g-(B-F)^2(\bar{X})\big)^{\mu\nu}(B-F)_{\nu\rho}(\bar{X})\bigg)
\nn\\
&&-\big(g-(B-F)^2(\bar{X})\big)^{\mu\nu}(B-F)_{\nu\rho}(\bar{X})\big(\partial^{\sigma}(B-F)^{\rho}{}_{\delta}(\bar{X})\big)
\nn\\
&&\times
\big(g-(B-F)^2(\bar{X})\big)^{\delta\gamma}(B-F)_{\gamma\sigma}(\bar{X})\bigg\rbrack.
\eea
When $\beta_{\nu}=0$, it exactly implies the equation of motion for the DBI theory with the constant dilaton
\bea
S_{\mathrm{DBI}}=-T_{D-1}\int d^Dx\ e^{-\phi}\sqrt{\det(g_{\mu\nu}+B_{\mu\nu}-F_{\mu\nu})}.
\label{DBI}
\eea
The DBI theory gives the effective action for the open-string sector, and the gravity part can be obtained by including the closed-string sector.

\subsection{T-Duality}
\noindent
We demonstrate how the T-duality transforms the DBI action \eqref{DBI} with one compact direction $y$.
The background fields and gauge fields are independent of the compact direction, and the integration along the direction shows
\bea
\int dy=2\pi\sqrt{\alpha^{\prime}},
\eea
which ensures
\bea
2\pi\sqrt{\alpha^{\prime}}T_{D-1}=\frac{1}{(2\pi)^{D-2}(\alpha^{\prime})^{(D-1)/2}}=T_{D-2}.
\eea
We assume that the gauge field with the component $y$ is a new scalar field $\Phi$.
The determinant can be expressed in a different form
\bea
&&
\det(g_{\mu\nu}+B_{\mu\nu}-F_{\mu\nu})
\nn\\
&=&
g_{yy}\det\bigg(g_{jk}+B_{jk}-F_{jk}-\frac{(g_{jy}+B_{jy}-\partial_j\Phi)(g_{yk}+B_{yk}+\partial_k\Phi)}{g_{yy}}\bigg)
\nn\\
&=&
g_{yy}\det\big(\tilde{g}_{jk}+\tilde{B}_{jk}-F_{jk}
\nn\\
&&
+\tilde{g}_{yy}(\partial_j\Phi)(\partial_k\Phi)
+(\partial_j\Phi)(\tilde{g}+\tilde{B})_{yk}
+(\partial_k\Phi)(\tilde{g}-\tilde{B})_{yj}\big),
\eea
in which we use the formula for the general matrix
\bea
\det(M_{\mu\nu})=M_{yy}\det\bigg(M_{jk}-\frac{M_{jy}M_{yk}}{M_{yy}}\bigg).
\eea
The dual fields under the T-duality is denoted as $\tilde{g}_{jk}$, $\tilde{B}_{jk}$.
When combining the dilaton field and the prefactor of the determinant $g_{yy}$, we obtain
\bea
e^{-\phi}\sqrt{g_{yy}}=e^{-\phi}e^{\ln\sqrt{g_{yy}}}=e^{-\tilde{\phi}}.
\eea
We also use the notation of the induced background fields
\bea
(\tilde{g}_{jk})_{\mathrm{induced}}=(\partial_j\Phi^{\mu})\tilde{g}_{\mu\nu}(\partial_k\Phi^{\nu}),
\eea
in which $\Phi^{j}=x^{j}$ for the non-compact direction, to simply the deterinenat
\bea
&&
\det\big(\tilde{g}_{jk}+\tilde{B}_{jk}-F_{jk}
+\tilde{g}_{yy}(\partial_j\Phi)(\partial_k\Phi)
+(\partial_j\Phi)(\tilde{g}+\tilde{B})_{yk}
+(\partial_k\Phi)(\tilde{g}-\tilde{B})_{yj}\big)
\nn\\
&=&
\det\big((\tilde{g}_{jk}+\tilde{B}_{jk})_{\mathrm{induced}}-F_{jk}\big).
\eea
Hence, the DBI action under the T-duality becomes
\bea
S_{\mathrm{DBIT}}=-T_{D-2}\int d^{D-1}x\ e^{-\tilde{\phi}}\sqrt{\det\big((\tilde{g}_{jk}+\tilde{B}_{jk})_{\mathrm{induced}}-F_{jk}\big)}.
\eea
For the closed string part, the procedure of the T-duality is similar.

\subsection{S-Duality}
\noindent
Let us begin from the 10D SO(32) Heterotic string effective action
\bea
S_{\mathrm{HET}}&\equiv&\frac{1}{16\pi G_N}\int d^{10}x\sqrt{|\det g_{\mu\nu}|}\  e^{-2\phi}\bigg(R+4\partial_{\rho}\phi\partial^{\rho}\phi-\frac{1}{12}H_{\sigma\delta\gamma}H^{\sigma\delta\gamma}\bigg)
\nn\\
&&-\frac{1}{4g_{\mathrm{YM}}^2}\int d^{10}x\sqrt{|\det g_{\mu\nu}|}\ e^{-2\phi}\mathrm{Tr}(F_{\mu\nu}F^{\mu\nu}),
\eea
where
\bea
H_{\mu\nu\rho}&\equiv&\partial_{\mu}B_{\nu\rho}+\partial_{\nu}B_{\rho\mu}+\partial_{\rho}B_{\mu\nu}
-\frac{\alpha^{\prime}}{4}(\omega_{\mu\nu\rho}^{\mathrm{YM}}-\omega_{\mu\nu\rho}^\mathrm{L})
\nn\\
F_{\mu\nu}&\equiv&\partial_{\mu}A_{\nu}-\partial_{\nu}A_{\mu}-i\lbrack A_{\mu}, A_{\nu}\rbrack.
\eea
The Yang-Mills Chern-Simons three-form is
\bea
\omega_{\mu\nu\rho}^{\mathrm{YM}}\equiv\mathrm{Tr}\bigg(A_{\mu}\partial_{\nu}A_{\rho}+\frac{2}{3}A_{\mu}A_{\nu}A_{\rho}\bigg),
\eea
and the Lorentz Chern-Simons three-form is
\bea
\omega_{\mu\nu\rho}^{\mathrm{L}}\equiv\mathrm{Tr}\bigg(\Omega_{\mu}\partial_{\nu}\Omega_{\rho}+\frac{2}{3}\Omega_{\mu}\Omega_{\nu}\Omega_{\rho}\bigg),
\eea
where $\Omega$ is the spin connection.
The $F_{\mu\nu}$ is the SO(32) field strength and the trace is in the adjoint representation.
The commutator is defined as follows
\bea
\lbrack A_{\mu}, A_{\nu}\rbrack\equiv A_{\mu}^aA_{\nu}^b(T^aT^b-T^bT^a),
\eea
where $A_{\mu}^a$ and $A_{\nu}^b$ are the components of the operators, and $T^a$ and $T^b$ are the corresponding generators.
$ a, b$ denotes the indices of a Lie algebra.
The gauge coupling $g_{\mathrm{YM}}$,
\bea
g^2_{\mathrm{YM}}=(2\pi)^7(\alpha^{\prime})^3,
\eea
can be determined by
\bea
\frac{8\pi G_N}{g^2_{\mathrm{YM}}}=\frac{\alpha^{\prime}}{2}.
\eea
We first transform the metric by the dilaton field
\bea
g_{\mu\nu}\rightarrow e^{\phi}g_{\mu\nu},
\eea
then we invert the string coupling:
\bea
g_s\equiv e^{\phi}\rightarrow \frac{1}{g_s},
\eea
and finally, we identify the NS-NS two-form field background as the R-R two-form field background.
\\

\noindent
This S-duality procedure gives the Type I string effective theory up to the two-derivative terms
\bea
S_{\mathrm{I}}&\equiv&\frac{1}{16\pi G_N}\int d^{10}x\sqrt{|\det g_{\mu\nu}|}\  e^{-2\phi}\bigg(R+4\partial_{\rho}\phi\partial^{\rho}\phi-\frac{e^{2\phi}}{12}H_{\sigma\delta\gamma}H^{\sigma\delta\gamma}\bigg)
\nn\\
&&-\frac{1}{4g_{\mathrm{YM}}^2}\int d^{10}x\sqrt{|\det g_{\mu\nu}|}\ e^{-\phi}\mathrm{Tr}(F_{\mu\nu}F^{\mu\nu}).
\eea
We redefine the DBI's field strength as $F_{\mu\nu}\rightarrow (2\pi\alpha^{\prime})F_{\mu\nu}$ for obtaining the Type I string effective action.
Now, the field strength of the R-R two-form field background, and we obtain the open string sector from $F_{\mu\nu}$.
Hence, the S-duality connects the closed string and open string sectors, which also generalizes the electromagnetic duality to the gauge theory.
Here, we also show that S-duality exchanges the NS-NS and R-R two-form field backgrounds.

\section{Non-Commutative Geometry of NS-NS D-Branes}
\label{sec:4}
\noindent
We first introduce open string parameters that we will use in the non-commutative theory from the Green's function \cite{Seiberg:1999vs}.
To have the consistent truncation to non-commutative Yang-Mills (YM) theory, containing the non-commutativity parameter expansion, we consider the following scaling limit \cite{Seiberg:1999vs}:
\bea
\alpha^{\prime}\sim\epsilon^{\frac{1}{2}}\rightarrow 0; \
g_{\dot{\alpha}\dot{\beta}}\sim\epsilon\rightarrow 0,
\eea
where we use $\dot{\alpha}, \dot{\beta}$ to denote the directions having the $B$-field background.
If the rank of the $B$-field background is $r$, the indices of the range is $\dot{\alpha}, \dot{\beta}=1, 2, \cdots, r$.
We introduce the generic concepts related to the relationship between commutative and non-commutative descriptions from the SW map and use the map to derive the non-commutative YM theory from the DBI theory for the U(1) case \cite{Seiberg:1999vs}.

\subsection{Open String Parameter}
\noindent
According to the result of the Green's function  on the boundary \cite{Fradkin:1985qd}
\bea
&&\frac{1}{4\pi\alpha^{\prime}}G^{\mu\nu}\bigg|_{z=-\bar{z}, z^{\prime}=-\bar{z^{\prime}}}
\nn\\
&=&-\frac{1}{8\pi}g^{\mu\nu}\ln|z-z^{\prime}|^2-\frac{1}{8\pi}(g+B-F)^{\mu\rho}\big(g-(B-F)\big)_{\rho\sigma}g^{\sigma\nu}\ln(z+\bar{z}^{\prime})
\nn\\
&&-\frac{1}{8\pi}\big(g-(B-F)\big)^{\mu\rho}\big(g+(B-F)\big)_{\rho\sigma}g^{\sigma\nu}\ln(\bar{z}+z^{\prime})\bigg)\bigg|_{z=-\bar{z}, z^{\prime}=-\bar{z^{\prime}}},
\eea
we can show the two points of the target space without the gauge field:
\bea
&&\langle X^{\mu}(z) X^{\nu}(z^{\prime})\rangle\bigg|_{z=-\bar{z}, z^{\prime}=-\bar{z^{\prime}}}
\nn\\
&=&-\alpha^{\prime}\bigg(g^{\mu\nu}\ln|z-z^{\prime}|
+\frac{1}{2}(g+B)^{\mu\rho}\big(g-B\big)_{\rho\sigma}g^{\sigma\nu}\ln(z+\bar{z}^{\prime})
\nn\\
&&+\frac{1}{2}\big(g-B\big)^{\mu\rho}\big(g+B)\big)_{\rho\sigma}g^{\sigma\nu}\ln(\bar{z}+z^{\prime})\bigg)\bigg|_{z=-\bar{z}, z^{\prime}=-\bar{z^{\prime}}}
\nn\\
&=&-\alpha^{\prime}\bigg(g^{-1}\ln|z-z^{\prime}|
+\frac{1}{2}(g+B)^{-1}(g-B)g^{-1}(g-B)(g-B)^{-1}\ln(z+\bar{z^{\prime}})
\nn\\
&&+\frac{1}{2}(g-B)^{-1}(g+B)g^{-1}(g+B)(g+B)^{-1}\ln(\bar{z}+z^{\prime})\bigg)\bigg|_{z=-\bar{z}, z^{\prime}=-\bar{z^{\prime}}}.
\eea
We can show that the symmetrization of the inverse of $g+B$ is \cite{Seiberg:1999vs}:
\bea
\bigg(\frac{1}{g+B}\bigg)_\mathrm{S}&\equiv&
\frac{1}{2}\big((g+B)^{-1}+(g-B)^{-1}\big)
\nn\\
&=&(g+B)^{-1}g(g-B)^{-1}
=(g-B)^{-1}g(g+B)^{-1},
\eea
and the asntisymmetrized one is \cite{Seiberg:1999vs}:
\bea
\bigg(\frac{1}{g+B}\bigg)_\mathrm{A}&\equiv&
\frac{1}{2}\big((g+B)^{-1}-(g-B)^{-1}\big)
\nn\\
&=&-(g+B)^{-1}B(g-B)^{-1}
=-(g-B)^{-1}B(g+B)^{-1}.
\eea
Therefore, we exchange $(g+B)^{-1}$ and $(g-B)^{-1}$, which do not change the result when they are in the left- and right-most places.
\\

\noindent
We can simplify the two points of the target space further \cite{Seiberg:1999vs}:
\bea
&&\langle X^{\mu}(z) X^{\nu}(z^{\prime})\rangle\bigg|_{z=-\bar{z}, z^{\prime}=-\bar{z^{\prime}}}
\nn\\
&=&-\alpha^{\prime}\bigg(g^{-1}\ln|z-z^{\prime}|
+\frac{1}{2}(g+B)^{-1}(g-B)g^{-1}(g-B)(g-B)^{-1}\ln(z+\bar{z^{\prime}})
\nn\\
&&+\frac{1}{2}(g-B)^{-1}(g+B)g^{-1}(g+B)(g+B)^{-1}\ln(\bar{z}+z^{\prime})\bigg)\bigg|_{z=-\bar{z}, z^{\prime}=-\bar{z^{\prime}}}
\nn\\
&=&-\alpha^{\prime}\bigg(g^{-1}\ln|z-z^{\prime}|
+(g+B)^{-1}(g+Bg^{-1}B)(g-B)^{-1}\ln|z+\bar{z^{\prime}}|
\nn\\
&&
-(g+B)^{-1}B(g-B)^{-1}\ln\frac{z+\bar{z^{\prime}}}{\bar{z}+z^{\prime}}\bigg)\bigg|_{z=-\bar{z}, z^{\prime}=-\bar{z^{\prime}}}
\nn\\
&=&-\alpha^{\prime}\bigg(g^{-1}\ln|z-z^{\prime}|
-g^{-1}\ln|z+\bar{z^{\prime}}|
\nn\\
&&
+G^{-1}\ln|z+\bar{z^{\prime}}|^2+\frac{\theta^{-1}}{2\pi\alpha^{\prime}}\ln\frac{z+\bar{z^{\prime}}}{\bar{z}+z^{\prime}}\bigg)\bigg|_{z=-\bar{z}, z^{\prime}=-\bar{z^{\prime}}}
\nn\\
&=&-\alpha^{\prime}\bigg(
G^{-1}\ln|z-z^{\prime}|^2+\frac{\theta^{-1}}{2\pi\alpha^{\prime}}\ln\frac{z+\bar{z^{\prime}}}{\bar{z}+z^{\prime}}\bigg)\bigg|_{z=-\bar{z}, z^{\prime}=-\bar{z^{\prime}}}
\nn\\
&=&-\alpha^{\prime}\bigg(
G^{-1}\ln|\sigma^0-\sigma^{0\prime}|^2-\frac{i}{2\alpha^{\prime}}\theta^{-1}\sgn(\sigma^0-\sigma^{0\prime})\bigg),
\label{2pt1}
\eea
where the open string metric is \cite{Seiberg:1999vs}
\bea
G_{\mu\nu}=(g-Bg^{-1}B)_{\mu\nu},
\eea
the inverse of the open string metric is \cite{Seiberg:1999vs}
\bea
G^{\mu\nu}=\big((g+B)^{-1}g(g-B)^{-1}\big)^{\mu\nu},
\eea
the non-commutativity parameter is \cite{Seiberg:1999vs}
\bea
\theta^{\mu\nu}=-2\pi\alpha^{\prime}\big((g+B)^{-1}B(g-B)^{-1}\big)^{\mu\nu},
\eea
and the sign function is
\bea
\sgn(\lambda)= \left\{\begin{array}{ll}
1, & \mbox{if $\lambda\ge 0$}; \\
-1, & \mbox{if $\lambda<0$}.
\end{array} \right.
\eea
We can calculate the commutator of the target space by using the time-ordering operator ${\cal T}$ \cite{Seiberg:1999vs},
\bea
\lbrack X^{\mu}(\sigma^0), X^{\nu}(\sigma^0)\rbrack
={\cal T}\big(X^{\mu}(\sigma^0)X^{\nu}(\sigma^{0}_-)-X^{\mu}(\sigma^0)X^{\nu}(\sigma^0_+)\big)
=i\theta^{\mu\nu},
\eea
where $\sigma^0_{\pm}\equiv\sigma^0\pm\delta$ ($\delta>0$).
Consequently, the target space on the D-brane becomes non-commutative when the $B$-field background is turned on \cite{Seiberg:1999vs}.
\\

\noindent
When considering the scaling limit \cite{Seiberg:1999vs}:
\bea
\alpha^{\prime}\sim\epsilon^{\frac{1}{2}}; \ g_{\alpha\beta}\sim\epsilon^0; \ g_{\dot\alpha\dot\beta}\sim\epsilon; \
B_{\dot\alpha\dot\beta}\sim\epsilon^{\frac{1}{2}},
\eea
where we denote the indices for the directions not along the $B$-field background directions by $\alpha, \beta$, and the indices for the directions along the  $B$-field background directions by $\dot{\alpha}, \dot{\beta}$, the dominant term of the open string parameters is \cite{Seiberg:1999vs}
\bea
G_{\dot{\alpha}\dot{\beta}}&\rightarrow& -(Bg^{-1}B)_{\dot\alpha\dot\beta}\sim\epsilon^0;
\nn\\
G^{\dot{\alpha}\dot{\beta}}&\rightarrow&-(B^{-1}gB^{-1})^{\dot\alpha\dot\beta}\sim\epsilon^0;
\nn\\
\theta^{\dot\alpha\dot\beta}&\rightarrow&2\pi\alpha^{\prime}\bigg(\frac{1}{B}\bigg)^{\dot\alpha\dot\beta}\sim\epsilon^0.
\eea
Other components of the open string metric are equivalent to the closed string metric \cite{Seiberg:1999vs}.
The non-commutativity parameter vanishes for other components \cite{Seiberg:1999vs}.
Hence, the open string metric and the non-commutativity parameter are finite under this limit \cite{Seiberg:1999vs}.
Under this scaling limit, the two-point target space becomes \cite{Seiberg:1999vs}
\bea
\langle X^{\mu}(\sigma^0) X^{\nu}(0)\rangle=\frac{i}{2}\theta^{\mu\nu}\sgn(\sigma^0).
\eea
We can also keep the $B$-field background as $\epsilon^0$ by redefining $B\rightarrow 2\pi\alpha^{\prime}B$ \cite{Seiberg:1999vs}.

\subsection{Non-Commutative YM}
\noindent
The action of the non-commutative YM theory is \cite{Tseytlin:1997csa,Terashima:2000ej,Seiberg:1999vs,Cornalba:1999hn}
\bea
S_{\mathrm{NYM}}=-\frac{1}{4g_{D-1}^2}\int d^Dx\ \mathrm{Str}(\hat{F}_{\mu\nu}\divideontimes\hat{F}^{\mu\nu}),
\eea
where
\bea
\mathrm{Str}({\cal O}_1{\cal O}_2\cdots{\cal O}_n)&\equiv&
\mathrm{Tr}\big(\mathrm{Sym}({\cal O}_1{\cal O}_2\cdots{\cal O}_n)\big);
\nn\\
\mathrm{Sym}({\cal O}_1{\cal O}_2\cdots{\cal O}_n)&\equiv&\frac{1}{n!}({\cal O}_1{\cal O}_2\cdots {\cal O}_n+ \mathrm{all\ permutations}).
\eea
The field strength associated with the non-commutative non-Abelian gauge field is
\bea
\hat{F}_{\mu\nu}=\partial_{\mu}\hat{A}_{\nu}-\partial_{\nu}\hat{A}_{\mu}-i\lbrack \hat{A}_{\mu}, \hat{A}_{\nu}\rbrack_{\divideontimes},
\eea
in which the Moyal product $\divideontimes$ is defined as:
\bea
\lbrack{\cal O}_1, {\cal O}_2\rbrack_{\divideontimes}&\equiv&{\cal O}_1\divideontimes{\cal O}_2-{\cal O}_2\divideontimes{\cal O}_1;
\nn\\
{\cal O}_1\divideontimes{\cal O}_2&\equiv& {\cal O}_1\exp\bigg(\frac{i}{2}\theta^{\mu\nu} \overleftarrow{\partial}_{\mu}\overrightarrow{\partial}_{\nu}\bigg){\cal O}_2.
\eea
The gauge field takes a value in a Lie algebra
\bea
A_{\mu}\equiv A_{\mu}^aT^a,
\eea
and the commutator of the Lie algebra satisfies
\bea
\lbrack T^a, T^b\rbrack= if^{abc} T^c,
\eea
where $f^{abc}$ us a structural constant.
The Moyal product deforms the commutative algebra, and the first-order deviation reproduces the Poisson bracket
\bea
\lbrack {\cal O}_1, {\cal O}_2\rbrack_{\divideontimes}=i\theta^{\mu\nu}(\partial_{\mu}{\cal O}_1)(\partial_{\nu}{\cal O}_2)
+{\cal O}(\theta^3).
\eea
The Poisson bracket is
\bea
\{{\cal O}_1, {\cal O}_2\}\equiv\theta^{\mu\nu}(\partial_{\mu}{\cal O}_1)(\partial_{\nu}{\cal O}_2).
\eea
The deformation is not commutative but preserves the associativity
\bea
({\cal O}_1\divideontimes {\cal O}_2)\divideontimes {\cal O}_3={\cal O}_1\divideontimes({\cal O}_2\divideontimes{\cal O}_3).
\eea
The associativity ensures the deformation of the Poisson bracket, Moyal product, to satisfy the first fundamental identity, or the Jacobi identity
\bea
\lbrack{\cal O}_1, \lbrack{\cal O}_2, {\cal O}_3\rbrack_{\divideontimes}\rbrack_{\divideontimes}
+\lbrack{\cal O}_3, \lbrack{\cal O}_1, {\cal O}_2\rbrack_{\divideontimes}\rbrack_{\divideontimes}
+\lbrack{\cal O}_2, \lbrack{\cal O}_3, {\cal O}_1\rbrack_{\divideontimes}\rbrack_{\divideontimes}
=0,
\eea
which is useful to construct the closed gauge transformation.
The difference between the Moyal and ordinary products is the total derivative term
\bea
\int d^Dx\ {\cal O}_1\divideontimes{\cal O}_2\cdots \divideontimes{\cal O}_n=\int d^Dx\ {\cal O}_1{\cal O}_2\cdots{\cal O}_n.
\eea
The Moyal product ensures the non-commutative coordinates
\bea
\lbrack x^{\mu}, x^{\nu}\rbrack_{\divideontimes}=i\theta^{\mu\nu}.
\eea
\\

\noindent
The gauge transformation is:
\bea
\hat{\delta}_{\hat{\lambda}}\hat{A}_{\mu}=\partial_{\mu}\hat{\lambda}+i\lbrack\hat{\lambda}, \hat{A}_{\mu}\rbrack_{\divideontimes}
=\lbrack \hat{D}_{\mu}, \hat{\lambda}\rbrack_{\divideontimes},
\eea
where the covariant product is given by
\bea
\hat{D}_{\mu}\equiv\partial_{\mu}-i\hat{A}_{\mu}.
\eea
The field strength is covariant under the gauge transformation:
\bea
&&\hat{\delta}_{\hat{\lambda}}\hat{F}_{\mu\nu}
\nn\\
&=&i\partial_{\mu}\big(\lbrack\hat{\lambda}, \hat{A}_{\nu}\rbrack_{\divideontimes}\big)-i\partial_{\nu}\big(\lbrack\hat{\lambda}, \hat{A}_{\mu}\rbrack_{\divideontimes}\big)
-i\lbrack\lbrack\hat{D}_{\mu}, \hat{\lambda}\rbrack_{\divideontimes}, \hat{A}_{\nu}\rbrack_{\divideontimes}
-i\lbrack \hat{A}_{\mu}, \lbrack\hat{D}_{\nu}, \hat{\lambda}\rbrack_{\divideontimes}\rbrack_{\divideontimes}
\nn\\
&=&i\lbrack\partial_{\mu}\hat{\lambda}, \hat{A}_{\nu}\rbrack_{\divideontimes}+i\lbrack\hat{\lambda}, \partial_{\mu}\hat{A}_{\nu}\rbrack_{\divideontimes}
-i\lbrack\partial_{\nu}\hat{\lambda}, \hat{A}_{\mu}\rbrack_{\divideontimes}-i\lbrack\hat{\lambda}, \partial_{\nu}\hat{A}_{\mu}\rbrack_{\divideontimes}
\nn\\
&&
-i\lbrack\partial_{\mu}\hat{\lambda}, \hat{A}_{\nu}\rbrack_{\divideontimes}
-\lbrack\lbrack\hat{A}_{\mu}, \hat{\lambda}\rbrack_{\divideontimes}, \hat{A}_{\nu}\rbrack_{\divideontimes}
-i\lbrack\hat{A}_{\mu}, \partial_{\nu}\hat{\lambda}\rbrack_{\divideontimes}
-\lbrack\hat{A}_{\mu}, \lbrack\hat{A}_{\nu}, \hat{\lambda}\rbrack_{\divideontimes}\rbrack_{\divideontimes}
\nn\\
&=&i\lbrack\hat{\lambda}, \hat{F}_{\mu\nu}\rbrack_{\divideontimes}.
\eea
Hence, the non-commutative YM theory is gauge invariant under the transformation.

\subsection{SW Map}
\begin{figure}[h]
\hspace{-2cm}
%\begin{center}
\includegraphics[width=1.4\textwidth]{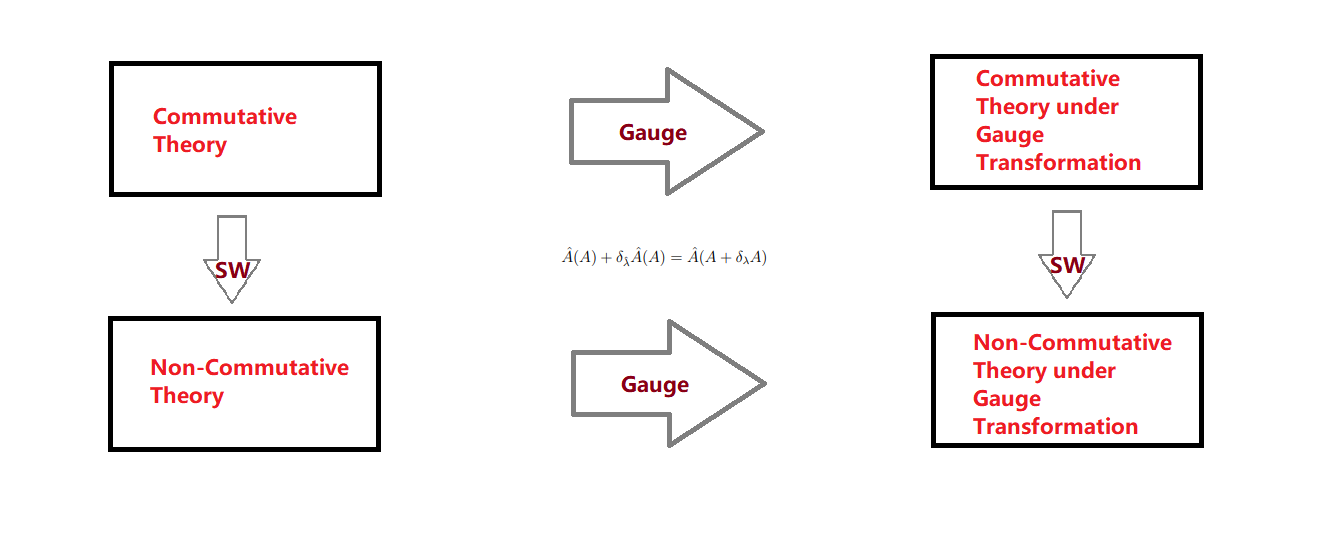}
%\end{center}
\caption{The SW map and gauge transformation commute. }
\label{SW.png}
\end{figure}

\noindent
We will now discuss the connection between non-commutative gauge theory and commutative gauge theory through the Seiberg-Witten map \cite{Seiberg:1999vs}.
In general, the SW map commutes with the gauge transformation \cite{Seiberg:1999vs} as in Fig. \ref{SW.png}.
Starting from the commutative gauge theory, we perform the gauge transformation
\bea
A+\delta_{\lambda}A,
\eea
and then the SW map maps the commutative gauge theory to the non-commutative gauge theory
\bea
\hat{A}(A+\delta_{\lambda}A).
\eea
Another equivalent way is to perform the SW map first
\bea
\hat{A}(A)
\eea
and then perform the gauge transformation
\bea
\hat{A}(A+\delta_{\lambda}A)
\eea
if the SW map and gauge transformation can form a commute diagram.
\\

\noindent
Since the gauge transformation is a redundant symmetry, it does not change a theory.
Hence, the existence of the solution for an SW map
\bea
\hat{A}(A+\delta_{\lambda}A)=\hat{A}(A)+\hat{\delta}_{\hat{\lambda}}\hat{A}
\label{SW}
\eea
implies that the non-commutative description is equivalent to the commutative description, which is invariant under the following gauge transformation:
\bea
\delta_{\lambda} A_{\mu}=\partial_{\mu}\lambda+i\lbrack \lambda, A_{\mu}\rbrack=\lbrack D_{\mu}, \lambda\rbrack.
\eea
In general, we can obtain a solution by the perturbation method.
We can collete all effects in the non-commutativity prameter in $\tilde{A}$ and $\tilde{\lambda}$,
\bea
\hat{A}=A+\tilde{A}(A); \ \hat{\lambda}=\lambda+\tilde{\lambda}(\lambda, A).
\eea
Hence, the leading-order of Eq. \eqref{SW} is trivially satisfied:
\bea
\hat{A}(A+\delta_{\lambda}A)=A+\delta_{\lambda}A+\cdots; \ \hat{A}(A)=A+\cdots; \ \hat{\delta}_{\hat{\lambda}}\hat{A}=\delta_{\lambda} A+\cdots.
\eea
The higher-order terms of the non-commutativity parameter should satisfy that
\bea
\tilde{A}(A+\delta_{\lambda}A)=\tilde{A}(A)+\partial_{\mu}\tilde{\lambda}
+i\lbrack\tilde{\lambda}, A\rbrack_{\divideontimes}
+i\lbrack\lambda, \tilde{A}\rbrack_{\divideontimes}
+i\lbrack\tilde{\lambda}, \tilde{A}\rbrack_{\divideontimes}
+i\lbrack\lambda, A\rbrack_{\divideontimes}
-i\lbrack\lambda, A\rbrack
\eea
Up to the first order in the non-commutativity parameter, it shows that
\bea
&&
\tilde{A}(A+\delta_{\lambda}A)
\nn\\
&=&\tilde{A}(A)+\partial_{\mu}\tilde{\lambda}
+i\lbrack\tilde{\lambda}, A\rbrack
+i\lbrack\lambda, \tilde{A}\rbrack
-\frac{1}{2}\theta^{\mu\nu}\big((\partial_{\mu}\lambda)(\partial_{\nu}A)+(\partial_{\nu}A)(\partial_{\mu}\lambda)\big)+{\cal O}(\theta^2).
\label{SW1}
\nn\\
\eea
The solution of the SW map up to the first order in $\theta$ is given by \cite{Seiberg:1999vs}:
\bea
\hat{\lambda}&=&\lambda+\frac{1}{4}\theta^{\mu\nu}(\partial_{\mu}\lambda\cdot A_{\nu}+A_{\nu}\cdot\partial_{\mu}\lambda)+{\cal O}(\theta^2);
\nn\\
\hat{A}_{\mu}&=&A_{\mu}-\frac{1}{4}\theta^{\nu\rho}\big(A_{\nu}(\partial_{\rho}A_{\mu}+F_{\rho\mu})
+(\partial_{\rho}A_{\mu}+F_{\rho\mu})A_{\nu}\big)+{\cal O}(\theta^2);
\nn\\
\hat{F}_{\mu\nu}&=&F_{\mu\nu}
\nn\\
&&
+\frac{1}{4}\theta^{\rho\sigma}\bigg(2F_{\mu\rho}F_{\nu\sigma}
+2F_{\nu\sigma}F_{\mu\rho}
\nn\\
&&
-A_{\rho}\big((D_{\sigma}F_{\mu\nu})+(\partial_{\sigma}F_{\mu\nu})\big)
-\big((D_{\sigma}F_{\mu\nu})+(\partial_{\sigma}F_{\mu\nu})\big)A_{\rho}\bigg)+{\cal O}(\theta^2).
\label{SWF}
\eea
\\

\noindent
Let us notice Eq. \eqref{SW1}, and
\bea
-\frac{1}{2}\theta^{\mu\nu}\big((\partial_{\mu}\lambda)(\partial_{\nu}A)+(\partial_{\nu}A)(\partial_{\mu}\lambda)\big)
\eea
implies \cite{Seiberg:1999vs}
\bea
\frac{\partial}{\partial\theta^{\mu\nu}}({\cal O}_1\divideontimes{\cal O}_2)\bigg|_{\theta=0}
=\frac{i}{2}(\partial_{\mu}{\cal O}_1)\divideontimes(\partial_{\nu}{\cal O}_2) \bigg|_{\theta=0}.
\eea
Nonetheless, this property can be extended to any finite $\theta$ \cite{Seiberg:1999vs}.
Therefore, the SW map should be generalized to include any finite $\theta$ version, rather than just the transformation between commutative and non-commutative gauge theory.
When assuming the commutative gauge field as that
\bea
A\equiv \hat{A}(\theta=0),
\eea
and thinking that the non-commutative gauge field is the small deviation of $\theta$,
\bea
\hat{A}(A)\equiv \hat{A}(\delta\theta),
\eea
we can rewrite the SW map as that
\bea
\hat{A}\big(\hat{A}(0)+\hat{\delta}_{\hat{\lambda}}\hat{A}(0)\big)=
\hat{A}(\delta\theta)+\hat{\delta}_{\hat{\lambda}}\hat{A}(\delta\theta).
\eea
For a finite-$\theta$ generalization version, the SW map becomes \cite{Seiberg:1999vs}
\bea
\hat{A}\big(\hat{A}(\theta)+\hat{\delta}_{\hat{\lambda}}\hat{A}(\theta)\big)=
\hat{A}(\theta+\delta\theta)+\hat{\delta}_{\hat{\lambda}}\hat{A}(\theta+\delta\theta).
\eea
Because we have
\bea
\frac{\partial}{\partial\theta^{\mu\nu}}({\cal O}_1\divideontimes{\cal O}_2)
=\frac{i}{2}(\partial_{\mu}{\cal O}_1)\divideontimes(\partial_{\nu}{\cal O}_2)
\eea
for any finite $\theta$, we can write a solution similar to Eq. \eqref{SWF} \cite{Seiberg:1999vs}:
\bea
\delta\hat{\lambda}&=&\frac{1}{4}\delta\theta^{\mu\nu}(\partial_{\mu}\hat{\lambda}\divideontimes \hat{A}_{\nu}+\hat{A}_{\nu}\divideontimes\partial_{\mu}\hat{\lambda});
\nn\\
\delta\hat{A}_{\mu}&=&-\frac{1}{4}\delta\theta^{\nu\rho}\big(\hat{A}_{\nu}\divideontimes(\partial_{\rho}\hat{A}_{\mu}+\hat{F}_{\rho\mu})
+(\partial_{\rho}\hat{A}_{\mu}+\hat{F}_{\rho\mu})\divideontimes\hat{A}_{\nu}\big);
\nn\\
\delta\hat{F}_{\mu\nu}&=&
\frac{1}{4}\delta\theta^{\rho\sigma}\bigg(2\hat{F}_{\mu\rho}\divideontimes\hat{F}_{\nu\sigma}
+2\hat{F}_{\nu\sigma}\divideontimes\hat{F}_{\mu\rho}
\nn\\
&&
-\hat{A}_{\rho}\divideontimes\big((\hat{D}_{\sigma}\hat{F}_{\mu\nu})+(\partial_{\sigma}\hat{F}_{\mu\nu})\big)
-\big((\hat{D}_{\sigma}\hat{F}_{\mu\nu})+(\partial_{\sigma}\hat{F}_{\mu\nu})\big)\divideontimes\hat{A}_{\rho}\bigg).
\eea
\\

\noindent
The SW map with a finite $\theta$ is generally challenging to solve.
However, for a constant field strength case, we can have the simplified equation \cite{Seiberg:1999vs}:
\bea
\delta\hat{F}=-\hat{F}\delta\theta\hat{F}; \ \delta\bigg(\frac{1}{\hat{F}}\bigg)=\delta\theta.
\eea
Therefore, we can do the integration to get the exact solution \cite{Seiberg:1999vs}
\bea
\hat{F}=\frac{1}{1+F\theta}F.
\eea
This solution has a singularity happening at $F\theta=-1$, which implies the loss of the non-commutative description \cite{Seiberg:1999vs}.
Thus, this solution serves as an example showing that the SW map cannot be generalized to any finite $\theta$ \cite{Seiberg:1999vs}.

\subsection{Scaling Limit}
\noindent
Let us first show the Lagrangian in the non-commutative description for the U(1) case \cite{Seiberg:1999vs}
\bea
{\cal L}_{\mathrm{NDBI}}=\frac{1}{\hat{G}_s(2\pi)^p(\alpha^{\prime})^{\frac{p+1}{2}}}
\sqrt{\det\big(\hat{G}+2\pi\alpha^{\prime}(\hat{F}+\hat{\Phi})\big)},
\label{noncDBI}
\eea
which has a similar generalization to the non-Abelian case by the $\mathrm{Str}$ \cite{Tseytlin:1997csa}.
This is a natural assumption because the structure resembles the commutative DBI theory.
However, it can also be traced back to the commutative description by setting the non-commutativity parameter to zero, $\theta=0$, which implies \cite{Seiberg:1999vs}:
\bea
\hat{G}_{\mu\nu}=g_{\mu\nu}; \ G_s=g_s; \ \hat{\Phi}_{\mu\nu}=B_{\mu\nu}; \ \hat{F}_{\mu\nu}=F_{\mu\nu}.
\eea
When we turn off the field strength $F$ in the commutative description, the field strength in the non-commutative description $\hat{F}$ also vanishes \cite{Seiberg:1999vs}.
Therefore, the equivalence up to the classical Lagrangian level up to a total derivative term helps determine the parameters between the commutative and non-commutative descriptions \cite{Seiberg:1999vs}.
The relation between $g_s$ and $G_s$ can be determiend when turing off the field strengths \cite{Seiberg:1999vs}
\bea
\hat{G}_s=g_s\bigg(\frac{\det(\hat{G}+2\pi\alpha^{\prime}\hat{\Phi})}{\det(g+2\pi\alpha^{\prime}B)}\bigg)^{\frac{1}{2}}.
\label{opensc}
\eea
According to the result of the propagator \eqref{2pt1}, we assume the open-closed string relation for the components along the $B$-field background directions \cite{Seiberg:1999vs}
\bea
\frac{1}{\hat{G}+2\pi\alpha^{\prime}\hat{\Phi}}=
-\frac{\theta}{2\pi\alpha^{\prime}}
+\frac{1}{g+2\pi\alpha^{\prime}B}.
\label{openclosed}
\eea
We consider the scaling limit \cite{Seiberg:1999vs}:
\bea
\alpha^{\prime}\sim\epsilon^{\frac{1}{2}}; \ g_{\alpha\beta}\sim\epsilon^0; \ g_{\dot\alpha\dot\beta}\sim\epsilon; \ B_{\dot\alpha\dot\beta}\sim\epsilon^0,
\eea
and the open string parameters are all finite $\sim\epsilon^0$ to do the expansion \cite{Seiberg:1999vs}
\bea
&&
\frac{1}{\hat{G}}-2\pi\alpha^{\prime}\frac{1}{\hat{G}}\hat{\Phi}\frac{1}{\hat{G}}
\nn\\
&=&-\frac{\theta}{2\pi\alpha^{\prime}}
+\frac{1}{2\pi\alpha^{\prime}}\frac{1}{B^{(0)}}
\nn\\
&&
-\frac{1}{(2\pi\alpha^{\prime})^2}\frac{1}{B^{(0)}}g\frac{1}{B^{(0)}}
\nn\\
&&
-\frac{\epsilon}{2\pi\alpha^{\prime}}\frac{1}{B^{(0)}}B^{(1)}\frac{1}{B^{(0)}}
+\frac{1}{(2\pi\alpha^{\prime})^3}\frac{1}{B^{(0)}}g\frac{1}{B^{(0)}}g\frac{1}{B^{(0)}}+{\cal O}(\epsilon),
\eea
where
\bea
B=B^{(0)}+\epsilon B^{(1)}+{\cal O}(\epsilon^2).
\eea
\\

\noindent
Equating $\epsilon^{-1/2}$-order, we get \cite{Seiberg:1999vs}
\bea
-\frac{\theta}{2\pi\alpha^{\prime}}
+\frac{1}{2\pi\alpha^{\prime}}\frac{1}{B^{(0)}}={\cal O}(\epsilon),
\eea
which implies \cite{Seiberg:1999vs}
\bea
\theta=\frac{1}{B^{(0)}}+{\cal O}(\epsilon).
\eea
Equating $\epsilon^0$-order, we get \cite{Seiberg:1999vs}
\bea
\frac{1}{\hat{G}}
=
-\frac{1}{(2\pi\alpha^{\prime})^2}\frac{1}{B^{(0)}}g\frac{1}{B^{(0)}}+{\cal O}(\epsilon),
\eea
which implies \cite{Seiberg:1999vs}
\bea
\hat{G}=-(2\pi\alpha^{\prime})^2B^{(0)}\frac{1}{g}B^{(0)}+{\cal O}(\epsilon).
\eea
Equating $\epsilon^{1/2}$-order, we get \cite{Seiberg:1999vs}
\bea
-2\pi\alpha^{\prime}\frac{1}{\hat{G}}\hat{\Phi}\frac{1}{\hat{G}}
=
-\frac{\epsilon}{2\pi\alpha^{\prime}}\frac{1}{B^{(0)}}B^{(1)}\frac{1}{B^{(0)}}
+\frac{1}{(2\pi\alpha^{\prime})^3}\frac{1}{B^{(0)}}g\frac{1}{B^{(0)}}g\frac{1}{B^{(0)}}+{\cal O}(\epsilon),
\eea
which implies \cite{Seiberg:1999vs}
\bea
\hat{\Phi}=-B^{(0)}+\epsilon(2\pi\alpha^{\prime})^2B^{(0)}\frac{1}{g}B^{(1)}\frac{1}{g}B^{(0)}+{\cal O}(\epsilon).
\eea
By using the above scaling limit to the open string parameters, we can also get the scaling limit for $\hat{G}_s$ \cite{Seiberg:1999vs},
\bea
\hat{G}_s=g_s\det\bigg(2\pi\alpha^{\prime}B^{(0)}\frac{1}{g}\bigg)^{\frac{1}{2}}\big(1+{\cal O}(\epsilon)\big).
\eea
The prefactor (gauge coupling with constant $\hat{G}_s$) for the non-commutative D$p$-brane or the non-commutative YM theory is:
\bea
\frac{1}{g_p^2}=\frac{(2\pi\alpha^{\prime})^2}{\hat{G}_s(2\pi)^p(\alpha^{\prime})^{\frac{p+1}{2}}}=\frac{(\alpha^{\prime})^{\frac{3-p}{2}}}{(2\pi)^{p-2}\hat{G}_s}.
\eea
To have a finite prefactor, the scaling limit for $g_s$ and $\hat{G}_s$ is \cite{Seiberg:1999vs}:
\bea
g_s\sim\epsilon^{\frac{3-p+r}{4}}; \ \hat{G}_s\sim\epsilon^{\frac{3-p}{4}},
\eea
where $r$ is the rank of the $B$ field background.

\subsection{Non-Commutative YM from DBI}
\noindent
Now, we proceed to show the equivalence between the commutative and non-commutative descriptions in the DBI action only up to a total derivative term \cite{Seiberg:1999vs}.
The open-closed string relation \eqref{openclosed} \cite{Seiberg:1999vs} implies
\bea
\delta(\hat{G}+2\pi\alpha^{\prime}\hat{\Phi})=(\hat{G}+2\pi\alpha^{\prime}\hat{\Phi})\frac{\delta\theta}{2\pi\alpha^{\prime}}(\hat{G}+2\pi\alpha^{\prime}\hat{\Phi})
\label{variationGP}
\eea
According to Eq. \eqref{opensc} \cite{Seiberg:1999vs}, the variation of $\hat{G}_s$ is
\bea
\delta\hat{G}_s=\frac{1}{2}\hat{G}_s\mathrm{Tr}\bigg(\frac{1}{\hat{G}+2\pi\alpha^{\prime}\hat{\Phi}}
\delta(\hat{G}+2\pi\alpha^{\prime}\hat{\Phi})\bigg).
\eea
We apply the result \eqref{variationGP} to $\delta\hat{G}_s$ and then show that \cite{Seiberg:1999vs}:
\bea
\delta\hat{G}_s=\frac{1}{4\pi\alpha^{\prime}}\hat{G}_s\mathrm{Tr}\big((\hat{G}+2\pi\alpha^{\prime}\hat{\Phi})\delta\theta\big)
=\frac{1}{4\pi\alpha^{\prime}}\hat{G}_s\mathrm{Tr}(2\pi\alpha^{\prime}\hat{\Phi}\delta\theta).
\eea
\\

\noindent
We now do the variation to ${\cal L}_{\mathrm{NDBI}}$ \eqref{noncDBI} \cite{Seiberg:1999vs}:
\bea
&&
\delta\big((2\pi)^p(\alpha^{\prime})^{\frac{p+1}{2}}{\cal L}_{\mathrm{DBI}}\big)
\nn\\
&=&
\delta\bigg(\frac{1}{\hat{G}_s}\det\big(\hat{G}+2\pi\alpha^{\prime}(\hat{\Phi}+\hat{F})\big)^{\frac{1}{2}}\bigg)
\nn\\
&=&
\frac{\det\big(\hat{G}+2\pi\alpha^{\prime}(\hat{\Phi}+\hat{F})\big)^{\frac{1}{2}}}{\hat{G}_s}
\nn\\
&&\times
\bigg\lbrack-\frac{\delta\hat{G}_s}{\hat{G}_s}
+\frac{1}{2}\mathrm{Tr}\bigg(\frac{1}{\hat{G}+2\pi\alpha^{\prime}(\hat{\Phi}+\hat{F})}
\delta\big(\hat{G}+2\pi\alpha^{\prime}(\hat{\Phi}+\hat{F})\big)\bigg)\bigg\rbrack.
\nn\\
\eea
Considering the variation of $\hat{F}$ that we obtiained \cite{Seiberg:1999vs}:
\bea
\delta\hat{F}_{\mu\nu}&=&
\frac{1}{4}\delta\theta^{\rho\sigma}\bigg(2\hat{F}_{\mu\rho}\divideontimes\hat{F}_{\nu\sigma}
+2\hat{F}_{\nu\sigma}\divideontimes\hat{F}_{\mu\rho}
\nn\\
&&
-\hat{A}_{\rho}\divideontimes\big((\hat{D}_{\sigma}\hat{F}_{\mu\nu})+(\partial_{\sigma}\hat{F}_{\mu\nu})\big)
-\big((\hat{D}_{\sigma}\hat{F}_{\mu\nu})+(\partial_{\sigma}\hat{F}_{\mu\nu})\big)\divideontimes\hat{A}_{\rho}\bigg)
\nn\\
&=&\delta\theta^{\rho\sigma}\bigg(\hat{F}_{\mu\rho}\hat{F}_{\nu\sigma}
-\frac{1}{2}\hat{A}_{\rho}(\hat{D}_{\sigma}+\partial_{\sigma})\hat{F}_{\mu\nu}\bigg)+{\cal O}\big((\partial\hat{F})^2\big),
\eea
the variation of ${\cal L}_{\mathrm{NDBI}}$ becomes \cite{Seiberg:1999vs}
\bea
&&
\delta\big((2\pi)^p(\alpha^{\prime})^{\frac{p+1}{2}}{\cal L}_{\mathrm{DBI}}\big)
\nn\\
&=&
\frac{1}{2}\frac{\det\big(\hat{G}+2\pi\alpha^{\prime}(\hat{\Phi}+\hat{F})\big)^{\frac{1}{2}}}{\hat{G}_s}
\nn\\
&&\times
\bigg\lbrack-\frac{1}{2\pi\alpha^{\prime}}\mathrm{Tr}\big(\delta\theta(\hat{G}+2\pi\alpha^{\prime}\hat{\Phi})\big)
\nn\\
&&
+\mathrm{Tr}\bigg(\frac{1}{\hat{G}+2\pi\alpha^{\prime}(\hat{\Phi}+\hat{F})}
(\hat{G}+2\pi\alpha^{\prime}\hat{\Phi})\frac{\delta\theta}{2\pi\alpha^{\prime}}(\hat{G}+2\pi\alpha^{\prime}\hat{\Phi})\bigg)
\nn\\
&&+2\pi\alpha^{\prime}
\bigg(\frac{1}{\hat{G}+2\pi\alpha^{\prime}(\hat{\Phi}+\hat{F})}\bigg)^{\nu\mu}
\delta\theta^{\rho\sigma}\bigg(\hat{F}_{\mu\rho}\hat{F}_{\nu\sigma}
-\frac{1}{2}\hat{A}_{\rho}(\hat{D}_{\sigma}+\partial_{\sigma})\hat{F}_{\mu\nu}\bigg)\bigg\rbrack
\nn\\
&&
+{\cal O}\big((\partial\hat{F})^2\big).
\eea
Using the following formula:
\bea
&&
\frac{1}{2\pi\alpha^{\prime}}\partial_l\det\big(\hat{G}+2\pi\alpha^{\prime}(\hat{\Phi}+\hat{F})\big)^{\frac{1}{2}}
\nn\\
&=&-\frac{1}{2}\det\big(\hat{G}+2\pi\alpha^{\prime}(\hat{\Phi}+\hat{F})\big)^{\frac{1}{2}}\mathrm{Tr}\bigg(\frac{1}{\hat{G}+2\pi\alpha^{\prime}(\hat{\Phi}+\hat{F})}\bigg)\partial_l\hat{F};
\nn\\
&&
\frac{1}{2\pi\alpha^{\prime}}\hat{D}_l\det\big(\hat{G}+2\pi\alpha^{\prime}(\hat{\Phi}+\hat{F})\big)^{\frac{1}{2}}
\nn\\
&=&-\frac{1}{2}\det\big(\hat{G}+2\pi\alpha^{\prime}(\hat{\Phi}+\hat{F})\big)^{\frac{1}{2}}\mathrm{Tr}\bigg(\frac{1}{\hat{G}+2\pi\alpha^{\prime}(\hat{\Phi}+\hat{F})}\bigg)\hat{D}_l\hat{F},
\eea
the variation of ${\cal L}_{\mathrm{NDBI}}$ becomes \cite{Seiberg:1999vs}:
\bea
&&
\delta\big((2\pi)^p(\alpha^{\prime})^{\frac{p+1}{2}}{\cal L}_{\mathrm{DBI}}\big)
\nn\\
&=&
\frac{1}{2}\frac{\det\big(\hat{G}+2\pi\alpha^{\prime}(\hat{\Phi}+\hat{F})\big)^{\frac{1}{2}}}{\hat{G}_s}
\nn\\
&&\times
\bigg\lbrack -\mathrm{Tr}\bigg(\frac{1}{\hat{G}+2\pi\alpha^{\prime}(\hat{\Phi}+\hat{F})}\hat{F}\delta\theta\big(\hat{G}+2\pi\alpha^{\prime}\hat{\Phi}\big)\bigg)
\nn\\
&&
-2\pi\alpha^{\prime}\mathrm{Tr}\bigg(\frac{1}{\hat{G}+2\pi\alpha^{\prime}(\hat{\Phi}+\hat{F})}\hat{F}\delta\theta\hat{F}\bigg)
\nn\\
&&
-\det\big(\hat{G}+2\pi\alpha^{\prime}(\hat{\Phi}+\hat{F})\big)^{-\frac{1}{2}}
\delta\theta^{\rho\sigma}\hat{A}_{\rho}(\partial_{\sigma}+\hat{D}_{\sigma})\det\big(\hat{G}+2\pi\alpha^{\prime}(\hat{\Phi}+\hat{F})\big)^{\frac{1}{2}}
\bigg\rbrack
\nn\\
&&
+{\cal O}\big((\partial\hat{F})^2\big)
\nn\\
&=&
\frac{1}{2}\frac{\det\big(\hat{G}+2\pi\alpha^{\prime}(\hat{\Phi}+\hat{F})\big)^{\frac{1}{2}}}{\hat{G}_s}
\nn\\
&&\times
\bigg\lbrack-\mathrm{Tr}\bigg(\frac{1}{\hat{G}+2\pi\alpha^{\prime}(\hat{\Phi}+\hat{F})}\hat{F}\delta\theta\big(\hat{G}+2\pi\alpha^{\prime}\hat{\Phi}\big)\bigg)
\nn\\
&&
-2\pi\alpha^{\prime}\mathrm{Tr}\bigg(\frac{1}{\hat{G}+2\pi\alpha^{\prime}(\hat{\Phi}+\hat{F})}\hat{F}\delta\theta\hat{F}\bigg)
+\mathrm{Tr}(\delta\theta\hat{F})
\bigg\rbrack
+{\cal O}\big((\partial\hat{F})^2\big)+\mathrm{Total\ Derivative}
\nn\\
&=&{\cal O}\big((\partial\hat{F})^2\big)+\mathrm{Total\ Derivative}.
\eea
Hence, we show the classical equivalence between the commutative and non-commutative descriptions in the DBI action up to the total derivative and higher-order derivative corrections for the U(1) case \cite{Seiberg:1999vs}.
Only a total derivative term in the non-commutative DBI theory depends on a non-commutativity parameter that leads to the difference between the two descriptions of the DBI theory.
However, the derivative correction \cite{Okawa:1999cm} and the U($N$) generalization \cite{Tseytlin:1997csa,Terashima:2000ej} do not affect the equivalence.
\\

\noindent
We can expand the non-commutative DBI under the scaling limit and obtain the non-commutative YM theory at leading order \cite{Seiberg:1999vs}.
The same result can also be achieved using a commutative description.
However, it needs to expand a non-polynomial term from the small field strength $F$ compared to the $B$-field background, which is more complicated in the non-commutative theory \cite{Seiberg:1999vs}.
Because the non-commutative theory with finite open string parameters encompasses all $\alpha^{\prime}$ effects from the leading-order effective action through the Moyal product \cite{Seiberg:1999vs}, it is analogous to the resummation situation, facilitating the exploration of the strongly coupled regime.
It is more convenient to find the gauge symmetry structure compared to the commutative theory's $\alpha^{\prime}$ expansion, particularly when considering finite closed string parameters.
Indeed, fixing different parameters yields different effective actions from the expansion.
To see how the Moyal product collects all $\alpha^{\prime}$ effects, we need to transform it to a dimensionless parameter
\bea
\tilde{\theta}\sim\frac{\theta}{\alpha^{\prime}},
\eea
and we can then find that the expansion in terms of the non-commutativity parameter induces a high-order derivative term from the higher-order $\alpha^{\prime}$ after using the dimensionless parameter $\tilde{\theta}$.

\section{Non-Commutative Geometry of M5-Brane}
\label{sec:5}
\begin{figure}[h]
\begin{center}
\includegraphics[width=1.\textwidth]{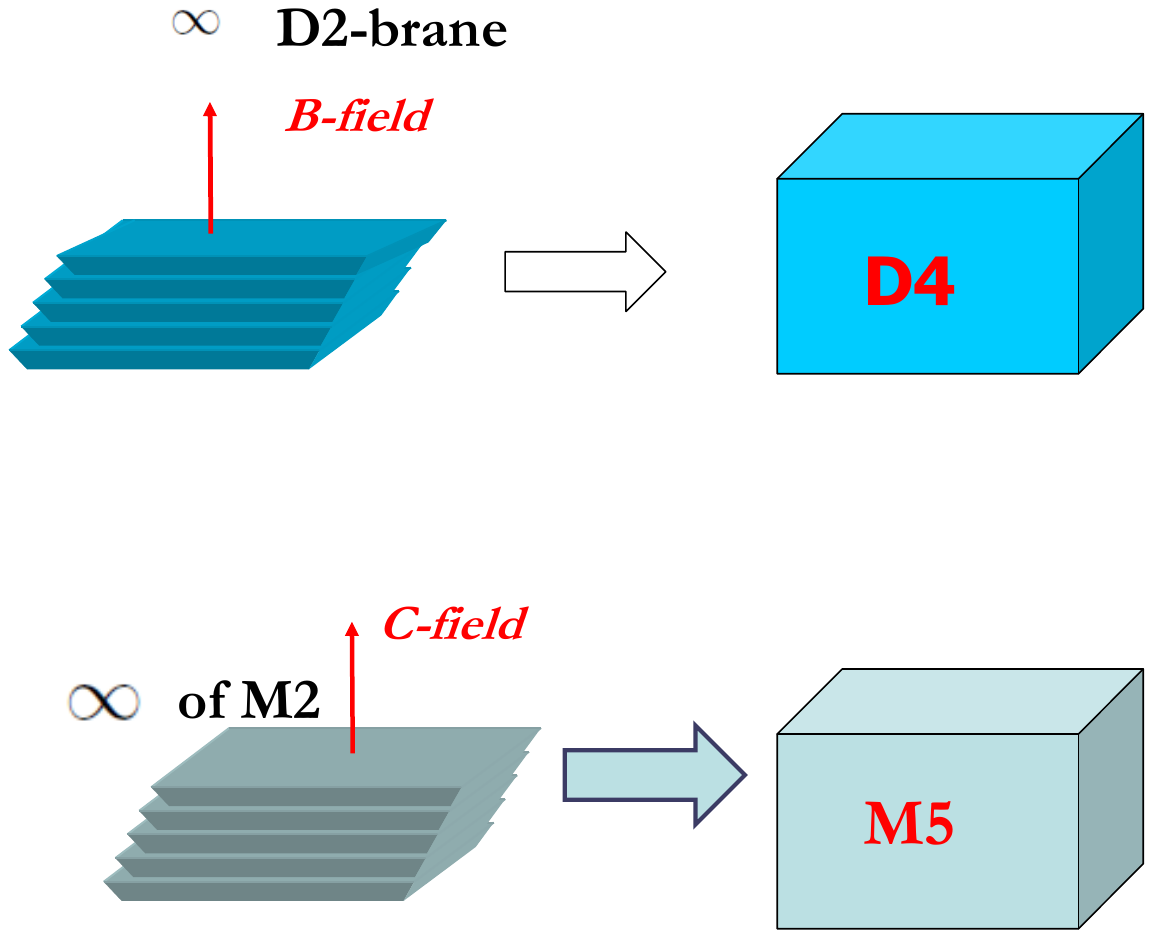}
\end{center}
\caption{The infinite M2-brane forms a single M5-brane analogous to the D-branes story.}
\label{M52.pdf}
\end{figure}

\noindent
We discuss the generalization of the non-commutative geometry to a single M5-brane theory \cite{Ho:2008nn} in this section.
The approach is to start from the infinite M2-branes with the transversal scalar fields satisfying the non-commutative relation, and it should be equivalent to a non-commutative M5-brane \cite{Ho:2008nn}.
A similar scenario has occurred with multiple D$p$-branes (Fig. \ref{M52.pdf}), which are effectively described by the U($N$) Yang-Mills theory in ($p+1$) dimensions.
When only turning on the tranversal scalar fields $X^I$, and they are static, the effective description only leaves a term $\mathrm{Str}(\lbrack X^I, X^J\rbrack\lbrack X^I, X^J\rbrack)$ in the Lagrangian.
The equation of motion for the transverse scalar fields is
\bea
\lbrack X^I, \lbrack X^J, X^K\rbrack\rbrack=0.
\eea
One simple solution to identifying the non-commutative geometry is
\bea
\lbrack X^I, X^J\rbrack=i\theta^{IJ}
\label{cX}
\eea
when $\theta^{IJ}$ is a constant.
To be compatible with a trace acting on a Lie algebra index for Eq. \eqref{cX}, it requires that $N$ is infinite.
Hence, the dynamics of the multiple D$p$-branes is the same as the single D($p+2$)-brane description.
\\

\noindent
To introduce this approach, we begin with the BLG model \cite{Bagger:2006sk,Gustavsson:2007vu,Bagger:2007jr,Bagger:2007vi} and then describe how a non-commutative single M5-brane with a large-$C$ field background, or an NP M5-brane, appears from an infinite number of M2-branes, similar to the D-branes case \cite{Ho:2008nn}.
The dynamics of the NP M5-brane are described using the Nambu-Poisson bracket, which also generates the VPD gauge symmetry \cite{Ho:2008nn}.
In the end, we compactify one direction along the large $C$-field background to show the D4-brane theory with a large NS-NS $B$-field background \cite{Ho:2008ve}, in which the gauge theory is the non-commutative U(1) gauge theory.

\subsection{BLG Model}
\noindent
The worldvolume theory of M2-branes in a flat spacetime background possesses an SO(8) rotational symmetry for the eight transverse scalars $X^I$.
Due to supersymmetry constraints, the dynamic degrees of freedom consist only of scalar fields.
In other words, the gauge sectors are dynamically similar to the Chern-Simons theory.
The BLG model introduces the Lie 3-algebra as gauge symmetry to realize the Lagrangian formulation \cite{Bagger:2006sk,Gustavsson:2007vu,Bagger:2007jr,Bagger:2007vi}.
We first introduce the Lie 3-algebra and then the Lagrangian, along with its corresponding gauge symmetry.

\subsubsection{Lie 3-Algebra}
\noindent
The Lie 3-algebra is a generalization of the Lie algebra through a totally antisymmetric 3-linear map
\bea
\lbrack T^a, T^b, T^c\rbrack=f^{abc}{}_dT^d,
\eea
where $T^a$ is the basis of an algebra, and $f^{abc}{}_d$ is a structure constant.
The indices of the algebra are defined as $a = 1, 2, \cdots, N$, where $N$ represents the dimension of the algebra, not the number of M2-branes, which is $N^2$ in the large-$N$ limit.
The Lie 3-algebra satisfies the first fundamental identity
\bea
&&
\lbrack T^a, T^b\lbrack T^c, T^d, T^e\rbrack\rbrack
\nn\\
&=&
\lbrack\lbrack T^a, T^b, T^c\rbrack, T^d, T^e\rbrack
+\lbrack T^c, \lbrack T^a, T^b, T^d\rbrack, T^e\rbrack
+\lbrack T^c, T^d, \lbrack T^a, T^b, T^e\rbrack\rbrack.
\eea
The trace form provides a metric to raise algebraic indices
\bea
h^{ab}\equiv\mathrm{Tr}(T^a, T^b).
\eea
The trace form also satisfies the following identity
\bea
\mathrm{Tr}(\lbrack T^a, T^b, T^c\rbrack, T^d\rbrack)
+\mathrm{Tr}(\lbrack T^c, \lbrack T^a, T^b, T^d\rbrack)=0.
\eea
When the BLG scalar field
\bea
X^I\equiv X^{I, a}T^a
\eea
is transformed as
\bea
\delta _{\Lambda} X^I=\Lambda_{ab}\lbrack T^a, T^b, X^I\rbrack,
\eea
which is a closed transformation implied by the first fundamental identity.
Hence, the Lie 3-algebra is a convenient way to construct the gauge symmetry.

\subsubsection{Lagrangian and Gauge Symmetry}
\noindent
The gauge transformation of the transverse scalars and worldvolume gauge fields is  \cite{Bagger:2006sk,Gustavsson:2007vu,Bagger:2007jr,Bagger:2007vi}:
\bea
\delta _{\Lambda} X^I&=&\theta\Lambda_{ab}\lbrack T^a, T^b, X^I\rbrack;
\nn\\
\delta_{\Lambda}\tilde{A}_{\mu}&\equiv&
\delta_{\Lambda}\tilde{A}_{\mu}{}^b{}_a
=\partial_{\mu}\tilde{\Lambda}^b{}_a
-\theta\tilde{\Lambda}^b{}_c\tilde{A}_{\mu}{}^c{}_a+\theta\tilde{A}_{\mu}{}^b{}_c\tilde{\Lambda}^c{}_a
\equiv\partial_{\mu}\tilde{\Lambda}-\theta\lbrack\tilde{\Lambda}, \tilde{A}_{\mu}\rbrack
\equiv D_{\mu}\tilde{\Lambda},
\eea
where
\bea
\tilde{A}_{\mu}{}^b{}_a\equiv f^{cdb}{}_aA_{\mu, cd}.
\eea
The corresponding covariant objects are:
\bea
D_{\mu}X^I\equiv\partial_{\mu}X^I-\theta\tilde{A}_{\mu}X^I; \
\tilde{F}_{\mu\nu}\equiv\partial_{\mu}\tilde{A}_{\nu}-\partial_{\nu}\tilde{A}_{\mu}+\theta\lbrack\tilde{A}_{\mu}, \tilde{A}_{\nu}\rbrack.
\eea
The bosonic sector of the BLG model is  \cite{Bagger:2006sk,Gustavsson:2007vu,Bagger:2007jr,Bagger:2007vi}
\bea
S_{\mathrm{BLG}}&=&T_{\mathrm{M}2}\int d^3x\ \bigg(-\frac{1}{2}\mathrm{Tr}(D_{\mu}X^I, D^{\mu}X^I)
-\frac{\theta^4}{12}\mathrm{Tr}(\lbrack X^I, X^J, X^K\rbrack, \lbrack X^I, X^J, X^K\rbrack)
\nn\\
&&
+\frac{1}{2}\epsilon^{\mu\nu\lambda}f^{abcd}A_{\mu, ab}\partial_{\nu}A_{\lambda, cd}
+\frac{1}{3}\theta f^{cda}{}_gf^{efgb}A_{\mu, ab}A_{\nu, cd}A_{\lambda, ef}\bigg).
\eea
The tension of the M2-branes is
\bea
T_{\mathrm{M}2}\equiv\frac{1}{(2\pi)^2l_p^3}.
\eea
By using the correspondence between the M-theory and type IIA theory, the 11D Planck scale and the compactification radius can be rewritten as the type IIA parameters:
\bea
l_p=g_s^{\frac{1}{3}}l_s; \ \frac{R_2}{\sqrt{g_{22}}}=R_2=g_sl_s.
\eea
The $T_{\mathrm{M}2}$ conincides with the D2-brane's case:
\bea
T_{\mathrm{M}2}=\frac{1}{(2\pi)^2g_sl_s^3}=\frac{T_2}{g_s}.
\eea
When compactifying the $x^{\mu=2}$ direction, the tension of the M2-branes becomes the string tension:
\bea
T_{\mathrm{M}2}(2\pi R_2)=\frac{1}{2\pi\alpha^{\prime}}=T_s.
\eea

\subsection{NP M5-Brane from Infinite M2-Branes}
\noindent
We consider the infinite M2-branes in the context of an algebra with infinite dimension.
For any manifold with an NP bracket, one can define a positive-definite Lie 3-algebra with infinite dimension.
We first introduce the NP bracket and then the NP M5-brane theory.

\subsubsection{NP Bracket}
\noindent
We consider the NP bracket as a totally antisymmetric 3-linear map on the three-dimensional internal manifolds, which are parametrized by the coordinates $y^{\dot\mu}$,
\bea
\{{\cal O}_1, {\cal O}_2, {\cal O}_3\}\equiv \epsilon^{\dot\mu\dot\nu\dot\lambda}
(\partial_{\dot{\mu}}{\cal O}_1)(\partial_{\dot{\nu}}{\cal O}_2)(\partial_{\dot{\lambda}}{\cal O}_3),
\eea
where
\bea
\partial_{\dot{\mu}}\equiv\frac{\partial}{\partial y^{\dot\mu}}.
\eea
The NP bracket satisfies the first fundamental identity
\bea
&&
\{{\cal O}_1, {\cal O}_2, \{{\cal O}_3, {\cal O}_4, {\cal O}_5\}\}
\nn\\
&=&
\{\{{\cal O}_1, {\cal O}_2, {\cal O}_3\}, {\cal O}_4, {\cal O}_5\}
+\{{\cal O}_1, \{{\cal O}_2, {\cal O}_3, {\cal O}_4\}, {\cal O}_5\}
+\{{\cal O}_3, {\cal O}_4, \{{\cal O}_1, {\cal O}_2, {\cal O}_5\}\}.
\nn\\
\eea
Hence, we can define the NP bracket as the Lie 3-algebra
\bea
\{\chi^a(y), \chi^b(y), \chi^c(y)\}= f^{abc}{}_d\chi^d(y).
\eea
We define the trace form as the integration
\bea
\mathrm{Tr}(\chi_1, \chi_2)=\frac{T_{\mathrm{M}2}}{2\pi\theta^2}\int d^3y\ \chi_1(y)\chi_2(y),
\eea
where $\theta$ is the dimensionless non-commutativity parameter.
According to the convention of the trace form, the summation of the basis of the algebra is
\bea
\chi^a(y)\chi_a(\tilde{y})=\frac{2\pi\theta^2}{T_{\mathrm{M}2}}\delta^3(y-\tilde{y}).
\eea
The coefficient $T_{\mathrm{M}2}/(2\pi\theta^2)$ is chosen because we define that the field backgrounds couple to the worldvolume of corresponding branes with a charge of 1 through the couplings $\int_{\mathrm{F1}}B$, where F1 indicates the fundamental string worldsheet, and $\int_{\mathrm{M2}}C$.
\\

\noindent
We now explain how the coefficient appears.
Because the NP M5-brane is in a large $C$-field background along $y^{\dot\mu}$ directions, the corresponding NS-NS D4-brane should be under a large $B$-field background.
We define the dimensionless parameter in terms of the large $B$-field background in D4-brane as:
\bea
\frac{1}{B_{\dot1\dot2}}=\theta^{\dot1\dot2}\equiv\frac{\theta}{T_s}.
\eea
The target space corresponding to the 3D internal manifold in M5-brane theory is derived from the BLD model's transverse scalar field, denoted as $X^{\dot{\mu}}$.
This can be expressed in terms of $y^{\dot{3}}$ as follows:
\bea
X^{\dot{\mu}} \sim \frac{1}{\theta} y^{\dot{3}}.
\eea
When compactifying $y^{\dot 3}$, the compactification radius of the target space is:
\bea
X^{\dot3}\sim X^{\dot 3}+2\pi \frac{R_{\dot 3}}{\sqrt{g_{\dot 3\dot 3}}}=X^{\dot 3}+2\pi R_{\dot 3},
\eea
which implies the compactification radius of the $y^{\dot 3}$ is
\bea
y^{\dot3}\sim y^{\dot 3}+2\pi\theta R_{\dot 3}.
\eea
The large $C$-field background of the NP M5-brane is given by
\bea
C_{\dot 1\dot 2\dot 3}=\frac{T_{\mathrm{M}2}}{\theta^2},
\label{lC}
\eea
which can be determined from the equation
\bea
C_{\dot1\dot 2\dot 3}2\pi R_{\dot{3}}\theta=B_{\dot 1\dot 2}.
\eea
Hence, the density of a single M2-brane in the internal manifold is $T_{\mathrm{M}2}/(2\pi\theta^2)$.
The small dimensionless non-commutativity parameter in the NP M5-brane controls the strength of the large $C$-field background.

\subsubsection{NP M5 from BLG}
\noindent
The supersymmetry restricts the bosonic sector of the M5-brane to only 8 degrees of freedom.
Since the transverse scalars contain 5 degrees of freedom, the gauge fields should only capture 3 degrees of freedom. However, the conventional 2-form gauge fields in 6D have 6 degrees of freedom.
However, the 2-form gauge fields are self-dual, which eliminates half the degrees of freedom.
We introduce the basis of the algebra $\chi^a$ to the fields of the BLG model for obtaining the NP M5-brane theory \cite{Ho:2008nn}:
\bea
X^I(x, y)\equiv X^{I}{}_a(x)\chi^a(y); \ A_{\alpha}(x, y, \tilde{y})=A_{\alpha}^{ab}(x)\chi_a(y)\chi_b(\tilde{y}).
\eea
We then decompose the transverse directions of the M2-branes as a direct product of the 3D worldvolume directions of the M5-brane, denoted by the index $\dot{\mu}$, and the transverse directions of the M5-brane, denoted by the index $\tilde{I}$.
Therefore, the NP M5-brane's worldvolume is a direct product of the 3D BLG worldvolume, denoted by the index $\alpha$, and another three directions given by the transverse directions of the M2-branes.
The NP M5-brane's worldvolume gauge fields can be identified as the following \cite{Ho:2008nn}:
\bea
A_{\alpha}(x, y, \tilde{y})=a_{\alpha}(x, y)+\hat{b}_{\alpha\dot\mu}(x, y)\delta y^{\dot{\mu}}+\cdots; \
X^{I=\dot{\mu}}=\hat{X}^{\dot{\mu}}=\frac{y^{\dot{\mu}}}{\theta}+\hat{b}^{\dot{\mu}},
\eea
where
\bea
\delta y^{\dot{\mu}}\equiv \tilde{y}^{\dot\mu}-y^{\dot{\mu}}; \
\hat{b}_{\alpha\dot\mu}(x, y)=\frac{\partial}{\partial\tilde{y}^{\dot\mu}}A_{\alpha}(x, y, \tilde{y})\bigg|_{\tilde{y}=y}; \
\hat{b}^{\dot\mu}=\frac{1}{2}\epsilon^{\dot\mu\dot\nu\dot\rho}\hat{b}_{\dot\nu\dot\lambda}.
\eea
The NP M5-brane's transverse scalar fields are identified as those
\bea
X^{I=\tilde{I}}=\hat{X}^{\tilde{I}}.
\eea
We will apply the above decomposition and the identification to introduce the gauge symmetry \cite{Ho:2008nn}, the SW map \cite{Ho:2008ve}, and the Lagrangian \cite{Ho:2008nn,Ho:2008ve} and show that the 2-form gauge fields for the NP-M5-brane are self-dual \cite{Ho:2008nn}.

\subsubsection{Gauge Transformation}
\noindent
We can rewrite the transverse scalars' gauge transformation of the BLG model as \cite{Ho:2008nn}:
\bea
\delta_{\Lambda} X^I&=&\theta\Lambda_{ab}(x)f^{abc}{}_d X^I{}_c(x)\chi^d(y)
=\theta\Lambda_{ab}(x)\{\chi^a, \chi^b, X^I\}
\nn\\
&=&\theta\Lambda_{ab}(x)\epsilon^{\dot\mu\dot\nu\dot\rho}\big(\partial_{\dot{\mu}}\chi^a(y)\big)\big(\partial_{\dot\nu}\chi^b(y)\big)\big(\partial_{\dot\rho}X^I(x, y)\big).
\eea
We define the gauge parameter as \cite{Ho:2008nn}
\bea
\hat{\kappa}^{\dot{\rho}}\equiv\Lambda_{ab}\epsilon^{\dot\mu\dot\nu\dot\rho}(\partial_{\dot{\mu}}\chi^a)(\partial_{\dot\nu}\chi^b),
\eea
which is divergenceless \cite{Ho:2008nn}
\bea
\partial_{\dot\rho}\hat{\kappa}^{\dot{\rho}}=0.
\eea
By using the decomposition of the index, we obtain the gauge transformation of the NP M5-brane \cite{Ho:2008nn}:
\bea
\hat{\delta}_{\hat{\Lambda}}\hat{X}^{\dot{\mu}}=\delta_{\Lambda}X^{\dot{\mu}}=\theta\hat{\kappa}^{\dot\rho}\partial_{\dot\rho}\hat{X}^{\dot\mu}; \
\hat{\delta}_{\hat{\Lambda}}\hat{X}^{\tilde{I}}=\delta_{\Lambda}X^{\tilde{I}}=\theta\hat{\kappa}^{\dot\rho}\partial_{\dot\rho}\hat{X}^{\tilde{I}}.
\eea
The first gauge transformation implies \cite{Ho:2008nn}
\bea
\hat{\delta}_{\hat{\Lambda}}\hat{b}^{\dot{\mu}}=\hat{\kappa}^{\dot{\mu}}+\theta\hat{\kappa}^{\dot\rho}\partial_{\dot\rho}\hat{b}^{\dot{\mu}}.
\eea
\\

\noindent
We now discuss another gauge transformation of the BLG worldvolume gauge fields \cite{Ho:2008nn}:
\bea
\delta_{\Lambda}\tilde{A}_{\alpha}{}^b{}_a=
f^{cdb}{}_a\delta_{\Lambda} A_{\alpha, cd}
=f^{cdb}{}_a\partial_{\alpha}\Lambda_{cd}
-\theta f^{efb}{}_cf^{ghc}{}_a\Lambda_{ef}A_{\alpha, gh}
+\theta f^{deb}{}_cf^{fgc}{}_aA_{\alpha, de}\Lambda_{fg}.
\eea
We then multiply $\chi^a(y)$ to extract the worldvolume gauge fields' gauge transformation:
\bea
&&
\big(\delta_{\Lambda}\tilde{A}_{\alpha}{}^b{}_a(x)\big)\chi^a(y)
\nn\\
&=&f^{cdb}{}_a\big(\delta_{\Lambda} A_{\alpha, cd}(x)\big)\chi^a(y)
=\{\chi^c, \chi^d, \chi^b\}\delta A_{\alpha, cd}(x)
\nn\\
&=&\epsilon^{\dot\mu\dot\nu\dot\rho}
\frac{\partial^2\delta_{\Lambda} A_{\alpha}(x, y, \tilde{y})}{\partial y^{\dot\mu}\partial\tilde{y}^{\dot\nu}}\bigg|_{\tilde{y=y}}
\partial_{\dot\rho}\chi^b(y)
=\epsilon^{\dot\mu\dot\nu\dot\rho}\big(\partial_{\dot\mu}\hat{\delta}_{\hat{\Lambda}}\hat{b}_{\alpha\dot\nu}(x, y)\big)\big(\partial_{\dot\rho}\chi^b(y)\big)
\nn\\
&=&\hat{\delta}_{\hat{\Lambda}}\hat{B}_{\alpha}{}^{\dot\rho}(x, y)\partial_{\dot\rho}\chi^b(y),
\eea
where
\bea
\hat{B}_{\alpha}{}^{\dot\rho}\equiv \epsilon^{\dot\mu\dot\nu\dot\rho}\partial_{\dot\mu}\hat{b}_{\alpha\dot\nu};
\eea
\bea
f^{cdb}{}_a\big(\partial_{\alpha}\Lambda_{cd}(x)\big)\chi^a(y)
&=&\{\chi^c, \chi^d, \chi^b\}\partial_{\alpha}\Lambda_{cd}(x, y)
=\epsilon^{\dot\mu\dot\nu\dot\rho}\bigg(\partial_{\alpha}\frac{\partial^2\Lambda(x, y, \tilde{y})}{\partial y^{\dot\mu}\partial \tilde{y}^{\dot\nu}}\bigg|_{\tilde{y}=y}\bigg)\big(\partial_{\dot\rho}\chi^b(y)\big)
\nn\\
&=&(\partial_{\alpha}\hat{\kappa}^{\dot\rho}(x, y))(\partial_{\dot\rho}\chi^b(y));
\eea
\bea
&&
-\theta f^{efb}{}_cf^{ghc}{}_a\Lambda_{ef}(x)A_{\alpha, gh}(x)\chi^a(y)
\nn\\
&=&-\theta f^{efb}{}_c\{\chi^g, \chi^h, \chi^c\}\Lambda_{ef}(x)A_{\alpha, gh}(x)
\nn\\
&=&-\theta f^{efb}{}_c\epsilon^{\dot\mu\dot\nu\dot\rho}\frac{\partial^2 A_{\alpha}(x, y, \tilde{y})}{\partial y^{\dot\mu}\partial\tilde{y}^{\dot\nu}}\bigg|_{\tilde{y}=y}\big(\partial_{\dot\rho}\chi^c(y)\big)\Lambda_{ef}(x)
\nn\\
&=&-\theta\hat{B}_{\alpha}{}^{\dot\rho}(x, y)
\bigg(\big(\partial_{\dot\rho}\hat{\kappa}^{\dot\lambda}(x, y)\big)\big(\partial_{\dot\lambda}\chi^b(y)\big)
+\hat{\kappa}^{\dot\lambda}(x, y)\partial_{\dot\rho}\partial_{\dot\lambda}\chi^b(y)\bigg);
\eea
\bea
&&
\theta f^{deb}{}_cf^{fgc}{}_a A_{\alpha, de}(x)\Lambda_{fg}(x)\chi^a(y)
\nn\\
&=&\theta f^{deb}{}_c
\epsilon^{\dot\mu\dot\nu\dot\rho}\frac{\partial^2\Lambda(x, y, \tilde{y})}{\partial y^{\dot\mu}\partial y^{\dot\nu}}\bigg|_{\tilde{y}=y}
(\partial_{\dot\rho}\chi^c)A_{\alpha, de}(x)
=\theta f^{deb}{}_c\hat{\kappa}^{\dot\rho}(\partial_{\dot\rho}\chi^c)A_{\alpha, de}
\nn\\
&=&\theta\big(\hat{\kappa}^{\dot\rho}(\partial_{\dot\rho}\hat{B}_{\alpha}{}^{\dot\lambda})(\partial_{\dot\lambda}\chi^b)
+\hat{\kappa}^{\dot\lambda}\hat{B}_{\alpha}{}^{\dot\lambda}\partial_{\dot\rho}\partial_{\dot\lambda}\chi^b\big).
\eea
Hence, we obtain the gauge transformation of $\hat{B}_{\alpha}{}^{\dot\mu}$ \cite{Ho:2008nn},
\bea
\hat{\delta}_{\hat{\Lambda}}\hat{B}_{\alpha}{}^{\dot{\mu}}&=&\partial_{\alpha}\hat{\kappa}^{\dot{\mu}}+\theta\big(\hat{\kappa}^{\dot{\nu}}\partial_{\dot{\nu}}\hat{B}_{\alpha}{}^{\dot{\mu}}-(\partial_{\dot{\nu}}\hat{\kappa}^{\dot{\mu}})\hat{B}_{\alpha}{}^{\dot{\nu}}\big).
\eea
The covariant derivatives are:
\bea
\hat{D}_{\alpha}\hat{X}^{\tilde{I}}&=&D_{\alpha}X^{\tilde{I}}=\partial_{\alpha}\hat{X}^{\tilde{I}}-\theta\{\hat{b}_{\alpha\dot\nu}, y^{\dot\nu}, \hat{X}^{\tilde{I}}\}
=\partial_{\alpha}\hat{X}^{\tilde{I}}-\theta\hat{B}_{\alpha}{}^{\dot\mu}\partial_{\dot\mu}\hat{X}^{\tilde{I}};
\nn\\
\hat{D}_{\alpha}\hat{X}^{\dot\mu}&=&D_{\alpha}X^{\dot\mu}=\partial_{\alpha}\hat{X}^{\dot\mu}-\theta\{\hat{b}_{\alpha\dot\nu}, y^{\dot\nu}, \hat{X}^{\dot\mu}\}
=\partial_{\alpha}\hat{b}^{\dot\mu}-\hat{B}_{\alpha}{}^{\dot\mu}-\theta\hat{B}_{\alpha}{}^{\dot\rho}\partial_{\dot\rho}\hat{b}^{\dot\mu}
\nn\\
&=&\partial_{\alpha}\hat{b}^{\dot\mu}-\hat{V}_{\dot\rho}{}^{\dot\mu}\hat{B}_{\alpha}{}^{\dot\rho},
\nn\\
\eea
where
\bea
\hat{V}_{\dot\rho}{}^{\dot\mu}\equiv\partial_{\dot\rho}\hat{X}^{\dot\mu}=\delta_{\dot\rho}{}^{\dot\mu}+\theta\partial_{\dot\rho}\hat{b}^{\dot\mu}.
\eea
We find that the gauge transformation in the NP M5-brane corresponds to a gauge parameter $\hat{\kappa}^{\dot\mu}$ that is divergentless.
Indeed, this gauge parameter corresponds to the transformation that preserves volume invariance.

\subsubsection{VPD Gauge Symmetry}
\noindent
We show that the divergenceless gauge parameter corresponds to the VPD gauge symmetry.
The $n$-dimenaional volume is
\bea
\int d^nx =\int d^na\ {\cal J},
\eea
where
\bea
{\cal J}\equiv\det\bigg(\frac{\partial x}{\partial a}\bigg); \ J^{i_1}{}_{i_2}\equiv\partial_{a^{i_2}} x^{i_1}.
\eea
The variation of the Jacobian ${\cal J}$ is
\bea
\delta_V {\cal J}=\det(J)\mathrm{Tr}(J^{-1}\delta_V J),
\eea
where
\bea
\delta_V x^j=V^j.
\eea
The variation of $J$ is:
\bea
\delta_V J^{i_1}{}_{i_2}
=\partial_{a^{i_2}}\delta_V x^{i_1}
=\partial_{a^{i_2}}V^{i_1}
=(\partial_kV^{i_1})\partial_{a^{i_2}}x^k
=(\partial_kV^{i_1}) J^k{}_{i_2},
\eea
where
\bea
\partial_k\equiv \frac{\partial}{\partial x^k}.
\eea
Substituting the variation of $J$ to the variation of the Jacobian ${\cal J}$ shows that
\bea
\delta_V {\cal J}=\det(J)\mathrm{Tr}\bigg((J^{-1})^j{}_{i_1}(\partial_k V^{i_1})J^k{}_j\bigg)
=\det(J)(\partial_k V^{k}).
\eea
Hence, the invariant volume under the transformation requires that the gauge parameter is divergenceless
\bea
\partial_k V^k=0.
\eea
The general solution of the gauge parameter is
\bea
V^k=\epsilon^{k_1k_2\cdots k_{p-2}k}(\partial_{k_1} f_1)(\partial_{k_2} f_2)\cdots(\partial_{k_{p-2}} f_{p-2}).
\eea
In the NP M5-brane theory, the VPD gauge symmetry arises from the NP bracket, corresponding to a 3-bracket or $p=4$.
For the D$p$-brane in the large R-R ($p-1$)-form field background, it also has the VPD gauge symmetry, generated by the ($p-1$)-bracket \cite{Ho:2011yr,Ho:2013paa}.

\subsubsection{SW Map}
\noindent
In the NP M5-brane theory, the SW map satisfies the following relations:
\bea
\hat{b}^{\dot{\mu}}(b+\delta_{\Lambda}b)&=&\hat{\delta}_{\hat{\Lambda}}\hat{b}^{\dot{\mu}}(b)+\hat{b}^{\dot{\mu}}(b);
\nn\\
\hat{B}_{\alpha}{}^{\dot{\alpha}}(B+\delta_{\Lambda}B, b+\delta_{\Lambda}b)
&=&\hat{\delta}_{\hat{\Lambda}}\hat{B}_{\mu}{}^{\dot{\mu}}(B, b)+\hat{B}_{\alpha}{}^{\dot{\mu}}(B, b);
\nn\\
\hat{X}^{\tilde{I}}(X+\delta_{\Lambda}X, b+\delta_{\Lambda}b)&=&\hat{\delta}_{\hat{\Lambda}}\hat{X}^{\tilde{I}}(X, b)+\hat{X}^{\tilde{I}}(X, b),
\eea
where
\bea
\delta_{\Lambda}b^{\dot{\mu}}=\kappa^{\dot{\mu}}; \ \delta_{\Lambda}B_{\mu}{}^{\dot{\mu}}=\partial_{\mu}\kappa^{\dot{\mu}}; \delta_{\Lambda}X^{\tilde{I}}=0.
\eea
Up to the first order in $\theta$, the solution of the SW map \cite{Ho:2008ve} is given by:
\bea
&&\hat{b}^{\dot{\mu}}(b)
\nn\\
&=&b^{\dot{\mu}}+\theta\bigg(\frac{1}{2}b^{\dot{\nu}}\partial_{\dot{\nu}}b^{\dot{\mu}}
+\frac{1}{2}b^{\dot{\mu}}\partial_{\dot{\nu}}b^{\dot{\nu}}\bigg)+{\cal O}(\theta^2);
\nn\\
&&\hat{B}_{\alpha}{}^{\dot{\mu}}(B, b)
\nn\\
&=&B_{\alpha}{}^{\dot{\mu}}+\theta\bigg( b^{\dot{\nu}}\partial_{\dot{\nu}}B_{\alpha}{}^{\dot{\mu}}
-\frac{1}{2}b^{\dot{\nu}}\partial_{\alpha}\partial_{\dot{\nu}}b^{\dot{\mu}}
+\frac{1}{2}b^{\dot{\mu}}\partial_{\alpha}\partial_{\dot{\nu}}b^{\dot{\nu}}
\nn\\
&&
+(\partial_{\dot{\nu}}b^{\dot{\nu}})B_{\alpha}{}^{\dot{\mu}}
-(\partial_{\dot{\nu}}b^{\dot{\mu}})B_{\mu}{}^{\dot{\nu}}
-\frac{1}{2}(\partial_{\dot{\nu}}b^{\dot{\nu}})(\partial_{\mu}b^{\dot{\mu}})
+\frac{1}{2}(\partial_{\dot{\nu}}b^{\dot{\mu}})(\partial_{\alpha}b^{\dot{\nu}})\bigg)+{\cal O}(\theta^2);
\nn\\
&&\hat{X}^{\tilde{I}}(X, b)
\nn\\
&=&X^{\tilde{I}}+\theta b^{\dot{\mu}}\partial_{\dot{\mu}}X^{\tilde{I}}+{\cal O}(\theta^2);
\nn\\
&&\hat{\kappa}^{\dot{\mu}}(\kappa, b)
\nn\\
&=&
\kappa^{\dot{\mu}}+\theta\bigg(\frac{1}{2}b^{\dot{\nu}}\partial_{\dot{\nu}}\kappa^{\dot{\mu}}
+\frac{1}{2}(\partial_{\dot{\nu}}b^{\dot{\nu}})\kappa^{\dot{\mu}}
-\frac{1}{2}(\partial_{\dot{\nu}}b^{\dot{\mu}})\kappa^{\dot{\nu}}\bigg)+{\cal O}(\theta^2).
\eea

\subsubsection{Lagrangian}
\noindent
We now proceed using the decomposition and the identification to write the Lagrangian for the NP M5-brane theory \cite{Ho:2008nn,Ho:2008ve}.
The first term is the kinetic term of the transverse scalars of the BLG model \cite{Ho:2008nn,Ho:2008ve}:
\bea
&&
T_{\mathrm{M}2}\int d^3x\ \bigg(-\frac{1}{2}\mathrm{Tr}(D_{\alpha}X^I, D^{\alpha}X^I)\bigg)
\nn\\
&=&
T_{\mathrm{M}2}\int d^3x\ \bigg(-\frac{1}{2}\mathrm{Tr}(\hat{D}_{\alpha}\hat{X}^{\dot\mu}, \hat{D}^{\alpha}\hat{X}_{\dot\mu})
-\frac{1}{2}\mathrm{Tr}(\hat{D}_{\alpha}\hat{X}^{\tilde{I}}, \hat{D}^{\alpha}\hat{X}^{\tilde{I}})\bigg)
\nn\\
&=&
T_{\mathrm{M}2}\int d^3x\ \bigg(-\frac{1}{4}\mathrm{Tr}(\hat{{\cal H}}_{\alpha\dot\mu\dot\nu}, \hat{{\cal H}}^{\alpha\dot\mu\dot\nu})
-\frac{1}{2}\mathrm{Tr}(\hat{D}_{\alpha}\hat{X}^{\tilde{I}}, \hat{D}^{\alpha}\hat{X}^{\tilde{I}})
\bigg)
\nn\\
&=&\frac{T^2_{\mathrm{M}2}}{2\pi\theta^2}\int d^3xd^3y\ \bigg(-\frac{1}{4}\mathrm{Tr}(\hat{{\cal H}}_{\alpha\dot\mu\dot\nu}, \hat{{\cal H}}^{\alpha\dot\mu\dot\nu})
-\frac{1}{2}\mathrm{Tr}(\hat{D}_{\alpha}\hat{X}^{\tilde{I}}, \hat{D}^{\alpha}\hat{X}^{\tilde{I}})
\bigg)
\nn\\
&=&\frac{T_{\mathrm{M}5}}{\theta^2}\int d^3xd^3y\ \bigg(-\frac{1}{4}\hat{{\cal H}}_{\alpha\dot\mu\dot\nu}\hat{{\cal H}}^{\alpha\dot\mu\dot\nu}
-\frac{1}{2}\hat{D}_{\alpha}\hat{X}^{\tilde{I}}\hat{D}^{\alpha}\hat{X}^{\tilde{I}}
\bigg),
\eea
where the tension of the M5-brane is
\bea
T_{\mathrm{M}5}=\frac{T^2_{\mathrm{M}2}}{2\pi},
\eea
and the covariant field strength $\hat{{\cal H}}_{\alpha\dot\mu\dot\nu}$ is:
\bea
\hat{{\cal H}}_{\alpha\dot\mu\dot\nu}=\epsilon_{\dot\mu\dot\nu\dot\lambda}D_{\alpha}\hat{X}^{\dot\lambda}
=\epsilon_{\dot\mu\dot\nu\dot\lambda}(\partial_{\alpha}\hat{b}^{\dot\lambda}-\hat{B}_{\alpha}{}^{\dot\lambda}
-\theta\hat{B}_{\alpha}{}^{\dot\rho}\partial_{\dot\rho}\hat{b}^{\dot\lambda})
=\hat{H}_{\alpha\dot\mu\dot\nu}-\theta\epsilon_{\dot\mu\dot\nu\dot\lambda}\hat{B}_{\alpha}{}^{\dot\rho}\partial_{\dot\rho}\hat{b}^{\dot\lambda}.
\eea
Because the covariant field strength at the leading order for $\theta$ is the conventional abelian field strength
\bea
\hat{H}_{\alpha\dot\mu\dot\nu}=\partial_{\alpha}\hat{b}_{\dot\mu\dot\nu}+\partial_{\dot\mu}\hat{b}_{\dot\nu\alpha}+\partial_{\dot\nu}\hat{b}_{\alpha\dot\mu},
\eea
the covariant field strength can be viewed as a deformation of the commutative field strength induced by the non-commutativity parameter \cite{Ho:2008nn,Ho:2008ve}.
\\

\noindent
Another term of the transverse scalars in the BLG model is \cite{Ho:2008nn,Ho:2008ve}:
\bea
&&
T_{\mathrm{M}2}\int d^3x\ \bigg(-\frac{\theta^4}{12}\mathrm{Tr}(\lbrack X^I, X^J, X^K\rbrack, \lbrack X^I, X^J, X^K\rbrack)\bigg)
\nn\\
&=&\frac{T_{\mathrm{M}5}}{\theta^2}\int d^3xd^3y\ \bigg(-\frac{\theta^4}{12}(\{ X^{\dot\mu}, X^{\dot\nu}, X^{\dot\lambda}\}\{ X_{\dot\mu}, X_{\dot\nu}, X_{\dot\lambda}\}
\nn\\
&&
+3\{ X^{\dot\mu}, X^{\dot\nu}, X^{\tilde{I}}\}\{ X_{\dot\mu}, X_{\dot\nu}, X^{\tilde{I}}\}
+3\{ X^{\dot\mu}, X^{\tilde{I}}, X^{\tilde{J}}\}\{ X_{\dot\mu}, X^{\tilde{I}}, X^{\tilde{J}}\}
\nn\\
&&
+\{ X^{\tilde{I}}, X^{\tilde{J}}, X^{\tilde{K}}\}\{ X^{\tilde{I}}, X^{\tilde{J}}, X^{\tilde{K}}\}
)
\nn\\
&=&
\frac{T_{\mathrm{M}5}}{\theta^2}\int d^3xd^3y\ \bigg(
-\frac{1}{12}\hat{{\cal H}}_{\dot\mu\dot\nu\dot\rho}\hat{{\cal H}}^{\dot\mu\dot\nu\dot\rho}
-\frac{1}{2\theta^2}
-\frac{1}{2}(\hat{D}_{\dot\mu}\hat{X}^{\tilde{I}})(\hat{D}^{\dot\mu}\hat{X}^{\tilde{I}})
\nn\\
&&
-\frac{\theta^4}{4}\{ X^{\dot\mu}, X^{\tilde{I}}, X^{\tilde{J}}\}\{ X_{\dot\mu}, X^{\tilde{I}}, X^{\tilde{J}}\}
-\frac{\theta^4}{12}\{ X^{\tilde{I}}, X^{\tilde{J}}, X^{\tilde{K}}\}\{ X^{\tilde{I}}, X^{\tilde{J}}, X^{\tilde{K}}\}\bigg),
\eea
where the covariant field strength $\hat{{\cal H}}^{\dot\mu\dot\nu\dot\rho}$ is \cite{Ho:2008nn,Ho:2008ve}
\bea
\hat{{\cal H}}^{\dot\mu\dot\nu\dot\rho}=\theta^2\{\hat{X}^{\dot\mu}, \hat{X}^{\dot\nu}, \hat{X}^{\dot\rho}\}-\frac{\epsilon^{\dot\mu\dot\nu\dot\rho}}{\theta},
\eea
the covariant derivative of the transverse scalar fields is \cite{Ho:2008nn,Ho:2008ve}
\bea
\hat{D}_{\dot\mu}\hat{X}^{\tilde{I}}=\frac{\theta^2}{2}\epsilon_{\dot\mu\dot\nu\dot\rho}\{\hat{X}^{\dot\nu}, \hat{X}^{\dot\rho}, \hat{X}^{\tilde{I}}\}.
\eea
Because the covariant field strength $\hat{{\cal H}}^{\dot\mu\dot\nu\dot\rho}$ at the leading order of the non-commutativity parameter is equivalent to the conventional field strength, which is given by the expression
\bea
\hat{H}_{\dot\mu\dot\nu\dot\rho}=\partial_{\dot\mu}\hat{b}_{\dot\nu\dot\rho}+\partial_{\dot\nu}\hat{b}_{\dot\rho\dot\mu}+\partial_{\dot\nu}\hat{b}_{\dot\rho\dot\mu}.
\eea
Therefore, the covariant field strength can be viewed as a deformation of the commutative field strength due to the non-commutativity parameter \cite{Ho:2008nn,Ho:2008ve}.
The covariant derivative of the transverse scalar fields at the leading order for the non-commutativity theta is the ordinary derivative of scalar fields \cite{Ho:2008nn,Ho:2008ve}.
Therefore, the covariant derivative can also be seen as the deformation from the non-commutativity parameter \cite{Ho:2008nn,Ho:2008ve}.
Because the prefactor of boundary interaction in the D4-brane is $T_s$, the prefactor of the boundary interaction of the M2-brane is $T_{\mathrm{M}2}/\theta$ \cite{Ho:2008ve}.
It implies that the M2-brane action has the following term \cite{Ho:2008ve}
\bea
\int_{\mathrm{M}2}C+\frac{T_{\mathrm{M}2}}{\theta}\int_{\partial\mathrm{M}2}\hat{b}
=\int_{\mathrm{M}2}\bigg(C_3+\frac{T_{\mathrm{M}2}}{\theta}d\hat{b}\bigg).
\eea
Therefore, it shows that the gauge invariant object is \cite{Ho:2008ve}
\bea
d\hat{b}+\frac{\theta}{T_{\mathrm{M}2}}C=d\hat{b}+\frac{1}{\theta},
\eea
which uses Eq. \eqref{lC}.
Hence, it explains why the following VPD covariant object's leading order is the abelian field strength and the constant term \cite{Ho:2008ve}
\bea
\{\hat{X}^{\dot\mu}, \hat{X}^{\dot\nu}, \hat{X}^{\dot\rho}\}=\hat{H}^{\dot\mu\dot\nu\dot\rho}+\frac{\epsilon^{\dot\mu\dot\nu\dot\rho}}{\theta}.
\eea
\\

\noindent
The final term is the Chern-Simons term in the BLG model, which can be expressed as \cite{Ho:2008nn,Ho:2008ve}:
\bea
&&
T_{\mathrm{M}2}\int d^3x\ \epsilon^{\alpha\beta\gamma}\bigg(\frac{1}{2}f^{abcd}A_{\alpha, ab}\partial_{\beta}A_{\gamma, cd}
+\frac{\theta}{3}f^{cda}{}_gf^{efgb}A_{\alpha, ab}A_{\beta, cd}A_{\gamma, ef}\bigg)
\nn\\
&=&
T_{\mathrm{M}2}\int d^3x\ \epsilon^{\alpha\beta\gamma}\bigg(
\frac{1}{2}\mathrm{Tr}(\{\chi^a, \chi^b, \chi^c\}, \chi^d\}A_{\alpha, ab}\partial_{\beta}A_{\gamma, cd}
\nn\\
&&
+\frac{\theta}{3}\mathrm{Tr}(\{\chi^c, \chi^d, \chi^a\}, \chi^g)\mathrm{Tr}(\{\chi^e, \chi^f, \chi_g\}, \chi^b)
A_{\alpha, ab}A_{\beta, cd}A_{\gamma, ef}\bigg)
\nn\\
&=&
\frac{T_{\mathrm{M}5}}{\theta^2}\int d^3xd^3y\ \bigg(-\frac{1}{2}\epsilon^{\alpha\beta\gamma}\epsilon^{\dot\mu\dot\nu\dot\rho}\frac{\partial^2 A_{\alpha}}{\partial y^{\dot\mu}\tilde{y}^{\dot\nu}}\bigg|_{\tilde y=y}
\partial_{\beta}\frac{\partial A_{\gamma}}{\partial\tilde{y}^{\dot\rho}}\bigg|_{\tilde{y}=y}\bigg)
\nn\\
&&
+\frac{T_{\mathrm{M}5}}{\theta^2}\int d^3xd^3y\ \bigg(-\frac{\theta}{3}\epsilon^{\alpha\beta\gamma}\epsilon^{\dot\mu\dot\nu\dot\lambda}\epsilon^{\dot\rho\dot\sigma\dot\tau}
\frac{\partial^2 A_{\beta}}{\partial y^{\dot\mu}\partial\tilde{y}^{\dot\nu}}\bigg|_{\tilde{y}=y}
\frac{\partial^2 A_{\gamma}}{\partial y^{\dot\rho}\partial\tilde{y}^{\dot\sigma}}\bigg|_{\tilde{y}=y}
\frac{\partial^2 A_{\alpha}}{\partial y^{\dot\lambda}\partial\tilde{y}^{\dot\tau}}\bigg|_{\tilde{y}=y}\bigg)
\nn\\
&=&\frac{T_{\mathrm{M}5}}{\theta^2}\int d^3xd^3y\ \epsilon^{\alpha\beta\gamma}
\bigg(-\frac{1}{2}\epsilon^{\dot\mu\dot\nu\dot\rho}(\partial_{\dot\mu}\hat{b}_{\alpha\dot\nu})(\partial_{\beta}\hat{b}_{\gamma\dot\rho})
+\frac{\theta}{3}\epsilon^{\dot\mu\dot\nu\dot\lambda}\epsilon^{\dot\rho\dot\sigma\dot\tau}
(\partial_{\dot\mu}\hat{b}_{\beta\dot\nu})
(\partial_{\dot\sigma}\hat{b}_{\gamma\dot\rho})
(\partial_{\dot\lambda}\hat{b}_{\alpha\dot\tau})\bigg)
\nn\\
&=&\frac{T_{\mathrm{M}5}}{\theta^2}\int d^3xd^3y\ \epsilon^{\alpha\beta\gamma}
\bigg(-\frac{1}{2}\epsilon^{\dot\mu\dot\nu\dot\rho}(\partial_{\dot\mu}\hat{b}_{\alpha\dot\nu})(\partial_{\beta}\hat{b}_{\gamma\dot\rho})
\nn\\
&&
+\frac{\theta}{6}\epsilon^{\dot\mu\dot\nu\dot\lambda}\epsilon^{\dot\rho\dot\sigma\dot\tau}
(\partial_{\dot\mu}\hat{b}_{\beta\dot\nu})
(\partial_{\dot\sigma}\hat{b}_{\gamma\dot\rho})
(\partial_{\dot\lambda}\hat{b}_{\alpha\dot\tau}-\partial_{\dot\tau}\hat{b}_{\alpha\dot\lambda})\bigg).
\eea
Hence, we combine all terms that we obtain to show the Lagrangian formulation of the single M5-brane or the NP M5-brane theory in the bosonic sector
\bea
&&
{\cal L}_{\mathrm{NPM5}}
\nn\\
&=&\frac{T_{\mathrm{M}5}}{\theta^2}\int d^3xd^3y\
\bigg\lbrack-\frac{1}{12}\hat{{\cal H}}_{\dot\mu\dot\nu\dot\rho}\hat{{\cal H}}^{\dot\mu\dot\nu\dot\rho}
-\frac{1}{4}\hat{{\cal H}}_{\alpha\dot\mu\dot\nu}\hat{{\cal H}}^{\alpha\dot\mu\dot\nu}
\nn\\
&&+\epsilon^{\alpha\beta\gamma}
\bigg(-\frac{1}{2}\epsilon^{\dot\mu\dot\nu\dot\rho}(\partial_{\dot\mu}\hat{b}_{\alpha\dot\nu})(\partial_{\beta}\hat{b}_{\gamma\dot\rho})
\nn\\
&&
+\frac{\theta}{6}\epsilon^{\dot\mu\dot\nu\dot\lambda}\epsilon^{\dot\rho\dot\sigma\dot\tau}
(\partial_{\dot\mu}\hat{b}_{\beta\dot\nu})
(\partial_{\dot\sigma}\hat{b}_{\gamma\dot\rho})
(\partial_{\dot\lambda}\hat{b}_{\alpha\dot\tau}-\partial_{\dot\tau}\hat{b}_{\alpha\dot\lambda})\bigg)
\nn\\
&&
-\frac{1}{2}(\hat{D}_{\alpha}\hat{X}^{\tilde{I}})(\hat{D}^{\alpha}\hat{X}^{\tilde{I}})
-\frac{1}{2}(\hat{D}_{\dot\mu}\hat{X}^{\tilde{I}})(\hat{D}^{\dot\mu}\hat{X}^{\tilde{I}})
\nn\\
&&
-\frac{\theta^4}{4}\{ X^{\dot\mu}, X^{\tilde{I}}, X^{\tilde{J}}\}\{ X_{\dot\mu}, X^{\tilde{I}}, X^{\tilde{J}}\}
-\frac{\theta^4}{12}\{ X^{\tilde{I}}, X^{\tilde{J}}, X^{\tilde{K}}\}\{ X^{\tilde{I}}, X^{\tilde{J}}, X^{\tilde{K}}\}\bigg\rbrack.
\eea
The NP M5-brane theory is non-commutative, and the flat metric plays a role similar to that of the open-string metric.
We will show the self-duality from the leading-order equations of motion to demonstrate the consistent gauge degrees of freedom to the expectation of the single M5-brane \cite{Ho:2008nn}.

\subsubsection{Self-Duality}
\noindent
The Lagrangian formulation for the gauge sector up to the leading order with respect to the non-commutativity parameter is \cite{Ho:2008nn}
\bea
&&
{\cal L}_{\mathrm{NPM5G0}}
\nn\\
&=&
\frac{T_{\mathrm{M}5}}{\theta^2}\int d^3xd^3y\
\bigg(-\frac{1}{12}\hat{H}_{\dot\mu\dot\nu\dot\rho}\hat{H}^{\dot\mu\dot\nu\rho}
-\frac{1}{4}\hat{H}_{\alpha\dot\mu\dot\nu}\hat{H}^{\alpha\dot\mu\dot\nu}
-\frac{1}{2}\epsilon^{\alpha\beta\gamma}\epsilon^{\dot\mu\dot\nu\dot\rho}(\partial_{\dot\mu}\hat{b}_{\alpha\dot\nu})(\partial_{\beta}\hat{b}_{\gamma\dot\rho})
\bigg)
\nn\\
&=&-\frac{1}{4}\frac{T_{\mathrm{M}5}}{\theta^2}\int d^3xd^3y\
\bigg(\hat{H}_{\alpha\dot\mu\dot\nu}(\hat{H}-\tilde{\hat{H}})^{\alpha\dot\mu\dot\nu}
+\frac{1}{3}\hat{H}_{\dot\mu\dot\nu\dot\lambda}(\hat{H}-\tilde{\hat{H}})^{\dot\mu\dot\nu\dot\lambda}\bigg),
\eea
where
\bea
\tilde{\hat{H}}^{\alpha\dot\mu\dot\nu}=\frac{1}{2}\epsilon^{\alpha\beta\gamma}\epsilon^{\dot\mu\dot\nu\dot\lambda}\hat{H}_{\beta\gamma\dot\lambda}; \
\tilde{\hat{H}}^{\dot\mu\dot\nu\dot\lambda}=-\frac{1}{6}\epsilon^{\alpha\beta\gamma}\epsilon^{\dot\mu\dot\nu\lambda}\hat{H}_{\alpha\beta\gamma}.
\eea
Although the NP M5-brane theory initially does not have $\hat{b}_{\alpha\beta}$, it can appear in a total derivative term \cite{Ho:2008nn}.
The equations of motion are:
\bea
\partial_{\dot\lambda}(\hat{H}-\tilde{\hat{H}})^{\alpha\dot\mu\dot\lambda}=0; \
\partial_{\gamma}\hat{H}^{\gamma\dot\mu\dot\nu}+\partial_{\dot\lambda}\hat{H}^{\dot\lambda\dot\mu\dot\nu}=0.
\eea
which are derived from the variation of $\hat{b}_{\dot\mu\alpha}$ and $\hat{b}_{\dot\mu\dot\nu}$ \cite{Ho:2008nn}.
The equation of motion for $\hat{b}_{\dot\mu\alpha}$ at the local level implies that
\bea
(\hat{H}-\tilde{\hat{H}})^{\alpha\dot\mu\dot\nu}=-\frac{1}{2}\epsilon^{\alpha\beta\gamma}\epsilon^{\dot\lambda\dot\mu\dot\nu}\partial_{\dot\lambda}\hat{{\cal O}}_{1, \beta\gamma}.
\eea
We redefine the $\hat{b}_{\alpha\beta}$ as
\bea
\hat{b}_{\alpha\beta}\rightarrow\hat{b}_{\alpha\beta}+\hat{{\cal O}}_{1, \alpha\beta}
\eea
to get the first self-duality condition
\bea
\hat{H}_{\alpha\dot\mu\dot\nu}=\tilde{\hat{H}}_{\alpha\dot\mu\dot\nu},
\eea
which implies that the equation of motion for $\hat{b}_{\dot\mu\dot\nu}$ become
\bea
\partial^{\dot\lambda}\bigg(\hat{H}_{\dot\lambda\dot\mu\dot\nu}
+\frac{1}{2}\epsilon_{\alpha\beta\gamma}\epsilon_{\dot\mu\dot\nu\dot\lambda}\partial^{\alpha}\hat{b}^{\beta\gamma}\bigg)=0,
\eea
and the general solution is
\bea
\hat{H}_{\dot\lambda\dot\mu\dot\nu}
+\frac{1}{2}\epsilon_{\alpha\beta\gamma}\epsilon_{\dot\mu\dot\nu\dot\lambda}\partial^{\alpha}\hat{b}^{\beta\gamma}
=\epsilon^{\dot\lambda\dot\mu\dot\nu}\hat{{\cal O}}_2(x),
\eea
and then we can redefine $\hat{b}^{\beta\gamma}$ to get another self-dualty relation
\bea
\hat{H}_{\dot\mu\dot\nu\dot\lambda}=-\tilde{\hat{H}}_{\dot\mu\dot\nu\dot\lambda}.
\eea
Hence, we obtain all self-duality conditions to relate each field strength to another field strength as the expectation from a single M5-brane \cite{Ho:2008nn}.
In other words, the gauge degrees of freedom are 3 in the NP M5-brane theory \cite{Ho:2008nn}.

\subsection{NS-NS D4-Brane from NP M5-Brane}
\noindent
We compactify the $y^{\dot 3}$ direction in the NP M5-brane, and show that this theory is reuduced to the D4-brane theory in a large NS-NS $B$-field background along the $y^{\dot1}, y^{\dot 2}$ directions \cite{Ho:2008ve}.
Because the compactification radius on $y^{\dot 3}$ is $2\pi\theta R_{\dot 3}$, we obtain a new prefactor \cite{Ho:2008ve}
\bea
\frac{T_{\mathrm{M}5}}{\theta^2}2\pi\theta R_{\dot 3}=\frac{T_4}{\theta}.
\eea
We define the one-form gauge potential of the D4-brane as the following \cite{Ho:2008ve}:
\bea
\hat{a}_{\alpha}=\hat{b}_{\alpha\dot 3}; \ \hat{a}_{\dot\mu}=\hat{b}_{\dot\mu\dot 3},
\eea
identifiy the Levi-Civita symbol as \cite{Ho:2008ve}
\bea
\epsilon^{\dot\mu\dot\nu}=\epsilon^{\dot\mu\dot\nu\dot 3},
\eea
and also fix the gauge \cite{Ho:2008ve}
\bea
\hat{b}^{\dot 3}=0.
\eea
The field strengths become \cite{Ho:2008ve}:
\bea
\hat{{\cal H}}_{\dot 1\dot 2\dot 3}&=&\hat{F}_{\dot 1\dot 2}=\partial_{\dot 1}\hat{a}_{\dot 2}-\partial_{\dot 2}\hat{a}_{\dot 1}
+\theta\epsilon^{\dot\rho\dot\sigma}(\partial_{\dot\rho}\hat{a}_{\dot\mu})(\partial_{\dot\sigma}\hat{a}_{\dot\nu})
\equiv
\partial_{\dot 1}\hat{a}_{\dot 2}-\partial_{\dot 2}\hat{a}_{\dot 1}
+\theta\{\hat{a}_{\dot\mu}, \hat{a}_{\dot\nu}\};
\nn\\
\hat{{\cal H}}_{\alpha\dot 1\dot 2}&=&-\epsilon^{\dot\mu\dot\nu}\partial_{\dot\mu}\hat{b}_{\alpha\dot\nu}\equiv-\hat{\tilde{a}}_{\alpha};
\nn\\
\hat{{\cal H}}_{\alpha\dot\mu\dot 3}&=&\hat{F}_{\alpha\dot\mu}.
\eea
The Chern-Simons term becomes \cite{Ho:2008ve}
\bea
&&
\int d^3xd^3y\ \bigg\lbrack
\epsilon^{\alpha\beta\gamma}
\bigg(-\frac{1}{2}\epsilon^{\dot\mu\dot\nu\dot\rho}(\partial_{\dot\mu}\hat{b}_{\alpha\dot\nu})(\partial_{\beta}\hat{b}_{\gamma\dot\rho})
\nn\\
&&
+\frac{\theta}{6}\epsilon^{\dot\mu\dot\nu\dot\lambda}\epsilon^{\dot\rho\dot\sigma\dot\tau}
(\partial_{\dot\mu}\hat{b}_{\beta\dot\nu})
(\partial_{\dot\sigma}\hat{b}_{\gamma\dot\rho})
(\partial_{\dot\lambda}\hat{b}_{\alpha\dot\tau}-\partial_{\dot\tau}\hat{b}_{\alpha\dot\lambda})\bigg)\bigg\rbrack
\nn\\
&=&\int d^3x d^3y\ \bigg(\frac{1}{2}\epsilon^{\alpha\beta\gamma}\hat{F}_{\alpha\beta}\hat{\tilde{a}}_{\gamma}\bigg).
\eea
The covariant derivatives become \cite{Ho:2008ve}:
\bea
\hat{D}_{\alpha}\hat{X}^{\tilde{I}}=\partial_{\alpha}\hat{X}^{\tilde{I}}+\theta\{\hat{a}_{\alpha}, \hat{X}^{\tilde{I}}\}; \
\hat{D}_{\dot\mu}\hat{X}^{\tilde{I}}=\partial_{\dot\mu}\hat{X}^{\tilde{I}}+\theta\{\hat{a}_{\dot\mu}, \hat{X}^{\tilde{I}}\}; \
\hat{D}_{\dot3}\hat{X}^{\tilde{I}}=0.
\eea
Combinging the above results, we can obtain the action \cite{Ho:2008ve}
\bea
S_1&=&\frac{T_{4}}{\theta}\int d^3xd^2y\ \bigg(
-\frac{1}{4}\hat{F}_{\dot\mu\dot\nu}\hat{F}^{\dot\mu\dot\nu}
-\frac{1}{2}\hat{F}_{\alpha\dot\mu}\hat{F}^{\alpha\dot\mu}
-\frac{1}{2}\hat{\tilde{a}}_{\alpha}\hat{\tilde{a}}^{\alpha}
+\frac{1}{2}\epsilon^{\alpha\beta\gamma}\hat{F}_{\alpha\beta}\hat{\tilde{a}}_{\gamma}
\nn\\
&&
-\frac{1}{2}(\hat{D}_{\alpha}\hat{X}^{\tilde{I}})(\hat{D}^{\alpha}\hat{X}^{\tilde{I}})
-\frac{1}{2}(\hat{D}_{\dot\mu}\hat{X}^{\tilde{I}})(\hat{D}^{\dot\mu}\hat{X}^{\tilde{I}})
-\frac{\theta^2}{4}\{\hat{X}^{\tilde{I}}, \hat{X}^{\tilde{J}}\}\{\hat{X}^{\tilde{I}}, \hat{X}^{\tilde{J}}\}\bigg).
\eea
The $\hat{b}_{\alpha\dot\mu}$ is not a dynamical field \cite{Ho:2008ve}.
Therefore, we can integrate it out, and it is equivalent to using \cite{Ho:2008ve}
\bea
\partial_{\dot\mu}\bigg(\frac{1}{2}\epsilon_{\alpha\beta\gamma}\hat{F}_{\alpha\beta}-\hat{\tilde{a}}_{\gamma}\bigg)=0,
\eea
We redefine the gauge potential $\hat{\tilde{a}}_{\alpha}$ to get \cite{Ho:2008ve}
\bea
\frac{1}{2}\epsilon_{\alpha\beta\gamma}\hat{F}_{\alpha\beta}=\hat{\tilde{a}}_{\gamma}.
\eea
We also redefine the gauge potentials \cite{Ho:2008ve}:
\bea
\hat{a}_{\alpha}\rightarrow  (2\pi\alpha^{\prime})\hat{a}_{\alpha}; \
\hat{a}_{\dot\mu}\rightarrow(2\pi\alpha^{\prime})\hat{a}_{\dot\mu}
\eea
to have the proper coefficient of the boundary interaction $\int_{\partial\mathrm{F}1}\hat{a}$.
Hence, we obtain the NS-NS D4-brane theory \cite{Ho:2008ve}
\bea
S_{\mathrm{NSNSD4}}&=&\frac{T_{4}}{\theta}\int d^3d^2y\ \bigg(
-\frac{(2\pi\alpha^{\prime})^2}{4}\hat{F}_{\dot\mu\dot\nu}\hat{F}^{\dot\mu\dot\nu}
-\frac{(2\pi\alpha^{\prime})^2}{2}\hat{F}_{\alpha\dot\mu}\hat{F}^{\alpha\dot\mu}
-\frac{(2\pi\alpha^{\prime})^2}{4}\hat{F}_{\alpha\beta}\hat{F}^{\alpha\beta}
\nn\\
&&
-\frac{1}{2}(\hat{D}_{\alpha}\hat{X}^{\tilde{I}})(\hat{D}^{\alpha}\hat{X}^{\tilde{I}})
-\frac{1}{2}(\hat{D}_{\dot\mu}\hat{X}^{\tilde{I}})(\hat{D}^{\dot\mu}\hat{X}^{\tilde{I}})
-\frac{\theta^2}{4}\{\hat{X}^{\tilde{I}}, \hat{X}^{\tilde{J}}\}\{\hat{X}^{\tilde{I}}, \hat{X}^{\tilde{J}}\}\bigg).
\nn\\
\eea
The field strengths and covariant derivatives denoted by $\theta$ are replaced with $\theta^{\dot{1}\dot{2}}$ \cite{Ho:2008ve}.
The prefactor of the non-commutative D4-brane theory's gauge sector is \cite{Seiberg:1999vs}:
\bea
\frac{1}{2\pi\hat{G}_s}=\frac{1}{2\pi g_s}\frac{2\pi\alpha^{\prime}}{\theta^{\dot1\dot2}}=\frac{T_4(2\pi\alpha^{\prime})^2}{\theta},
\eea
which uses \cite{Seiberg:1999vs}:
\bea
\hat{G}_s= g_s\det\bigg(\frac{G\theta}{2\pi\alpha^{\prime} }\bigg)^{\frac{1}{2}}\big(1+{\cal O}(\epsilon)\big)
=g_s\det(G)^{\frac{1}{2}}\frac{\theta^{\dot 1\dot 2}}{2\pi\alpha^{\prime}}\big(1+{\cal O}(\epsilon)\big).
\eea

\subsection{Scaling Limit}
\noindent
We already get the D4-brane in the large NS-NS $B$-field background from the NP M5-brane \cite{Ho:2008ve}.
The NP M5-brane theory should effectively describe a single M5-brane under a scaling limit.
Therefore, now we analyze the effective action of the NP M5-brane behaves well under what scaling limit from the NS-NS D4-brane's scaling limit \cite{Seiberg:1999vs}:
\bea
g_s\sim\epsilon^{\frac{1}{4}}; \ l_s\sim\epsilon^{\frac{1}{4}}; \ g_{\alpha\beta}\sim\epsilon^0; \ g_{\dot\mu\dot\nu}\sim\epsilon; \
B_{\dot{\mu}\dot{\nu}}\sim\epsilon^0,
\eea
which corresponds to $p=4$ and $r=2$.
For the compactification direction, the metric $g_{\dot3\dot3}$ follows the same scaling limit as the components along the large $C$-field background directions.
According to the correspondence between M-theory and type IIA, we can determine the 11D Planck scale and the compactification radius:
\bea
l_p=g_s^{\frac{1}{3}}l_s\sim\epsilon^{\frac{1}{3}}; \ \frac{R_{\dot{3}}}{\sqrt{g_{\dot3\dot3}}}\sim\epsilon^0,
\eea
in which the scaling limit:
\bea
R_{\dot 3}=g_sl_s\sim\epsilon^{\frac{1}{2}}
\eea
is to have a finite compactification radius.
The finite compactification radius also determines the scaling limit of the large $C$-field background:
\bea
\frac{1}{2\pi\frac{R_{\dot 3}}{\sqrt{R_{\dot3\dot3}}}}B_{\dot1\dot2}=C_{\dot1\dot2\dot3}\sim\epsilon^0.
\eea
When we perform the T-duality exchanging the radius with the much smaller $R_{\dot{3}}$ compared to the string length, the particle description is dominant in the D4-brane.
We can also use the T-dual to obtain a D$p$-brane in a large NS-NS $B$-field background.
The much larger radius $R_{\alpha}$ compared to the string length truncates the winding modes in the NS-NS D$p$-brane theory.
Thus, the primary contribution arises from the zero momentum mode in the NS-NS D$p$-brane theory when considering the scaling limit.

\section{Duality Web}
\label{sec:6}
\noindent
We compactify the $x^{\alpha=2}$ direction to get the D4-brane in the large $C$-field background \cite{Ma:2012hc} and then generalize it to a D$p$-brane in the large ($p-1$)-form field background according to the following assumptions \cite{Ma:2023hgi}:
\begin{itemize}
\item{Partial Lorentz Symmetry: SO(1, 1)$\times$SO($p-1$)$\times$ SO($9-p$);}
\item{Gauge Symmetry: U(1) and VPD;}
\item{Field Content: 1-form, ($p-2$)-form gauge potentials living on the worldvolume and scalar fields living in the directions orthogonal to the worldvolume;}
\item{Duality: D$p$-brane is dual to D($p\pm 1$)-brane through T-duality.;}
\item{Leading-Order in Large R-R Field Background: effective description in the gauge sector agrees with the U(1) YM theory.}
\end{itemize}
We then demonstrate that the $p=3$ case is also consistent with the electromagnetic duality between the NS-NS D3-brane and R-R D3-brane theories supporting the S-duality \cite{Ho:2013opa,Ho:2015mfa}.

\subsection{R-R D4-Brane from NP M5-Brane}
\noindent
We now compactify $x^{\alpha=2}$ direction to the NP-M5-brane theory, and first identify that \cite{Ma:2012hc}:
\bea
\hat{{\cal H}}_{2\dot\mu\dot\nu}\equiv\hat{{\cal F}}_{\dot\mu\dot\nu}=
\hat{F}_{\dot\mu\dot\nu}
+\theta\big((\partial_{\dot\sigma}\hat{b}^{\dot\sigma})\hat{F}_{\dot\mu\dot\nu}-(\partial_{\dot\mu}\hat{b}^{\dot\sigma})\hat{F}_{\dot\sigma\dot\nu}-(\partial_{\dot\nu}\hat{b}^{\dot\sigma})\hat{F}_{\dot\mu\dot\sigma}\big); \
\epsilon^{\alpha\beta 2}=\epsilon^{\alpha\beta},
\eea
where
\bea
\hat{b}_{\dot\mu 2}=\hat{a}_{\dot\mu}; \ \hat{F}\equiv d\hat{a}.
\eea
The covariant derivative of the scalar field along the $x^{\alpha=2}$ direction becomes \cite{Ma:2012hc}
\bea
\hat{D}_2\hat{X}^{\tilde{I}}=\frac{\theta}{2}\epsilon^{\dot\mu\dot\nu\rho}\hat{F}_{\dot\nu\dot\rho}\partial_{\dot\mu}\hat{X}^{\tilde{I}}.
\eea
We cannot only use the identification to have the R-R D4-brane's gauge potential $\hat{a}_{\alpha}$ but introducing a new auxiliary field in the Lagrangian, $-\epsilon^{\alpha\beta}\hat{f}_{\beta\dot\mu}(\hat{\tilde{B}}_{\alpha}{}^{\dot\mu}-\epsilon^{\dot\mu\dot\nu\dot\rho}\partial_{\dot\nu}\hat{b}_{\alpha\dot\rho})$ \cite{Ma:2012hc}.
Integrating out the auxiliary field $\hat{f}$ can go back to the original Lagrangian, but now we integrate out the $\hat{b}_{\alpha\dot\mu}$, which is equivalent to using \cite{Ma:2012hc}
\bea
\epsilon^{\dot\mu\dot\nu\dot\rho}\partial_{\dot\mu}\hat{f}_{\alpha\dot\nu}=0.
\eea
The general solution up to the local level is \cite{Ma:2012hc}
\bea
\hat{f}_{\alpha\dot\mu}=\partial_{\dot\mu}\hat{a}_{\alpha}.
\eea
Now, the necessary gauge potential $\hat{a}_{\alpha}$ appears from the dual, and then we can identify another covariant field strength \cite{Ma:2012hc}
\bea
\hat{{\cal F}}_{\alpha\dot\mu}=\frac{1}{2}\epsilon_{\beta\alpha}\epsilon_{\dot\mu\dot\nu\dot\rho}
\hat{{\cal H}}^{\beta\dot\nu\dot\lambda}.
\eea
After integrating out the $\hat{\tilde{B}}_{\alpha}{}^{\dot\mu}$ and ignoring the non-trivial quantum correction in the path integration measure to obtain the dual theory, the $\hat{\tilde{B}}_{\alpha}{}^{\dot\mu}$ satisfies \cite{Ma:2012hc}
\bea
&&\hat{V}_{\dot{\mu}}{}^{\dot{\nu}}(\partial^{\alpha}\hat{b}_{\dot{\nu}}-\hat{V}^{\dot{\rho}}{}_{\dot{\nu}}\hat{\tilde{B}}^{\alpha}{}_{\dot{\rho}})
+\epsilon^{\alpha\beta}\hat{F}_{\beta\dot{\mu}}
+\theta\epsilon^{\alpha\beta}\hat{F}_{\dot{\mu}\dot{\nu}}\hat{\tilde{B}}_{\beta}{}^{\dot{\nu}}
+\theta\partial_{\dot{\mu}}\hat{X}^{\tilde{I}}\hat{D}^{\alpha}\hat{X}^{\tilde{I}}
=0,
\label{hatB}
\eea
where
\bea
\hat{D}_{\alpha}\hat{X}^{\tilde{I}}=\partial_{\alpha}X^{\tilde{I}}-\theta\hat{\tilde{B}}_{\alpha}{}^{\dot\mu}\partial_{\dot\mu}\hat{X}^{\tilde{I}}.
\eea
The compactification with the periodicity on $x^{\alpha=2}$ direction \cite{Ma:2012hc}
\bea
x^{\alpha=2}\sim x^{\alpha=2}+2\pi R_2
\eea
shows the prefactor to the R-R D4-brane theory
\bea
\frac{T_{\mathrm{M}5}(2\pi R_2)}{\theta^2}=\frac{T_4}{\theta^2}.
\eea
Combining the above result shows the D4-brane theory in a large $C$-field background \cite{Ma:2012hc}
\bea
&&
S_{\mathrm{RRD}4}
\nn\\
&=&\frac{T_4}{\theta^2}\int d^2xd^3y\ \bigg(-\frac{1}{2}\hat{{\cal H}}_{\dot1\dot2\dot3}\hat{{\cal H}}^{\dot1\dot2\dot3}
-\frac{1}{4}\hat{{\cal F}}_{\dot\nu\dot\rho}\hat{{\cal F}}^{\dot\nu\dot\rho}
+\frac{1}{2}\hat{{\cal F}}_{\beta\dot\mu}\hat{{\cal F}}^{\beta\dot\mu}
+\frac{1}{2\theta}\epsilon^{\alpha\beta}\hat{{\cal F}}_{\alpha\beta}
\nn\\
&&
-\frac{1}{2}(\hat{{\cal D}}_{\alpha}\hat{X}^{\tilde{I}})(\hat{{\cal D}}^{\alpha}\hat{X}^{\tilde{I}})
-\frac{1}{2}(\hat{{\cal D}}_{\dot\mu}\hat{X}^{\tilde{I}})(\hat{{\cal D}}^{\dot\mu}\hat{X}^{\tilde{I}})
-\frac{\theta^2}{8}\epsilon^{\dot\mu\dot\rho\dot\tau}\epsilon^{\dot\nu\dot\sigma\dot\delta}
\hat{F}_{\dot\rho\dot\tau}
\hat{F}_{\dot\sigma\dot\delta}
(\partial_{\dot\mu}\hat{X}^{\tilde{I}})
(\partial_{\dot\nu}\hat{X}^{\tilde{I}})
\nn\\
&&
-\frac{\theta^4}{4}\{\hat{X}^{\dot\mu}, \hat{X}^{\tilde{I}}, \hat{X}^{\tilde{J}}\} \{\hat{X}_{\dot\mu}, \hat{X}^{\tilde{I}}, \hat{X}^{\tilde{J}}\}
-\frac{\theta^4}{12}\{\hat{X}^{\tilde{I}}, \hat{X}^{\tilde{J}}, \hat{X}^{\tilde{K}}\} \{\hat{X}^{\tilde{I}}, \hat{X}^{\tilde{J}}, \hat{X}^{\tilde{K}}\}\bigg),
\eea
where
\bea
\hat{{\cal F}}_{\alpha\beta}=\hat{F}_{\alpha\beta}
+\theta(-\hat{F}_{\alpha\dot\mu}\hat{\tilde{B}}_{\beta}{}^{\dot\mu}
-\hat{F}_{\dot\mu\beta}\hat{\tilde{B}}_{\alpha}{}^{\dot\mu})
+\theta^2\hat{F}_{\dot\mu\dot\nu}
\hat{\tilde{B}}_{\alpha}{}^{\dot\mu}
\hat{\tilde{B}}_{\beta}{}^{\dot\nu}.
\eea
Since we only have a large $C$-field background, the $\epsilon^{\alpha\beta}\hat{{\cal F}}_{\alpha\beta}$ is the Wess-Zumino term that we expect \cite{Ma:2012hc}.
We will show how the U(1) YM theory appears in this D4-brane theory even with a strange coefficient of $\hat{{\cal F}}_{\beta\dot\mu}\hat{{\cal F}}^{\beta\dot\mu}$ \cite{Ma:2012hc}.

\subsection{U(1) YM}
\noindent
The R-R D4-brane Lagrangian up to the leading order is \cite{Ho:2011yr}
\bea
S_{\mathrm{RRD}40}=\frac{T_4}{\theta^2}\int d^2xd^3y\ \bigg(-\frac{1}{2}\hat{H}_{\dot 1\dot 2\dot 3}\hat{H}^{\dot1\dot2\dot3}
-\frac{1}{4}\hat{F}_{\dot{\mu}\dot{\nu}}\hat{F}^{\dot{\mu}\dot{\nu}}
-\frac{1}{2}\hat{F}_{\alpha\dot{\mu}}\hat{F}^{\alpha\dot{\mu}}
-\hat{F}_{01}\hat{H}_{\dot 1\dot2\dot 3}\bigg).
\eea
We use the leading-order solution of the $\hat{\tilde{B}}_{\alpha}{}^{\dot\mu}$ \cite{Ho:2011yr},
\bea
\hat{\tilde{B}}_{\alpha}{}^{\dot\mu}=\partial_{\alpha}\hat{b}^{\dot\mu}
+\epsilon_{\alpha\beta}\hat{F}^{\beta\dot\mu}+{\cal O}(\theta).
\eea
Because $\hat{b}$ does not have a time-derivative term, we can integrate it out, equivalent to using \cite{Ho:2011yr}
\bea
\hat{H}_{\dot1\dot2\dot3}=-\hat{F}_{01}+\hat{f},
\eea
where
\bea
\partial_{\dot{\mu}}\hat{f}=0.
\eea
Therefore, the $\hat{f}$ only depends on $x^{\alpha}$, which can be absorbed by $\hat{a}_{\alpha}$ through a field redefinition \cite{Ho:2011yr} .
Therefore, the leading-order term respects the U(1) YM theory \cite{Ho:2011yr}
\bea
S_{\mathrm{U(1)YM}}=\frac{T_4}{\theta^2}\int d^2xd^3y\ \bigg(-\frac{1}{4}\hat{F}_{\dot\mu\dot\nu}\hat{F}^{\dot\mu\dot\nu}
-\frac{1}{2}\hat{F}_{\alpha\dot\mu}\hat{F}^{\alpha\dot\mu}
-\frac{1}{4}\hat{F}_{\alpha\beta}\hat{F}^{\alpha\beta}\bigg).
\eea
It is easy to know that the R-R D$p$-brane's generalization always respects the U(1) YM theory at the leading-order for the non-commutativity parameter through the T-duality on the $y^{\dot\mu}$ direction \cite{Ho:2013paa}.

\subsection{R-R D$p$-Brane and T-duality}
\noindent
We first discuss the generalization of the gauge transformation to the D$p$-brane in a large R-R ($p+1$)-form field background \cite{Ho:2013paa} and then use the gauge transformation to obtain the solution of the SW map up to the first-order in $\theta$ \cite{Ma:2020msx}.
It should guarantee that the gauge transformation can have the non-commutative description of a Lagrangian formulation \cite{Ho:2013paa,Ma:2020msx}.
Finally, we show the Lagrangian under the requirement of the T-duality and the gauge symmetry, which leads to the ($p-1$)-bracket \cite{Ho:2013paa}.
The ($p-1$)-bracket describes the large R-R ($p-1$)-form field background and also generates the VPD symmetry \cite{Ho:2013paa}.
The leading-order gauge sector for $\theta$ is the U(1) YM gauge theory \cite{Ho:2013paa}.
We will use the index $I$ instead of $\tilde{I}$ to denote the D$p$-brane's transver scalar fields and also use the $\hat{B}_{\alpha}{}^{\dot\mu}$ to replace $\hat{\tilde{B}}_{\alpha}{}^{\dot\mu}$ in the D$p$-brane's case.

\subsubsection{Gauge Symmetry and SW Map}
\noindent
We generalize the gauge transformation of the R-R D$p$-brane theory as \cite{Ho:2013paa}:
\bea
\hat{\delta}_{\hat{\Lambda}}\hat{X}^I&=&\theta\hat{\kappa}^{\dot{\mu}}\partial_{\dot{\mu}}\hat{X}^I;
\nn\\
\hat{\delta}_{\hat{\Lambda}}\hat{b}^{\dot{\mu}}&=&\hat{\kappa}^{\dot{\mu}}
+\theta\hat{\kappa}^{\dot{\nu}}\partial_{\dot{\nu}}\hat{b}^{\dot{\mu}};
\nn\\
\hat{\delta}_{\hat{\Lambda}}\hat{B}_{\alpha}{}^{\dot{\mu}}&=&\partial_{\alpha}\hat{\kappa}^{\dot{\mu}}+\theta\big(\hat{\kappa}^{\dot{\nu}}\partial_{\dot{\nu}}\hat{B}_{\alpha}{}^{\dot{\mu}}-(\partial_{\dot{\nu}}\hat{\kappa}^{\dot{\mu}})\hat{B}_{\alpha}{}^{\dot{\nu}}\big);
\nn\\
\hat{\delta}_{\hat{\Lambda}}\hat{a}_{\alpha}&=&\partial_{\alpha}{\hat{\lambda}}+\theta(\hat{\kappa}^{\dot{\nu}}\partial_{\dot{\nu}}\hat{a}_{\alpha}
+\hat{a}_{\dot{\nu}}\partial_{\alpha}\hat{\kappa}^{\dot{\nu}});
\nn\\
\hat{\delta}_{\hat{\Lambda}}\hat{a}_{\dot{\mu}}&=&\partial_{\dot{\mu}}{\hat{\lambda}}+\theta(\hat{\kappa}^{\dot{\nu}}\partial_{\dot{\nu}}\hat{a}_{\dot{\mu}}
+\hat{a}_{\dot{\nu}}\partial_{\dot{\mu}}\hat{\kappa}^{\dot{\nu}}),
\eea
where $\hat{B}_{\alpha}{}^{\dot{\mu}}$ satisfies that \cite{Ho:2013paa}
\bea
&&\hat{V}_{\dot{\mu}}{}^{\dot{\nu}}(\partial^{\alpha}\hat{b}_{\dot{\nu}}-\hat{V}^{\dot{\rho}}{}_{\dot{\nu}}\hat{B}^{\alpha}{}_{\dot{\rho}})
+\epsilon^{\alpha\beta}\hat{F}_{\beta\dot{\mu}}
+\theta\epsilon^{\alpha\beta}\hat{F}_{\dot{\mu}\dot{\nu}}\hat{B}_{\beta}{}^{\dot{\nu}}
+\theta\partial_{\dot{\mu}}\hat{X}^I\hat{D}^{\alpha}\hat{X}^I=0.
\eea
The gauge parameter $\hat{\kappa}$ generates the VPD symmetry, and the $\hat{\lambda}$ generates the U(1) symmetry.
The SW map of the R-R D$p$-brane satisfies the following relations \cite{Ma:2020msx}:
\bea
\hat{b}^{\dot{\mu}}(b+\delta_{\Lambda}b)&=&\hat{\delta}_{\hat{\Lambda}}\hat{b}^{\dot{\mu}}+\hat{b}^{\dot{\mu}}(b);
\nn\\
\hat{B}_{\alpha}{}^{\dot{\mu}}(B+\delta_{\Lambda}B, b+\delta_{\Lambda}b)&=&\hat{\delta}_{\hat{\Lambda}}\hat{B}_{\alpha}{}^{\dot{\mu}}+\hat{B}_{\alpha}{}^{\dot{\mu}}(B, b)
\nn\\
\hat{a}_{\alpha}(a+\delta_{\Lambda}a, b+\delta_{\Lambda}b)&=&\hat{\delta}_{\hat{\Lambda}}\hat{a}_{\alpha}+\hat{a}_{\alpha}(a, b);
\nn\\
\hat{a}_{\dot\mu}(a+\delta_{\Lambda}a, b+\delta_{\Lambda}b)&=&\hat{\delta}_{\hat{\Lambda}}\hat{a}_{\dot\mu}+\hat{a}_{\dot\mu}(a, b);
\nn\\
\hat{X}^{I}(X+\delta_{\Lambda}X, b+\delta_{\Lambda}b)&=&\hat{\delta}_{\hat{\Lambda}}\hat{X}^{I}(X, b)+\hat{X}^{I}(X, b),
\eea
where
\bea
\delta_{\Lambda}b^{\dot{\mu}}=\kappa^{\dot{\mu}}; \
\delta_{\Lambda}B_{\alpha}{}^{\dot{\mu}}=\partial_{\alpha}\kappa^{\dot\mu}; \
\delta_{\Lambda}a_{\alpha}=\partial_{\alpha}\lambda; \
\delta_{\Lambda}a_{\dot\mu}=\partial_{\dot\mu}\lambda; \
\delta_{\Lambda}X^I=0.
\eea
In the R-R D-brane theory, the $B_{\alpha}{}^{\dot\mu}$ is not the same as in the NP-M5 brane case.
However, it is
\bea
B_{\alpha}{}^{\dot\mu}=\partial_{\alpha}b^{\dot\mu}+\epsilon_{\alpha\beta}F^{\beta\dot{\mu}}.
\eea
\\

\noindent
Because we can read the solution of the SW map about the R-R D4-brane from the NP M5-brane theory, and the gauge transformation has the same form for all D$p$-brane cases, we can obtain the SW map up to the first-order \cite{Ma:2020msx}:
\bea
&&\hat{b}^{\dot{\mu}}(b)
\nn\\
&=&b^{\dot{\mu}}+\theta\bigg(\frac{1}{2}b^{\dot{\nu}}\partial_{\dot{\nu}}b^{\dot{\mu}}
+\frac{1}{2}b^{\dot{\mu}}\partial_{\dot{\nu}}b^{\dot{\nu}}\bigg)+{\cal O}(\theta^2),
\nn\\
&&\hat{B}_{\alpha}{}^{\dot{\mu}}(B, b)
\nn\\
&=&B_{\alpha}{}^{\dot{\mu}}+\theta\bigg(b^{\dot{\nu}}\partial_{\dot{\nu}}B_{\alpha}{}^{\dot{\mu}}
-\frac{1}{2}b^{\dot{\nu}}\partial_{\alpha}\partial_{\dot{\nu}}b^{\dot{\mu}}
+\frac{1}{2}b^{\dot{\mu}}\partial_{\alpha}\partial_{\dot{\nu}}b^{\dot{\nu}}
\nn\\
&&
+(\partial_{\dot{\nu}}b^{\dot{\nu}})B_{\alpha}{}^{\dot{\mu}}
-(\partial_{\dot{\nu}}b^{\dot{\mu}})B_{\alpha}{}^{\dot{\nu}}
-\frac{1}{2}(\partial_{\dot{\nu}}b^{\dot{\nu}})(\partial_{\alpha}b^{\dot{\mu}})
+\frac{1}{2}(\partial_{\dot{\nu}}b^{\dot{\mu}})(\partial_{\alpha}b^{\dot{\nu}})\bigg)+{\cal O}(\theta^2),
\nn\\
&&
\hat{a}_{\alpha}(a, b)
\nn\\
&=&a_{
\alpha}+\theta(b^{\dot{\rho}}\partial_{\dot{\rho}}a_{\alpha}+a_{\dot{\rho}}\partial_{\alpha}b^{\dot{\rho}})+{\cal O}(\theta^2);
\nn\\
&&
\hat{a}_{\dot\mu}(a, b)
\nn\\
&=&a_{
\dot\mu}+\theta(b^{\dot{\rho}}\partial_{\dot{\rho}}a_{\dot\mu}+a_{\dot{\rho}}\partial_{\dot\mu}b^{\dot{\rho}})+{\cal O}(\theta^2);
\\
&&\hat{X}^I(X, b)
\nn\\
&=&X+\theta b^{\dot{\mu}}\partial_{\dot{\mu}}X^I+{\cal O}(\theta^2),
\nn\\
&&\hat{\kappa}^{\dot{\mu}}(\kappa, b)
\nn\\
&=&
\kappa^{\dot{\mu}}
+\bigg(\frac{\theta}{2}b^{\dot{\nu}}\partial_{\dot{\nu}}\kappa^{\dot{\mu}}
+\frac{1}{2}(\partial_{\dot{\nu}}b^{\dot{\nu}})\kappa^{\dot{\mu}}
-\frac{1}{2}(\partial_{\dot{\nu}}b^{\dot{\mu}})\kappa^{\dot{\nu}}\bigg)+{\cal O}(\theta^2);
\nn\\
&&
\hat{\lambda}(\lambda, b)
\nn\\
&=&\lambda+\theta b^{\dot{\rho}}\partial_{\dot{\rho}}\lambda+{\cal O}(\theta^2).
\eea
Hence, the gauge transformation provides a solution to the SW map up to first order, motivating us to explore the Lagrangian formulation derived from it.

\subsubsection{Lagrangian and T-Duality}
\noindent
We now generalize the R-R D4-brane to the D$p$-brane respecting the T-duality on $y^{\dot\mu}$ direction \cite{Ho:2013paa}.
The prefactor of the D$p$-brane can be determined, and it is given by the expression $T_p / \theta^{p+2}$.
Because we now have a large ($p-1$)-form field background in D$p$-brane theory, some field strengths need to be generalized to have the ($p-1$)-bracket structure \cite{Ho:2013paa}:
\bea
\hat{{\cal H}}^{\dot 1\dot 2\cdots\dot\mu_{p-1}}
&=&\theta^{p-2}\{\hat{X}^{\dot\mu_1}, \hat{X}^{\dot\mu_2}, \cdots, \hat{X}^{\dot\mu_{p-1}}\}_{p-1}-\frac{1}{\theta}\epsilon^{\dot\mu_1\dot\mu_2\cdots\dot\mu_{p-1}};
\nn\\
\hat{{\cal F}}_{\dot\mu\dot\nu}&=&\frac{\theta^{p-3}}{(p-3)!}\epsilon_{\dot\mu\dot\nu\dot\mu_1\dot\mu_2\cdots\dot\mu_{p-3}}
\{\hat{X}^{\dot\mu_1}, \hat{X}^{\dot\mu_2}, \cdots, \hat{X}^{\dot\mu_{p-3}}, \hat{a}_{\dot\rho}, y^{\dot\rho}\}_{(p-1)},
\eea
where
\bea
\{\hat{f}_1, \hat{f}_2 ,\cdots, \hat{f}_{p-1}\}_{(p-1)}\equiv \epsilon^{\dot{\mu}_1\dot{\mu}_2\cdots\dot{\mu}_{p-1}}(\partial_{\dot{\mu}_1}\hat{f}_1)(\partial_{\dot{\mu}_2}\hat{f}_2)\cdots(\partial_{\dot{\mu}_{p-1}}\hat{f}_{p-1}).
\eea
The generalization of the ($p-1$)-bracket provides a manifest demonstration of the VPD gauge symmetry \cite{Ho:2013paa}.
Because the ($p-1$)-bracket satisfies the generalized version of the Jacobi identity
\bea
&&
\{\hat{f}_1, \hat{f}_2, \cdots, \hat{f}_{p-2}, \{\hat{g}_1, \hat{g}_2, \cdots, \hat{g}_{p-1}\}_{(p-1)}\}_{(p-1)}
\nn\\
&=&\{\{ \hat{f}_1, \hat{f}_2, \cdots, \hat{f}_{p-2}, \hat{g}_1\}_{(p-1)}, \cdots, \hat{g}_{p-1}\}_{(p-1)}
\nn\\
&&
+\{\hat{g}_1, \{\hat{f}_1, \hat{f}_2, \cdots, \hat{f}_{p-2}, \hat{g}_2\}_{(p-1)}, \cdots, \hat{g}_{p-1}\}_{(p-1)}
+\cdots
\nn\\
&&
+\{ \hat{g}_1, \hat{g}_2, \cdots, \hat{g}_{p-2}, \{\hat{f}_1, \hat{f}_2, \cdots, \hat{f}_{p-2}, \hat{g}_{p-1}\}_{(p-1)}\}_{(p-1)},
\eea
it is easy to show that the closed property is preserved under the gauge transformation \cite{Ho:2013paa}.
\\

\noindent
To derive the R-R D($p-1$) brane theory from the R-R D$p$-brane theory, one can replace the term for the $y^{\dot\mu}=p$ direction using the following substitutions \cite{Ho:2013paa}:
\bea
\hat{a}_p \rightarrow \hat{X}^{I=p}; \quad \hat{b}^p \rightarrow 0.
\eea
Although $\hat{B}_{\alpha}{}^{\dot\mu}$ is complicated, the T-daulity rule can be derived by Eq. \eqref{hatB},
\bea
\hat{B}_{\alpha}{}^{\dot\mu}\rightarrow \hat{B}_{\alpha}{}^{\dot\mu}; \
\hat{B}_{\alpha}{}^p\rightarrow \epsilon_{\alpha\beta}\hat{D}^{\beta}\hat{X}^{I=p}.
\eea
We also have other T-duality transformations \cite{Ho:2013paa}:
\bea
\hat{D}_{\alpha}\hat{X}^I&\rightarrow&\hat{D}_{\alpha}\hat{X}^I;
\nn\\
\hat{D}_{\dot\mu}\hat{X}^I&\rightarrow&\hat{D}_{\dot\mu}\hat{X}^I;
\nn\\
\hat{D}_{p}\hat{X}^I&\rightarrow&0;
\nn\\
\hat{{\cal F}}_{\dot\mu\dot\nu}&\rightarrow&\hat{{\cal F}}_{\dot\mu\dot\nu};
\nn\\
\hat{{\cal F}}_{\dot\mu p}&\rightarrow&\hat{D}_{\dot\mu}\hat{X}^{I=p};
\nn\\
\hat{{\cal F}}_{\alpha\dot\mu}&\rightarrow&\hat{{\cal F}}_{\alpha\dot\mu};
\nn\\
\hat{{\cal F}}_{\alpha p}&\rightarrow&\hat{D}_{\alpha}\hat{X}^{I=p};
\nn\\
\frac{1}{2}\hat{{\cal F}}_{\alpha\beta}&\rightarrow&\frac{1}{2}\hat{{\cal F}}_{\alpha\beta}
-\theta\big(\hat{D}_{\alpha}\hat{X}^{I=p}\big)^2;
\nn\\
\hat{{\cal H}}_{\dot\mu_1\dot\mu_2\cdots\dot\mu_{p-1}}&\rightarrow&\hat{{\cal H}}_{\dot\mu_1\dot\mu_2\cdots\dot\mu_{p-2}}.
\eea
The R-R D$p$-brane theory respecting the T-duality is \cite{Ho:2013paa}
\bea
S_{\mathrm{RRDP}}=S_1+S_2,
\eea
where
\bea
&&
S_1
\nn\\
&=&\frac{T_p}{\theta^{p-2}}\int d^2xd^{p-1}y\ \bigg(-\frac{1}{2}(\hat{D}_{\alpha}\hat{X}^I)(\hat{D}^{\alpha}\hat{X}^I)
+\frac{1}{2}\hat{{\cal F}}_{\alpha\dot\mu}\hat{{\cal F}}^{\alpha\dot\mu}
+\frac{1}{2\theta}\epsilon^{\alpha\beta}\hat{{\cal F}}_{\alpha\beta}\bigg);
\nn\\
&&
S_2
\nn\\
&=&\frac{T_p}{\theta^{p-2}}\int d^2xd^{p-1}y\
\sum_{n, m, l\in S}\bigg(-\frac{1}{2}\frac{\theta^{2(p-2-m)}}{(n!)(m!)^2(l!)}
\nn\\
&&\times
(\{\hat{X}^{\dot\mu_1}, \hat{X}^{\dot\mu_2}, \cdots, \hat{X}^{\dot\mu_n}, \hat{a}_{\dot\nu_1}, \hat{a}_{\dot\nu_2}, \cdots, \hat{a}_{\dot\nu_m}, y^{\dot\nu_1}, y^{\dot\nu_2}, \cdots, y^{\dot\nu_m}, \hat{X}^{I_1}, \hat{X}^{I_2}, \cdots, \hat{X}^{I_l}\})^2\bigg).
\nn\\
\eea
The set is
\bea
S\equiv\{(n, m, l)| n, m, l\ge 0; \ n+2m+l=p-1\}.
\eea
The $S_1$ and $S_2$ each have the invariant form under the T-duality transformation \cite{Ho:2013paa}.
When $p=4$, the theory is reduced to the R-R D4-brane theory derived from the NP M5-brane theory \cite{Ho:2013paa}.

\subsection{Scaling Limit}
\noindent
We can induce the R-R D$p$-brane's scaling limit from the NP M5-brane's scaling limit \cite{Ho:2011yr,Ho:2013paa}.
According to the parameter correspondence between M-theory and Type IIA theory, the 11D Planck scale and a finite compactification radius are:
\bea
l_p=(g_s^{RR})^{\frac{1}{3}}l^{RR}_s\sim\epsilon^{\frac{1}{3}}; \ \frac{R_2}{\sqrt{g_{22}}}\sim\epsilon^0,
\eea
in which the scaling limit:
\bea
R_{2}=g^{RR}_sl^{RR}_s\sim\epsilon^{0}.
\eea
Hence, we can determine the scaling limit to string coupling and string length as \cite{Ho:2011yr}:
\bea
g^{RR}_s\sim\epsilon^{-\frac{1}{2}}; \ l^{RR}_s\sim\epsilon^{\frac{1}{2}}.
\eea
Since the D$p$-brane is dual to D($p\pm 1$)-brane through the T-duality on the $y^{\dot\mu}$-direction, the scaling limit of the large R-R field background is \cite{Ma:2012hc,Ho:2013paa}
\bea
C_{\dot{\mu}_1\dot{\mu}_2\cdots\dot{\mu}_{p-1}}\sim\epsilon^0.
\eea
Hence, the scaling limit of the R-R D$p$-brane is \cite{Ho:2011yr,Ho:2013paa}:
\bea
g^{RR}_s\sim\epsilon^{-\frac{1}{2}}; \ l^{RR}_s\sim\epsilon^{\frac{1}{2}}; \ g_{\alpha\beta}\sim\epsilon^0; \ g_{\dot\mu\dot\nu}\sim\epsilon; \
C_{\dot{\mu}_1\dot{\mu}_2\cdots\dot{\mu}_{p-1}}\sim\epsilon^0.
\eea
After performing the T-duality to the NP M5-brane in the direction $x^{\alpha=2}$, we can tune the finite radius $R_2$ to be small enough to let the zero mode be the leading-order contribution to the R-R D4-brane.
We can perform the T-duality on the direction $x^{\dot\mu}$ to the R-R D4-brane to show the R-R D$p$-brane \cite{Ho:2013paa}.
Since the ratio $R_{\dot\mu}/l_s^{RR}$ is finite, we can tune it as a small finite number by letting the zero momentum be the leading-order contribution to the R-R D$p$-brane.
When $p=3$, we should have the S-dualty to the NS-NS D3-brane theory, which has the scaling limit \cite{Seiberg:1999vs}
\bea
g_s\sim\epsilon^{\frac{1}{2}}; \ l_s\sim\epsilon^{\frac{1}{4}}; \ g_{\alpha\beta}\sim\epsilon^0; \ g_{\dot\mu\dot\nu}\sim\epsilon; \
B_{\dot 1\dot 2}\sim\epsilon^0.
\eea
In the NS-NS D3-brane theory and R-R D3-brane theory, the string length is determined by different compactification radii, leading to different scaling limits of the string length.
The scaling law of the string coupling is inverted between the R-R D3-brane theory and the NS-NS D3-brane theory, as expected from S-duality in the case of Heterotic SO(32) string \cite{Ho:2013opa}.
However, the prefactor of the NS-NS D3-brane theory only depends on $\hat{G}_s$,
\bea
\frac{1}{g_3^2}=\frac{1}{2\pi\hat{G}_s}.
\eea
We can invert $g_3$ to implement the S-duality in the brane theory \cite{Ho:2015mfa}
\bea
g_3\rightarrow\frac{1}{g_3},
\eea
analogous to the inversion of the string coupling in the case of Heterotic SO(32) string.
We will later demonstrate the electromagnetic duality that connects these two theories \cite{Ho:2015mfa}.

\subsection{S-Duality}
\noindent
To invert the prefactor of the NS-NS D3-brane from the S-duality, we can rescale the fields \cite{Ho:2013opa}:
\bea
\hat{b}^{\dot\mu}\rightarrow Q\hat{b}^{\dot\mu}; \
\hat{B}_{\alpha}{}^{\dot\mu}\rightarrow Q\hat{B}_{\alpha}{}^{\dot\mu}; \
\hat{a}_{\alpha}\rightarrow Q\hat{a}_{\alpha}; \
\hat{a}_{\dot\mu}\rightarrow Q\hat{a}_{\dot\mu}.
\eea
The non-commutativity parameter can be linked to the scaling \cite{Ho:2013opa}.
This scaling can be determined by examining a specific relationship between the NS-NS and R-R parameters \cite{Ho:2013opa}.
We apply the electromagnetic duality to implement the S-duality from the R-R D3-brane theory to the NS-NS D3-brane theory up to the first order of the non-commutativity parameter \cite{Ho:2013opa}.
The second-order theory reproduces the complete NS-NS D3-brane theory. However, the computation method is similar to that of the first-order \cite{Ho:2013opa}.
Therefore, we only show the computation details up to the first order \cite{Ho:2013opa}.

\subsubsection{Electromagnetic Duality}
\noindent
We first show how to perform the electromagnetic duality up to the first order in a general YM-type theory \cite{Yang:1954ek} by perturbation.
If the action is given by
\bea
S_1 = g_{3}^2 \int d^4x \ \bigg(-\frac{1}{4}F_{AB}F^{AB} + \theta^{\dot{1}\dot{2}} Q_1(F_{AB})\bigg),
\eea
where the spacetime indices are denoted by $A$ and $B$.
We can introduce another term
\bea
S_2=\int d^4x\ \bigg(-\frac{g_{3}^2}{4}F_{AB}F^{AB}+g_{3}^2\theta^{\dot1\dot2} Q_1(F_{AB})+\frac{1}{2}\tilde{G}_{AB}F^{AB}\bigg),
\eea
where
\bea
G_{AB}=\frac{1}{2}\epsilon_{ABCD}\tilde{G}^{CD},
\eea
to promote $F_{AB}$ to an unconstrained field.
Integrating $G=dB$ yields $dF=0$.
Therefore, we solve $dF=0$ to obtain $F=dA$ and return to the original theory.
We vary $F_{AB}$ to obtain
\bea
F_{AB}=\frac{1}{g_{YM}^2}\tilde{G}_{AB}+\theta^{\dot1\dot2}\frac{\delta Q_1}{\delta F^{AB}}+{\cal O}\big((\theta^{\dot1\dot2})^2\big).
\eea
Hence, it provides
\bea
S_3=\int d^4x\ \bigg\lbrack\frac{1}{4g_{3}^2}\tilde{G}_{AB}\tilde{G}^{AB}
+g_{3}^2\theta^{\dot1\dot2}Q_1\bigg(\frac{1}{g_{3}^2}\tilde{G}_{AB}\bigg)
\bigg\rbrack,
\eea
in which we use
\bea
Q_1(F_{AB})= Q_1\bigg(\frac{1}{g_{YM}^2}\tilde{G}_{AB}\bigg)+{\cal O}(\theta^{\dot 1\dot 2}).
\eea
We write the action in terms of $G_{AB}$ at the zeroth order
\bea
S_3=\int d^4x~\bigg( -\frac{1}{4g_{3}^2}G_{AB}G^{AB}
\bigg).
\eea

\subsubsection{First-Order Correction}
\noindent
We use the above electromagnetic duality formula to go from the R-R D3-brane to the NS-NS D3-brane \cite{Ho:2015mfa}.
This method relies on the fact that the action depends only on the abelian field strength.
For this goal, we fix \cite{Ho:2015mfa}:
\bea
B^{\dot 1}=b^{\dot 2}=0.
\eea
To have the YM term in the zeroth order in the R-R D3-brane, we integrate out the $\hat{b}^{\dot\mu}$ before performing the electromagnetic duality \cite{Ho:2011yr}.
This result of integration is equivalent to setting \cite{Ho:2011yr}
\bea
\hat{H}_{\dot1\dot2}=-\hat{F}_{01}+{\cal O}(\theta^{\dot1\dot2}).
\eea
\\

\noindent
The R-R D3-brane up the first order is given by \cite{Ho:2015mfa}:
\bea
&&
S_{\mathrm{RRD31}}
\nn\\
&=&
g_{3}^2\int d^4xd^2y~\bigg(\epsilon^{\alpha\beta}\hat{H}_{\dot1\dot2}\hat{F}_{\alpha\dot1}\partial_{\beta}\hat{b}^{\dot1}
-\frac{1}{2}\epsilon_{\alpha\beta}\hat{F}_{\dot\mu\dot\nu}\hat{F}^{\alpha\dot\mu}\hat{F}^{\beta\dot\nu}
+\hat{F}_{\alpha\dot\mu}\hat{F}^{\alpha}{}_{\dot1}\partial^{\dot\mu}\hat{b}^{\dot1}
-\hat{F}_{\alpha\dot\mu}\hat{F}^{\dot\mu\dot1}\partial^{\alpha}\hat{b}_{\dot1}\bigg).
\nn\\
\eea
Due to the gauge fixing, we can rewrite the gauge potentials in terms of the abelian field strength \cite{Ho:2015mfa}
\bea
\hat{b}^{\dot1}=\partial^{\dot1}\partial_{\dot1}^{-2}\hat{H}_{\dot1\dot2}.
\eea
After the electromagnetic duality, the action at the first order as \cite{Ho:2015mfa}
\bea
&&
S_{\mathrm{RRD31D}}
\nn\\
&=&\frac{1}{g_{3}^2}\int d^2xd^2y~\bigg(\epsilon^{\alpha\beta}\tilde{\hat{G}}_{01}\tilde{\hat{G}}_{\alpha\dot 1}\partial_{\beta}\partial^{\dot 1}\partial_{\dot 1}^{-2}\tilde{\hat{G}}_{01}
-\frac{1}{2}\epsilon_{\alpha\beta}\tilde{\hat{G}}_{\dot\mu\dot\nu}\tilde{\hat{G}}^{\alpha\dot\mu}\tilde{\hat{G}}^{\beta\dot\nu}
\nn\\
&&
-\tilde{\hat{G}}_{\alpha\dot\mu}\tilde{\hat{G}}^{\alpha}{}_{\dot 1}\partial^{\dot\mu}\partial^{\dot 1}\partial_{\dot 1}^{-2}\tilde{\hat{G}}_{01}
+\tilde{\hat{G}}_{\alpha\dot\mu}\tilde{\hat{G}}^{\dot\mu\dot 1}\partial^{\alpha}\partial_{\dot 1}\partial_{\dot 1}{}^{-2}\tilde{\hat{G}}_{01}\bigg).
\eea
We can rewrite all terms as in the following \cite{Ho:2015mfa}:
\bea
&&
\frac{1}{g_{3}^2}\int d^2xd^2y~\bigg(-\epsilon^{\alpha\beta}\epsilon_{\alpha\gamma}\hat{G}^{\dot1\dot2}
\hat{G}^{\gamma\dot2}\partial_{\beta}\partial^{\dot 1}\partial_{\dot 1}^{-2}\hat{G}^{\dot1\dot2}\bigg)
\nn\\
&=&\frac{1}{g_{3}^2}\int  d^2xd^2y~\bigg(-\epsilon^{\alpha\beta}\epsilon_{\alpha\gamma}
(\partial_{\dot1}\hat{B}_{\dot2})
\hat{G}^{\gamma\dot2}
(\partial_{\beta}\partial^{\dot 1}\partial_{\dot 1}^{-2}\partial_{\dot1}\hat{B}_{\dot2})\bigg)
\nn\\
&=&\frac{1}{g_{3}^2}\int d^2xd^2y~\bigg(\hat{G}^{\beta\dot2}
(\partial_{\dot1}\hat{B}_{\dot2})
(\partial_{\beta}\hat{B}_{\dot2})\bigg);
\nn\\
&&
\frac{1}{g_{3}^2}\int d^2xd^2y~\bigg(-\epsilon_{\alpha\beta}\tilde{G}_{\dot1\dot2}\tilde{G}^{\alpha\dot1}
\tilde{G}^{\beta\dot2}\bigg)
\nn\\
&=&\frac{1}{g_{3}^2}\int d^2xd^2y~\bigg(\epsilon^{\beta\delta}\hat{G}^{01}\hat{G}^{\beta}{}_{\dot2}\hat{G}_{\delta\dot1}\bigg)
\nn\\
&=&\frac{1}{g_{3}^2}\int d^2xd^2y~\bigg(-\frac{1}{2}\hat{G}^{\alpha\beta}\{\hat{B}_{\alpha}, \hat{B}_{\beta}\}
-\epsilon^{\beta\delta}\hat{G}^{01}(\partial_{\dot 1}\hat{B}_{\delta})(\partial_{\beta}\hat{B}_{\dot2})\bigg);
\nn\\
&&
\frac{1}{g_{3}^2}\int d^2xd^2y~\bigg(\epsilon_{\alpha\dot\mu\beta\dot\nu}\epsilon^{\alpha\dot1\gamma\dot2}
\hat{G}^{\beta\dot\nu}\hat{G}_{\gamma\dot2}
\partial^{\dot\mu}\partial^{\dot1}\partial_{\dot1}^{-2}\hat{G}^{\dot1\dot2}\bigg)
\nn\\
&=&\frac{1}{g_{3}^2}\int d^2xd^2y~\bigg(\epsilon_{\dot\mu\dot\nu}\epsilon_{\alpha\beta}\epsilon^{\alpha\gamma}\hat{G}^{\beta\dot\nu}\hat{G}_{\gamma\dot2}
\partial^{\dot\mu}\partial^{\dot1}\partial_{\dot1}^{-2}\hat{G}^{\dot1\dot2}\bigg)
\nn\\
&=&\frac{1}{g_{3}^2}\int d^2xd^2y~\bigg(-(\partial^{\dot 1}\hat{B}^{\gamma})\hat{G}_{\gamma\dot2}(\partial^{\dot2}\hat{B}_{\dot2})
-
\hat{G}^{\gamma\dot2}\hat{G}_{\gamma\dot2}\partial^{\dot1}\hat{B}^{\dot2}\bigg);
\nn\\
&&
\frac{1}{g_{3}^2}\int d^2xd^2y~\bigg(\tilde{\hat{G}}_{\alpha\dot2}\tilde{\hat{G}}^{\dot2\dot1}\partial^{\alpha}\partial_{\dot1}
\partial_{\dot1}^{-2}\tilde{\hat{G}}_{01}\bigg)
\nn\\
&=&\frac{1}{g_{3}^2}\int d^2xd^2y~\bigg(-\tilde{\hat{G}}_{\alpha\dot2}\hat{G}^{01}\partial^{\alpha} \hat{B}_{\dot2}\bigg)
\nn\\
&=&\frac{1}{g_{3}^2}\int d^2xd^2y~\bigg(\epsilon_{\alpha\beta}\hat{G}^{01}(\partial_{\dot1}\hat{B}^{\beta})(\partial^{\alpha}\hat{B}_{\dot2})\bigg).
\eea
We then combine all four terms to obtain \cite{Ho:2015mfa}
\bea
S_{\mathrm{NSNSD31}}=\frac{1}{g_{3}^2}\int d^2xd^2y\ \bigg(-\hat{G}^{\alpha\dot2}\{\hat{B}_{\alpha}, \hat{B}_{\dot2}\}
-\frac{1}{2}\hat{G}^{\alpha\beta}\{\hat{B}_{\alpha}, \hat{B}_{\beta}\}\bigg).
\eea
This is the first-order correction of the NS-NS D3-brane theory, based on our chosen gauge fixing \cite{Ho:2015mfa}.
This computation of the electromagnetic duality is particularly interesting because it establishes the S-duality connection in the non-commutative theory without using the SW map \cite{Ho:2015mfa}.

\section{R-R D-Branes}
\label{sec:7}
\noindent
We extend the U(1) symmetry to U($N$) symmetry for generalizing from a single R-R D$p$-brane to R-R D$p$-branes \cite{Ma:2023hgi}.
Since the VPD gauge symmetry governs the large R-R field background, we keep it as the U(1) parameter as in the single D-brane case \cite{Ma:2023hgi,Ma:2020msx}.
Because the D4-brane's 2-form gauge potential is dual to the 1-form gauge potential with the U($N$) gauge group, we generalize the gauge group of the ($p-2$)-form gauge potential to the U($N$) gauge group \cite{Ma:2023hgi,Ma:2020msx}.
We first show the gauge symmetry \cite{Ma:2023hgi,Ma:2020msx}, and then the Lagrangian formulation \cite{Ma:2023hgi} and also how the U($N$) YM gauge theory appears \cite{Ma:2023hgi}.

\subsection{Gauge Symmetry}
\noindent
We generalize the gauge transformation of the R-R D$p$-branes theory as \cite{Ma:2023hgi}:
\bea
\hat{\delta}_{\hat{\Lambda}}\hat{b}^{\dot{\mu}}&=&\hat{\kappa}^{\dot{\mu}}
+i\lbrack\hat{\lambda}, \hat{b}^{\dot\mu}\rbrack
+\theta\hat{\kappa}^{\dot{\nu}}\partial_{\dot{\nu}}\hat{b}^{\dot{\mu}};
\nn\\
\hat{\delta}_{\hat{\Lambda}}\hat{B}_{\alpha}{}^{\dot{\mu}}&=&\partial_{\alpha}\hat{\kappa}^{\dot{\mu}}
+i\lbrack\hat{\lambda}, \hat{B}_{\alpha}{}^{\dot\mu}\rbrack
+\theta\big(\hat{\kappa}^{\dot{\nu}}\partial_{\dot{\nu}}\hat{B}_{\alpha}{}^{\dot{\mu}}-(\partial_{\dot{\nu}}\hat{\kappa}^{\dot{\mu}})\hat{B}_{\alpha}{}^{\dot{\nu}}\big);
\nn\\
\hat{\delta}_{\hat{\Lambda}}\hat{a}_{\alpha}&=&\partial_{\alpha}{\hat{\lambda}}
+i\lbrack\hat{\lambda}, \hat{a}_{\alpha}\rbrack
+\theta(\hat{\kappa}^{\dot{\nu}}\partial_{\dot{\nu}}\hat{a}_{\alpha}
+\hat{a}_{\dot{\nu}}\partial_{\alpha}\hat{\kappa}^{\dot{\nu}});
\nn\\
\hat{\delta}_{\hat{\Lambda}}\hat{a}_{\dot{\mu}}&=&\partial_{\dot{\mu}}{\hat{\lambda}}
+i\lbrack\hat{\lambda}, \hat{a}_{\dot\mu}\rbrack
+\theta(\hat{\kappa}^{\dot{\nu}}\partial_{\dot{\nu}}\hat{a}_{\dot{\mu}}
+\hat{a}_{\dot{\nu}}\partial_{\dot{\mu}}\hat{\kappa}^{\dot{\nu}}),
\eea
where $\hat{B}_{\alpha}{}^{\dot{\mu}}$ satisfies that \cite{Ma:2023hgi}
\bea
&&\hat{V}_{\dot{\mu}}{}^{\dot{\nu}}(\lbrack\hat{D}^{\alpha}, \hat{b}_{\dot{\nu}}\rbrack-\hat{V}^{\dot{\rho}}{}_{\dot{\nu}}\hat{B}^{\alpha}{}_{\dot{\rho}})
+\epsilon^{\alpha\beta}\hat{F}_{\beta\dot{\mu}}
+\theta\epsilon^{\alpha\beta}\hat{F}_{\dot{\mu}\dot{\nu}}\hat{B}_{\beta}{}^{\dot{\nu}}=0.
\eea
The $\hat{V}_{\dot\nu}{}^{\dot\mu}$ is generalized as \cite{Ma:2023hgi}
\bea
\hat{V}_{\dot\mu}{}^{\dot\mu}=\delta_{\dot\nu}{}^{\dot\mu}
+\theta\lbrack\hat{D}_{\dot\nu}, \hat{b}^{\dot\mu}\rbrack,
\eea
where
\bea
\lbrack\hat{D}_{\dot\nu}, \hat{b}^{\dot\mu}\rbrack=\partial_{\dot\nu}\hat{b}^{\dot\mu}
-i\lbrack\hat{a}_{\dot\nu}, \hat{b}^{\dot\mu}\rbrack.
\eea
The gauge parameter $\hat{\kappa}$ generates the VPD symmetry, and the $\hat{\lambda}$ generates the U($N$) symmetry.

\subsection{Lagrangian}
\noindent
To generalize from single R-R D-brane to the multiple D-branes, the ($p-1$)-bracket needs to be generalized to \cite{Ma:2023hgi}
\bea
\{\hat{f}_1, \hat{f}_2 ,\cdots, \hat{f}_{p-1}\}_{(p-1)}\equiv \epsilon^{\dot{\mu}_1\dot{\mu}_2\cdots\dot{\mu}_{p-1}}\lbrack\hat{D}_{\dot{\mu}_1}, \hat{f}_1\rbrack\lbrack\hat{D}_{\dot{\mu}_2}, \hat{f}_2\rbrack\cdots\lbrack\hat{D}_{\dot{\mu}_{p-1}}, \hat{f}_{p-1}\rbrack.
\eea
The Lagrangian formulation of the gauge sector is presented as follows \cite{Ma:2023hgi}
\bea
S_{\mathrm{RRDPS}}=S_1+S_2,
\eea
where
\bea
&&
S_1
\nn\\
&=&\frac{T_p}{\theta^{p-2}}\int d^2xd^{p-1}y\ \mathrm{Str}\bigg(\frac{1}{2}\hat{{\cal F}}_{\alpha\dot\mu}\hat{{\cal F}}^{\alpha\dot\mu}
+\frac{1}{2\theta}\epsilon^{\alpha\beta}\hat{{\cal F}}_{\alpha\beta}\bigg);
\nn\\
&&
S_2
\nn\\
&=&\frac{T_p}{\theta^{p-2}}\int d^2xd^{p-1}y\ \mathrm{Str}\bigg\lbrack
\sum_{n, m, l\in S}\bigg(-\frac{1}{2^{m+1}}\frac{\theta^{2(p-2-m)}}{(n!)(m!)^2}
\nn\\
&&\times
(\{\hat{X}^{\dot\mu_1}, \hat{X}^{\dot\mu_2}, \cdots, \hat{X}^{\dot\mu_n}, \hat{D}_{\dot\nu_1}, \hat{D}_{\dot\nu_2}, \cdots, \hat{D}_{\dot\nu_m}, y^{\dot\nu_1}, y^{\dot\nu_2}, \cdots, y^{\dot\nu_m}\})^2\bigg)\bigg\rbrack.
\eea
The set is \cite{Ma:2023hgi}
\bea
S\equiv\{(n, m)| n, m\ge 0; \ n+2m=p-1\}.
\eea
The covariant field strength that we still need to modify is \cite{Ma:2023hgi}
\bea
\hat{{\cal F}}_{\alpha\beta}=\hat{F}_{\alpha\beta}
-\theta(\hat{F}_{\alpha\dot{\mu}}\hat{B}_{\beta}{}^{\dot{\mu}}+\hat{F}_{\dot{\mu}\beta}\hat{B}_{\alpha}{}^{\dot{\mu}})
+\frac{\theta^2}{2}\hat{F}_{\dot{\mu}\dot{\nu}}(\hat{B}_{\alpha}{}^{\dot{\mu}}\hat{B}_{\beta}{}^{\dot{\nu}}+\hat{B}_{\beta}{}^{\dot{\nu}}\hat{B}_{\alpha}{}^{\dot{\mu}}).
\eea
The generalization is non-trivial because we lose the generalized Jacobi identity in the non-Abelian sector. However, we still have the closed gauge transformation \cite{Ma:2023hgi}.
We can begin with the D9-branes and then perform T-duality to introduce the transverse scalars; however, the expression cannot be written as a finite term for a generic $p$ \cite{Ma:2023hgi}.
In the next section, we will show how the U($N$) YM theory appears \cite{Ma:2023hgi}.

\subsection{U($N$) YM}
\noindent
The covariant field strengths up to the leading order for the non-commutativity parameter are \cite{Ma:2023hgi}:
\bea
\hat{{\cal H}}_{\dot\mu_1\dot\mu_2\cdots \dot\mu_{p-1}}&=&\hat{H}_{\dot\mu_1\dot\mu_2\cdots \dot\mu_{p-1}}+{\cal O}(\theta)\equiv \lbrack\hat{D}_{\dot{\mu}}, \hat{b}^{\dot{\mu}}\rbrack+{\cal O}(\theta);
\nn\\
\hat{{\cal F}}_{\dot{\mu}\dot{\nu}}&=&\hat{F}_{\dot{\mu}\dot{\nu}}+{\cal O}(\theta);
\nn\\
\hat{{\cal F}}_{\alpha\dot{\mu}}&=&\hat{F}_{\alpha\dot{\mu}}+{\cal O}(\theta);
\nn\\
\frac{1}{2\theta}\epsilon^{\alpha\beta}\hat{{\cal F}}_{\alpha\beta}&=&
\frac{1}{2\theta}\epsilon^{\alpha\beta}\hat{F}_{\alpha\beta}
-\epsilon^{\alpha\beta}\hat{F}_{\alpha\dot{\mu}}
(\lbrack\hat{D}_{\beta}, \hat{b}^{\dot{\mu}}\rbrack+\epsilon_{\beta\gamma}\hat{F}^{\gamma\dot{\mu}})
+{\cal O}(\theta),
\eea
where $\hat{F}$ is now the conventional non-Abelian field strength.
We observe a similar result to the single D-brane from the following \cite{Ma:2023hgi}:
\bea
&&
-\frac{T_p}{\theta^{p-2}}\int d^2xd^{p-1}y\ \mathrm{Str}(\epsilon^{\alpha\beta}\hat{F}_{\alpha\dot\mu}\lbrack\hat{D}_{\beta}, \hat{b}^{\dot\mu}\rbrack)
\nn\\
&=&-\frac{T_p}{\theta^{p-2}}\int d^2xd^{p-1}y\ \mathrm{Str}(\epsilon^{\alpha\beta}\lbrack\hat{D}_{\alpha}, \hat{D}_{\dot\mu}\rbrack\lbrack\hat{D}_{\beta}, \hat{b}^{\dot\mu}\rbrack)
\nn\\
&=&\frac{T_p}{\theta^{p-2}}\int d^2xd^{p-1}y\ \mathrm{Str}(\epsilon^{\alpha\beta}\lbrack\hat{D}_{\beta}, \lbrack\hat{D}_{\alpha}, \hat{D}_{\dot\mu}\rbrack\rbrack\hat{b}^{\dot\mu})
\nn\\
&=&-\frac{T_p}{\theta^{p-2}}\int d^2xd^{p-1}y\ \mathrm{Str}\bigg(\frac{1}{2}\epsilon^{\alpha\beta}\hat{F}_{\alpha\beta}\lbrack\hat{D}_{\dot\mu}, \hat{b}^{\dot\mu}\rbrack\bigg)
\nn\\
&=&-\frac{T_p}{\theta^{p-2}}\int d^2xd^{p-1}y\ \mathrm{Str}(\hat{F}_{01}\lbrack\hat{D}_{\dot\mu}, \hat{b}^{\dot\mu}\rbrack).
\eea
In the third equality, we not only use the partial integration by part but also uses \cite{Ma:2023hgi}
\bea
\lbrack\hat{D}_{\beta}, \lbrack\hat{D}_{\alpha}, \hat{D}_{\dot\mu}\rbrack\rbrack
+\lbrack\hat{D}_{\alpha}, \lbrack\hat{D}_{\dot\mu}, \hat{D}_{\beta}\rbrack\rbrack
+\lbrack\hat{D}_{\dot\mu}, \lbrack\hat{D}_{\beta}, \hat{D}_{\alpha}\rbrack\rbrack=0.
\eea
The Lagrangian description of R-R D-branes up to the leading order for the non-commutativity parameter is \cite{Ma:2023hgi}
\bea
&&
S_{\mathrm{RRDPS0}}
\nn\\
&=&
\frac{T_p}{\theta^{p-2}}
\int d^2xd^{p-1}y\ \mathrm{Str}\bigg(-\frac{1}{2}(\lbrack\hat{D}_{\dot\mu}, \hat{b}^{\dot\mu}\rbrack)^2
-\frac{1}{4}\hat{F}_{\dot{\mu}\dot{\nu}}\hat{F}^{\dot{\mu}\dot{\nu}}
-\frac{1}{2}\hat{F}_{\alpha\dot{\mu}}\hat{F}^{\alpha\dot{\mu}}
-\hat{F}_{01}\lbrack\hat{D}_{\dot\mu}, \hat{b}^{\dot\mu}\rbrack\bigg).
\nn\\
\eea
Integrating out the non-dynamical field $\hat{b}^{\dot\mu}$ is equivalent to substituting \cite{Ma:2023hgi}
\bea
\lbrack\hat{D}_{\dot\mu}, \hat{b}^{\dot\mu}\rbrack=-\hat{F}_{01}+\hat{f},
\eea
where
\bea
\hat{D}_{\dot{\mu}}\hat{f}=0.
\eea
By choosing a proper boundary condition at the infinite point, we can choose $\hat{f}=0$ \cite{Cronstrom:2005wt,Ma:2023hgi},
\bea
S_{\mathrm{U(N)YM}}=\frac{T_p}{\theta^{p-2}}\int d^2xd^{p-1}y\ \mathrm{Str}\bigg(-\frac{1}{4}\hat{F}_{\dot\mu\dot\nu}\hat{F}^{\dot\mu\dot\nu}
-\frac{1}{2}\hat{F}_{\alpha\dot\mu}\hat{F}^{\alpha\dot\mu}
-\frac{1}{4}\hat{F}_{\alpha\beta}\hat{F}^{\alpha\beta}\bigg).
\eea
Hence, the classical Lagrangian up to the leading order for the non-commutativity parameter respects the U($N$) YM description \cite{Ma:2023hgi}.

\section{Outlook and Future Directions}
\label{sec:8}
\noindent
In this review, we have discussed various aspects of D-branes and their deep connections to string theory.
One of the most intriguing open questions in this direction is the Lagrangian formulation of multiple M5-branes, particularly its relationship to D-branes from the perspective of string theory.
Such a formulation would provide important insight into the existence and structure of M-theory, viewed through symmetry principles analogous to those that give rise to electromagnetism from U(1) gauge symmetry.
\\

\noindent
The recent development of R–R D-branes \cite{Ma:2023hgi} brings us closer to understanding M5-branes by providing strong constraints through T-duality.
A complete Lagrangian formulation of M5-branes should naturally extend the existing duality web from the single-brane case \cite{Ho:2013opa,Ho:2015mfa,Ho:2013paa} to systems of coincident branes.
Although various no-go theorems have long seemed to forbid such a Lagrangian description, the loss of Lorentz symmetry in the large-field background limit offers a potential loophole.
Another traditional difficulty is the absence of a free parameter in M-theory.
However, in the large-field background limit, an additional non-commutativity parameter appears, providing a perturbative expansion parameter and circumventing the lack of free parameters.
\\

\noindent
Another promising avenue involves generalizing the Nambu–Poisson (NP) bracket.
In general, it is impossible to preserve both volume-preserving diffeomorphism (VPD) symmetry and the Jacobi identity simultaneously.
An exciting possibility is to retain the VPD symmetry while abandoning the Jacobi identity.
However, no nontrivial example of such a generalization has yet been constructed.
When generalizing R–R D-branes to coincident branes, the resulting theory loses the Jacobi identity but retains VPD symmetry \cite{Ma:2023hgi}.
Extending this line of work to more general settings, such as curved backgrounds, would be highly valuable.
This direction is relevant not only to string theory.
However, it may also have applications in condensed matter physics, especially in the study of the quantum Hall effect.
Furthermore, establishing modified algebraic structures of this kind could have interesting implications for representation theory.
\\

\noindent
Another important open question concerns the Dirac–Born–Infeld (DBI) description of R–R D-branes.
From the worldsheet formulation, one can compute the $\beta$-function to determine how the closed-string sector couples to the DBI action \cite{Fradkin:1985qd,Callan:1986bc}.
However, the analogue of the DBI action for R–R D-branes remains unknown.
By applying S-duality to the NS–NS D3-brane, one may be able to derive a consistent DBI description for R–R D-branes.
Such a result would clarify the open–closed string correspondence and enable perturbative expansions with finite parameters \cite{Seiberg:1999vs}.
Because the NS-NS D3-brane is only up to the second order of the non-commutativity parameter, the R-R D3-brane has an infinite order \cite{Ho:2013opa}.
Therefore, beyond the second-order for the non-commutativity parameter, the result of the S-duality should be exciting.
S-duality should give another constraint on the construction of the R-R D$p$-branes.
\\

\noindent
The extension to curved backgrounds is also of great interest, particularly for understanding how curvature effects appear in the BLG model \cite{Bagger:2006sk,Gustavsson:2007vu,Bagger:2007jr,Bagger:2007vi}.
This could determine whether the same number of supercharges is preserved as in the flat-space case.
Since we are dealing with BPS configurations, it is consistent to set the fermionic fields to zero and focus on the bosonic sector. Nevertheless, supersymmetry remains an essential consistency condition for R–R D-brane theories.
Currently, the supersymmetry transformation is explicitly known only for the R–R D4-brane \cite{Ma:2012hc}.
Extending it to general D$p$-branes via T-duality and, further, to coincident branes, represents a natural next step.
\\

\noindent
R–R D$p$-branes are characterized by the ($p-1$)-bracket structure \cite{Ma:2023hgi}, which may be realized as a Lie-($p-1$) algebra.
It should be possible to construct an effective action for D$q$–D$p$ brane systems using techniques similar to those applied to the infinite M2-brane construction \cite{Ho:2008nn}.
Such developments would deepen our understanding of effective theories in various large-field background limits.

\section*{Acknowledgments}
\noindent
The author thanks Nan-Peng Ma for his encouragement.
%CTM acknowledges the Nuclear Physics Quantum Horizons program through the Early Career Award (Grant No. DE-SC0021892); 
%YST Program of the APCTP;  
%Post-Doctoral International Exchange Program; 
%China Postdoctoral Science Foundation, Postdoctoral General Funding: Second Class (Grant No. 2019M652926); 
%Foreign Young Talents Program (Grant No. QN20200230017). 
%The author thanks the National Tsing Hua University, the Institute for Advanced Study at the Tsinghua University, and the Center for Quantum Science at the Sogang University. 
%Discussion during the workshops, ``East Asia Joint Workshop on Fields and Strings 2019'' and ``The 17th Italian-Korean Symposium for Relativistic Astrophysics'', was helpful to this work. 

%\appendix

\end{document}